\documentclass[a4paper,11pt]{article}
\usepackage{jcappub} 

\usepackage{lineno}
\usepackage[normalem]{ulem}
\usepackage{enumitem}

\DeclareMathOperator*{\SumInt}{%
\mathchoice%
  {\ooalign{$\displaystyle\sum$\cr\hidewidth$\displaystyle\int$\hidewidth\cr}}
  {\ooalign{\raisebox{.14\height}{\scalebox{.7}{$\textstyle\sum$}}\cr\hidewidth$\textstyle\int$\hidewidth\cr}}
  {\ooalign{\raisebox{.2\height}{\scalebox{.6}{$\scriptstyle\sum$}}\cr$\scriptstyle\int$\cr}}
  {\ooalign{\raisebox{.2\height}{\scalebox{.6}{$\scriptstyle\sum$}}\cr$\scriptstyle\int$\cr}}
}

\title{\boldmath{Multiple Soft Scatterings in Scalar Dark Matter Freeze-In }}

\author[a,b]{M. Becker}
\author[c]{\!\!\!, M. J. Fernández Lozano}
\author[c]{\!\!\!, J. Harz}
\author[c]{and C. Tamarit}
\affiliation[a]{Dipartimento di Fisica e Astronomia, Universit\`a degli Studi di Padova, Via Marzolo 8, 35131 Padova, Italy}
\affiliation[b]{
INFN, Sezione di Padova, Via Marzolo 8, 35131 Padova, Italy}
\affiliation[c]{PRISMA$^+$ Cluster of Excellence \& Mainz Institute for Theoretical Physics,\\ FB 08 - Physics,
Mathematics and Computer Science, Johannes Gutenberg-Universität Mainz, Staudingerweg 9, 550099 Mainz, Germany}

\emailAdd{mathias.becker@unipd.it}
\emailAdd{mfernnde@uni-mainz.de}  
\emailAdd{jharz@uni-mainz.de}    
\emailAdd{ctamarit@uni-mainz.de}

\abstract{
We present an improved calculation of the freeze-in production rate for scalar dark matter (DM) from a gauge-charged parent particle via a renormalizable interaction. 
Building on the previously developed 1PI-resummed framework to accurately capture the relevant regime $T \sim M$~\cite{Copello}, we expand the analysis to include the Landau-Pomeranchuk-Migdal (LPM) effect, which contributes at leading order $g^2 T$ to the interaction rate in the ultra-relativistic limit. To this end, we derive an equation for the LPM rate of a scalar particle for the first time and combine it with the previous 1PI results, providing a new state-of-the art calculation. 
In contrast to the 1PI results, the LPM treatment neglects vacuum mass scales such that a phenomenological switch-off function between the ultra-relativistic and non-relativistic regime is required. We propose a new function motivated by a thermal loop contribution and compare it to other approaches in the literature, quantifying the resulting uncertainty of this method.
Depending on the gauge coupling and mass splitting between DM and mediator particles, the LPM effect contributes between $1\%$ and $27\%$ to the relic density, with the impact increasing for larger gauge couplings and smaller mass splittings. 
Additionally, we compare our results to commonly used semi-classical Boltzmann approaches. For instance, when these include decays and scatterings regulated with thermal masses, we find deviations ranging from -30\% to +20\% depending on the mass splitting.
Finally, we compare to results based on hard-thermal-loop (HTL) approximations.
}

\begin{document}

\begin{flushright}
    MITP-25-046
\end{flushright}
\vskip-13.2pt
\vskip-13.2pt
\maketitle
\flushbottom

\FloatBarrier
\section{Introduction} \label{sec:intro}

The existence of Dark Matter (DM) \cite{Planck2018,Bertone_2018,Cirelli:2024ssz} remains one of the major open questions in modern physics. Since it cannot be explained within the Standard Model (SM), its presence points to the need for theories that extend beyond the SM framework. Over the past decades, the most commonly discussed candidate, the Weakly Interacting Massive Particle (WIMP), has been exhaustively searched for but has not been detected so far~\cite{Arcadi_2018,Roszkowski_2018,PhysRevLett.131.041003,PhysRevLett.131.041002}. 
This has induced a growing interest in alternatives, particularly feebly interacting massive particles (FIMPs)~\cite{Hall_2010,Bernal_2017}, which feature even weaker interactions with the SM such that a non-observation at direct detection experiments seems more consistent with expectations. 
Their production in the early universe from the SM plasma proceeds via the \emph{freeze-in} mechanism, in which DM remains out-of-equilibrium with the SM bath at all times \cite{Bernal_2017,Hall_2010,D_Eramo_2018}. 

The production dynamics of FIMPs take place at temperatures $T \gtrsim M$, where $M$ denotes the largest mass scale involved in the process. 
This production is especially relevant in the relativistic regime, when $T \sim M$. 
This is in contrast to the production mechanism of WIMPs, whose relic abundance is set by thermal \emph{freeze-out} at later stages in the universe’s thermal history, typically when the annihilating particles are non-relativistic.
Consequently, accurately predicting the relic abundance of FIMPs requires incorporating finite-temperature corrections to their production rates. 
These effects include, for example, the correct implementation of quantum statistics, such as Bose-Einstein or Fermi-Dirac distributions, as well as thermal masses induced by the plasma, which can open or close kinematically allowed decay and scattering channels or regularize infrared (IR) divergences that would otherwise appear in vacuum calculations. 
Thermal environments alter not only particle dispersion relations, but also modify interaction rates through coherence effects, screening, and collective excitations. 
Neglecting these corrections can lead to substantial inaccuracies in the estimation of dark matter production rate.

Recently, these effects have been extensively addressed in Refs.~\cite{Biondini,Copello}. 
Ref.~\cite{Copello} applied a first-principles treatment for the evaluation of the dark matter production rate in particular around temperatures $T \sim M$, the regime most relevant for freeze-in production of dark matter from renormalizable interactions.
This approach, which used 1PI-resummed propagators (where 1PI stands for one-particle-irreducible), consistently includes the vacuum mass scales throughout the calculation and hence remains valid in this important, intermediate regime  $T \sim M$, where commonly used resummation schemes such as the hard-thermal-loop (HTL)  approach \cite{PhysRevLett.63.1129,Braaten1990SoftAI,Kraemmer:2003gd} are not justified.  While 1PI-resummation techniques have been explored in the literature before for other early universe processes such as leptogenesis \cite{Prokopec:2003pj,Prokopec:2004ic,Garbrecht:2019zaa} and gravitino and axion production \cite{Rychkov:2007uq,Salvio:2013iaa}, the previous works did not consider a full one-loop resummation in all kinematic regimes, either using simplified propagators in the timelike region, or resorting to the HTL approximation. Ref.~\cite{Copello} was the first to apply a full 1PI resummation, while in the context of axion-like particles a similar treatment has been used in Ref.~\cite{Becker:2025yvb}.
An additional advantage of the 1PI approach is that it naturally accounts for the main $2\leftrightarrow2$ scattering processes, which become particularly relevant in the ultra-relativistic regime. However, while Ref.~\cite{Copello} considered so far only the leading-order contribution in the loop expansion of the 2PI-effective action, some contributions of the same order $\mathcal{O}(g^2T^2$) in the gauge coupling constant arising from higher loop orders are still missing. In particular, multiple soft scatterings  with gauge bosons in the plasma--the so-called Landau-Pomeranchuk-Midgal (LPM) effect \cite{Landau:1953gr, Landau:1953um, PhysRev.103.1811, Aurenche_1998, Aurenche_2000, Aurenche_2002, Arnold, Arnold_2001, Arnold_2002, Arnold_20012}--contribute at  $\mathcal{O}(g^2T^2$). While Ref.~\cite{Copello} gave simple estimates of their impact based on the results of Ref.~\cite{Biondini}, a dedicated computation was missing, which is the objective of this work.
On the other hand, Ref.~\cite{Biondini} had considered -- in the case of a fermionic dark matter model -- an approach based on the HTL to account for all leading order effects in the ultra-relativistic limit ($T \gg M$), i.e., $2 \leftrightarrow 2$ scatterings and the LPM effect. These contributions were manually switched off when transitioning to the non-relativistic regime, allowing for a unified evaluation of the interaction rate across all temperatures. 
In the intermediate regime $T \sim M$, however, the computation still relies on the HTL approximation, which itself is only accurate if $T \gg M$(see detailed discussion in Ref.~\cite{Copello}).

In view of the above, the aim of this paper is to extend the results of Ref.~\cite{Copello} for scalar dark matter by including the contribution of the LPM effect to the production rate calculated with 1PI-resummed propagators. 
This represents the first calculation of the LPM effect for a scalar FIMP, as Ref.~\cite{Biondini} focused on fermionic DM candidates. 
We present its impact on the relic density and compare these results with those based on other approaches frequently used in the literature.
Hereby, the calculation of the LPM contribution follows the same effective formalism used in Ref.~\cite{Biondini}, which is itself based on  works in the context of photon production from a quark-gluon plasma~\cite{Landau:1953gr, Landau:1953um, PhysRev.103.1811, Aurenche_1998, Aurenche_2000, Aurenche_2002, Arnold, Arnold_2001, Arnold_2002, Arnold_20012}, and subsequent applications in thermal leptogenesis~\cite{Anisimov_2011, Besak_2012, Ghisoiu_2014, Ghiglieri_2016, Besak_2010, Besak, Hutig:2013oka, Depta:2020zmy}.
We emphasize that the calculation of the LPM effect neglects vacuum mass scales and is strictly valid only in the ultra-relativistic regime; therefore, it must be manually switched off at lower temperatures. 
In particular, we estimate the theoretical uncertainty associated with this procedure and evaluate the relic density using three alternative switch-off prescriptions: the scheme previously used in the context of DM freeze-in~\cite{Biondini}, a scheme discussed in relation to dilepton production rates~\cite{Ghisoiu_2014}, and a new scheme we propose, inspired by a Daisy resummation in the scalar effective potential analysis~\cite{Ringwald:2020vei}. 
We then incorporate the LPM rate to our previous results from Ref.~\cite{Copello} based on 1PI-resummed propagators, to obtain the most accurate evaluation of the relic density for scalar DM freeze-in and compare with the results obtained from the HTL approximation or semi-classical approaches only.
A full implementation of the LPM effect beyond the phenomenological switch off, accurately accounting for both the parent and DM vacuum masses, is a challenging task and will be part of a follow-up work.

The article is structured as follows: In Sec.~\ref{sec:model}, we introduce the class of scalar FIMP models discussed. 
In Sec.~\ref{sec:ProductionRate}, we introduce DM interaction rate, sketch the derivation of the LPM contribution to the DM self-energy and state the integral equation necessary to obtain the LPM rate for a scalar FIMP. 
In Sec.~\ref{sec:results}, we present our numerical solutions to the DM interaction rate and summarize the implications for the DM relic density for various gauge couplings and dark sector mass splittings. 
Furthermore, we provide a simple fit function for the LPM induced production rate of a scalar FIMP in the ultra-relativistic regime. 
Finally, in Sec.~\ref{sec:conclusion}, we conclude with a summary of our results.

\FloatBarrier
\section{Vectorlike portal FIMP model} \label{sec:model}
A well-motivated class of dark matter models arises when the dark sector communicates with the SM through a portal: a renormalizable Yukawa interaction involving new mediator particles.
We are interested in a class of models in which DM is a scalar gauge singlet $s$, interacting with the SM sector through a DM mediator vectorlike fermion $F$, which has the same quantum numbers as the SM particle $f$. The particle $f$ is taken to have a definite chirality, e.g., $f=e_L,q_L$ for left-handed lepton and quark doublets, or $f=e_R,u_R,d_R$ for right-handed electrons, up quarks and down quarks.
These types of models have been discussed in the context of both freeze-out~\cite{Giacchino_2013, Giacchino_2016, Colucci_2018, Arina_2020} and freeze-in~\cite{B_langer_2019, becker2023confronting, Copello}. The Lagrangian is given by
\begin{align}
    \mathcal{L} &= \mathcal{L}_{\text{SM}} + \frac{1}{2} (\partial_{\mu}s)^2 - \frac{1}{2} m_{\mathrm{DM}}^2 s^2 - V(s,\phi) + \bar{F}(i\gamma^\mu D_\mu - m_{F,0})F - \left[y_{\text{DM}}\bar{F}fs + \text{h.c.}\right], \label{eq:lagrangian}
\end{align}
where $V(s,\phi)$ is the scalar potential of the DM, $D_\mu$ is the covariant derivative, and $y_{\text{DM}}$ is the portal Yukawa coupling. 
In this work, we neglect DM self-interactions and interactions with the Higgs in $V(s, \phi)$. 
To ensure that the DM is produced via freeze-in, we assume that the portal coupling $y_{\text{DM}} \ll 1$, so that the DM particle $s$ never reaches thermal equilibrium with the Standard Model bath.

Following the notation of Ref.~\cite{Copello}, we parametrize these models with four parameters: the portal gauge coupling $y_{\mathrm{DM}}$, the effective gauge coupling $G$, the mass splitting of the dark sector $\delta$, and the mediator particle vacuum mass $m_{F,0}$.
The effective gauge coupling $G$ parametrizes the gauge interaction of both vectorlike fermion $F$ and SM fermion $f$ via
\begin{equation}
    G = Y^2g_1^2+ C_2(\mathcal{R}_2)g_2^2 + C_2(\mathcal{R}_3)g_3^2, \label{eq:effective_gauge_coupling}
\end{equation}
where $g_1, g_2,$ and $g_3$ are the SM gauge couplings for the $\mathrm{U(1)}_Y, \mathrm{SU(2)}_L,$ and $\mathrm{SU(3)}_C$ gauge groups, respectively. 
In addition, $Y$ stands for the weak hypercharge of $f/F$, $\mathcal{R}_2$, $\mathcal{R}_3$ are the corresponding representations under $\mathrm{SU(2)}_L$ and $\mathrm{SU(3)}_C$, and  $C_2(\mathcal{R}_i)$ are the Casimir invariants.

In this work, we fix the value of the effective gauge coupling $G$ by evaluating the one-loop renormalization group equations at the energy scale corresponding to the mass of the parent particle $m_{F,0}$. 
This approach is well-motivated given that DM production occurs dominantly at $T \sim m_{F,0}$.
The effect of neglecting the running of the couplings for a scattering-like interaction (such as the LMP effect) was estimated in Ref.~\cite{Copello} to be up to $\mathcal{O}(1\%)$ on the level of the resulting relic density, thus making it appropriate to set the effective gauge coupling $G$ as a constant.

Lastly, the mass splitting of the dark sector is parametrized using the dimensionless variable 
\begin{equation}
    \delta = \frac{m_{F,0} - m_{\text{DM}}}{m_{\text{DM}}} \, .
\end{equation}
In the following, we will denote the five different realizations of the DM mediator as $e_L$, $q_L$, $e_R$, $u_R$ and $d_R$, like the corresponding SM fermion representations. 
With the model framework and particle content defined, we now focus on the dynamics of dark matter production in the early universe, governed by the interaction of the dark scalar with the thermal bath.

\FloatBarrier
\section{The DM production rate} \label{sec:ProductionRate}

The production rate of feebly-interacting DM is directly related to the spectral DM self-energy via \cite{Bodeker:2015exa, Copello}
\begin{align} \label{eq:RateDensity} \gamma_{\mathrm{DM}} &\equiv \frac{d}{dt} n_{s,\mathrm{ph}}(t) + 3Hn_{s,\mathrm{ph}} = \int \frac{d^3\vec{p}}{(2\pi)^3} \frac{\Pi_s^{\mathcal{A}}(\omega_p, |\vec{p}|)}{\omega_p} f_{-}(\omega_p), \end{align}
where $H$ is the Hubble parameter, $f_{\pm} (\omega) = \frac{1}{\exp(\omega/T) \pm 1}$ are the Fermi-Dirac (+) and Bose-Einstein (-) equilibrium distributions, $\omega_p = \sqrt{p^2 + m_{\mathrm{DM}}^2}$ is the on-shell DM energy, and $n_{s,ph}$ is the DM number density in physical (as opposed to comoving) coordinates. The quantity $\Pi_s^{\cal A}$ is the so-called spectral self-energy of the scalar field $s$, which coincides with the imaginary part of the retarded self-energy.
The expression in Eq.~\eqref{eq:RateDensity} is valid to all orders in the gauge coupling and to $\mathcal{O} (y_\text{DM}^2)$ in the feeble interaction coupling.
However, in practice, one must truncate the expansion of the spectral self-energy, which compromises the all-order accuracy in the gauge coupling. 
The scalar FIMP production rate calculated in Ref.~\cite{Copello} considered leading-order contributions to the DM self-energy from the two-particle irreducible (2PI) effective action. 
This calculation, however, neglected interference contributions from s- and t-channel processes in two-particle scatterings and the LPM effect. 
While Ref.~\cite{Copello} estimated the interference terms to be subleading (approximately 10\%) relative to the squared $s$- and $t$-channel scattering contributions, results from Ref.~\cite{Biondini} suggest that the LPM effect could alter the scalar FIMP relic density by up to $30\%$ on the level of the relic density. 
Therefore, we aim to incorporate the calculation of the LPM effect into the 2PI-based framework used in Ref.~\cite{Copello}, and to quantify its impact on the resulting dark matter relic abundance.

\begin{figure}[ht]
\centering
\includegraphics[width=0.7\textwidth]{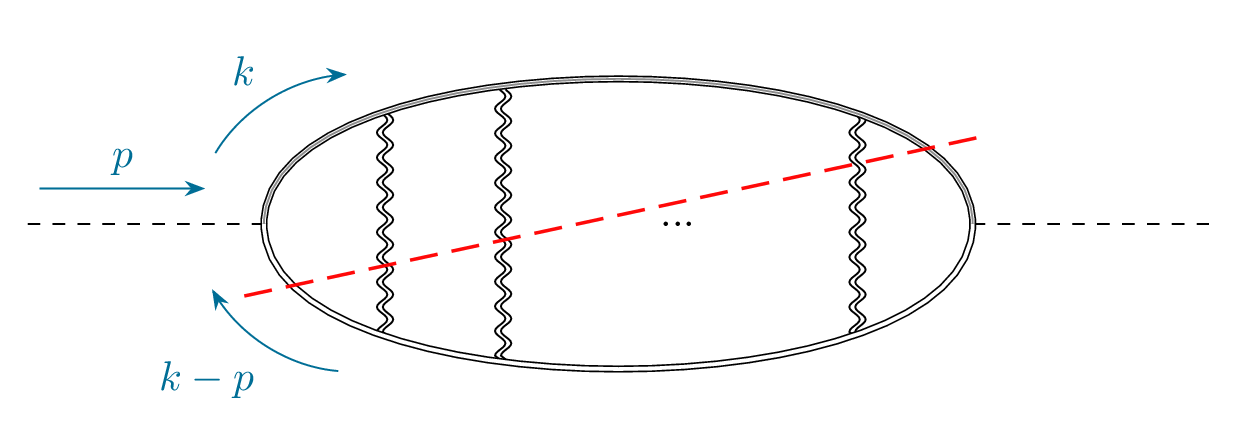}
\caption{The ladder diagram represents a self-energy correction with an infinite series of gauge boson insertions. The orientation of the fermion lines has been omitted. When cutting the self-energy along the dashed line, one obtains the interference of the scattering processes of Fig.~\ref{fig:SoftScatterings}. This cutting procedure is equivalent to evaluating the spectral self-energy $\Pi_s^{\cal A}$ appearing in Eq.~\eqref{eq:RateDensity} for the DM production rate.}
\label{fig:ladderDiagram}
\end{figure}

In the following, we address the contribution of the LPM effect, which arises from resumming an infinite series of ladder diagrams. 
In the scalar FIMP model, these diagrams involve gauge bosons inserted between the two fermion propagators in the leading-order DM self-energy, as shown in Fig.~\ref{fig:ladderDiagram}. 
To our knowledge, no equation has been derived previously to describe the LPM effect for a scalar particle interacting with fermions via Yukawa couplings. 
In this section, we derive and present such an equation.

\FloatBarrier
\subsection{The LPM Effect in Scalar DM Production}

\begin{figure}[ht]
     \centering
        \includegraphics[width=0.8\textwidth]{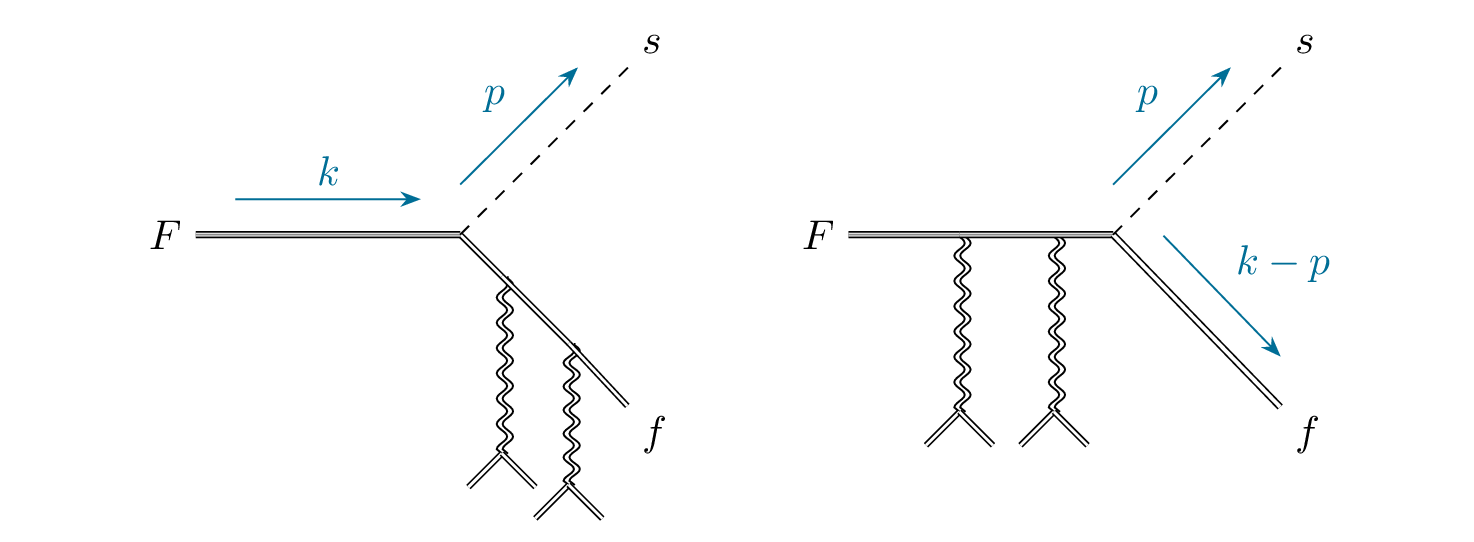}
        \caption{Scatterings with the thermal plasma after and before the decay of the dark mediator particle. The interference of these two diagrams is a leading-order contribution to the production of the dark matter particle $s$.}
        \label{fig:SoftScatterings}
\end{figure}

The resummation of the self-energy, as illustrated in Fig.~\ref{fig:ladderDiagram}, becomes necessary when the formation time of the decay $F \rightarrow f s$ is large compared to the typical timescales of gauge scatterings in the thermal plasma. 
In this regime, the interference between the scattering diagrams shown in Fig.~\ref{fig:SoftScatterings}, which correspond to the cuts of the self-energy diagram in Fig.~\ref{fig:ladderDiagram}, contributes significantly to the total rate.
This phenomenon, known as the Landau-Pomeranchuk-Migdal (LPM) effect, was originally studied in the context of bremsstrahlung and pair production in a medium~\cite{{PhysRev.103.1811,Landau:1953gr,Landau:1953um}}, and arises from the quantum interference among multiple soft scatterings that occur during the formation of the final state. 

Contributions to the self-energy from ladder diagrams as in Fig.~\ref{fig:ladderDiagram} become leading-order in a systematic power counting scheme when the four-momenta of the gauge-charged particles are nearly light-like and collinear across the rails of the ladder~\cite{Arnold}.
This implies that the gauge boson rungs are soft, and that the external momentum $p$ is also approximately light-like and collinear with the momenta along the rails. With all the momenta of the matter fields being light-like and collinear, the equation that describes the LPM effect is typically derived under the assumption of collinear kinematics, leading to a hierarchy of scales:  
\begin{align}
p_{\parallel} \sim T, \quad p_{\perp} \sim gT, \quad p^2\sim g^2 T^2, \label{eq:collinear_kinematics}
\end{align}
where  $p_{\parallel} = p_\mu v^\mu$ and $p_\perp = 1 - p_{\parallel}$ are light-cone coordinates defined with respect to a light-like four-momentum $v_\mu = (1, \hat{v})$. The scaling in Eq.~\eqref{eq:collinear_kinematics} are meant to apply not only to the external momentum $p$ of the ladder diagrams but also to the momenta running on the rails of the ladder, that is, $k$ and $k-p$ in Fig.~\ref{fig:ladderDiagram}. Defining $v_\mu$ to be aligned with the external momentum $p$, we can set $p_\perp=0$, and we take $\hat{v}=(0,0,1)$.

A systematic power counting for the LPM effect in thermal field theory was first established in Ref.~\cite{Arnold}. 
The key result is that, despite involving arbitrarily many soft gauge boson exchanges, this type of ladder diagrams remain of the same parametric order as the one-loop self-energy contribution in the kinematic region satisfying Eq.~\eqref{eq:collinear_kinematics}. 
Specifically, each additional gauge boson exchange introduces a suppression of $g^2$ from the vertices and $g^3$ from the soft phase space of the gauge boson momentum. 
However, this suppression is exactly compensated by an $1/g^3$ enhancement from the  gauge boson propagator and a further $1/g^2$ enhancement from the integration along pinching poles belonging to a pair of retarded and advanced matter propagators adjacent to the gauge boson rung, which can be approximately simultaneously on-shell when their momenta are light-like and collinear. 
As a result, all such ladder diagrams contribute at the same leading order, and a resummation is needed to correctly capture the effect of multiple soft scatterings in the medium.

\begin{figure}[ht]
     \centering
        \includegraphics[width=0.8\textwidth]{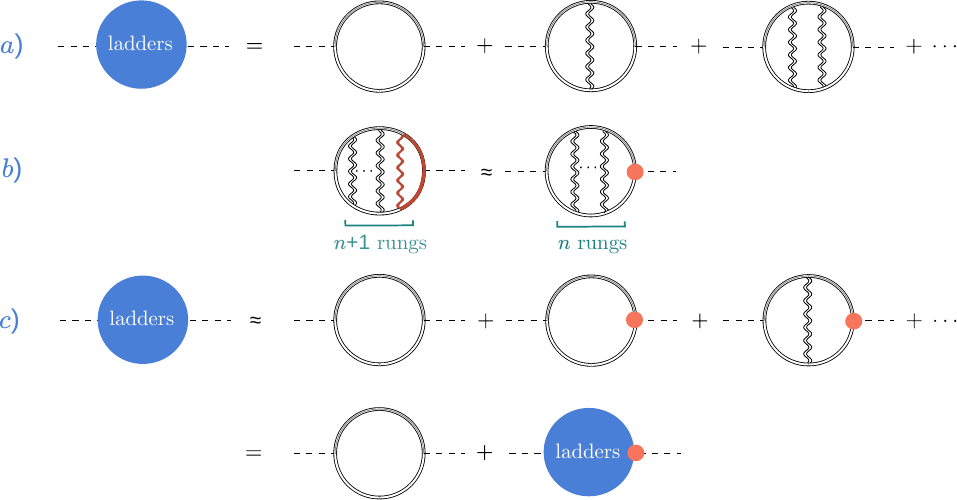}
        \caption{Diagrammatic representation of the recursive relation used to calculate the LPM effect.}
        \label{fig:LPM_Equation}
\end{figure}

To derive the expression for the LPM rate, we build upon the approach developed in Ref.~\cite{Besak_2010}. 
In that work, the authors obtained a recursion relation for one-loop diagrams involving external momenta near the light-cone and in collinear configurations, with an arbitrary number of attached soft gauge fields. 
The soft fields were subsequently integrated out, leading to the integral equation that captures the LPM effect. The resulting equation can be understood diagramatically, as shown in Fig.~\ref{fig:LPM_Equation}. The starting point is the sum of all ladder diagrams, as indicated by Fig.~\ref{fig:LPM_Equation}\,$a)$. One then applies the recursion relation mentioned earlier, which—within the relevant kinematic regime—relates a diagram with 
$n+1$ rungs  to a diagram with $n$ rungs, albeit with a modified vertex, according to Fig.~\ref{fig:LPM_Equation}\,$b)$. Repeated application of this relation leads to an identity for the sum of all ladder diagrams, as shown in Fig.~\ref{fig:LPM_Equation}\,$c)$.

The recursion relation was formulated in Ref.~\cite{Besak_2010}  within the imaginary-time formalism\footnotemark{}. 
The resulting LPM equation was derived for both right-handed neutrinos—corresponding to a fermion self-energy—and for photons in the quark-gluon plasma—corresponding to a photon self-energy.
\footnotetext{An equivalent expression can also be derived using the approach of Refs.~\cite{Arnold_20012, Arnold, Jeon_1995}, where the LPM effect was investigated in the context of photon emission from an ultra-relativistic plasmas using the real-time formalism.}

In our case, we modify the spin-statistics to account for a scalar particle as the external state. 
Adapting the notation from Ref.~\cite{Besak_2010}, the self-energy of the scalar particle can be expressed in terms of a reduced self-energy $\hat{\Pi}(p,\vec{k})$ as
\begin{equation}\label{eq:Pi_Pihat}
    \Pi^{\mathcal{A},\mathrm{LPM}}_s (p)  =-
    \int \frac{d^3k}{(2\pi)^3} \, \mathrm{Im}\left[\Phi(k,k-p) \, \hat{\Pi}(p,\vec{k})\right].
\end{equation}
Above, $\vec{k}$ denotes the loop momentum running along the top of the ladder diagrams (see Fig.~\ref{fig:ladderDiagram}), and $\Phi$ represents a vertex factor associated with the right external leg, which is common to all diagrams. By convention, the vertex factors are not taken from the standard Feynman rules, instead, they are defined to include spinor structures that arise from the numerators of the fermionic propagators attached to the vertices. In the kinematic regime where ladder diagrams are enhanced—corresponding to nearly light-like momenta—the fermion propagator can be approximated as (generalizing the treatment of Ref.~\cite{Besak})
\begin{align}
   \slashed{S}(k)\approx -\frac{2k_\parallel}{k^2-m^2_\infty}\left[\begin{array}{cc}0 & \eta(k)\eta(k)^\dagger\\
    \xi(k)\xi(k)^\dagger & 0
    \end{array}\right].
\end{align}
In this expression, $\xi$ and $\eta$ are appropriate Weyl spinors given in Appendix~\ref{app:Weyl_spinors}, and $m_\infty$ denotes the asymptotic thermal mass. The hard vertex is defined then as\vskip0.4cm
\begin{minipage}{0.3\textwidth}\hskip0.5cm\includegraphics[height=2.2cm]{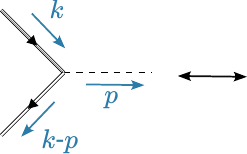}\end{minipage}\begin{minipage}{0.65\textwidth}
\begin{align}\begin{aligned}
     \hskip-1cm\Phi(k,k-p)=&\,y_{\rm DM}\,\eta^\dagger(k-p)\xi(k)\\
    =&\,y_{\rm DM}(k_1 + i k_2) \left( \frac{1}{2k_{\parallel}} - \frac{1}{2(k_{\parallel} - p_{\parallel})} \right),
    \label{eq:hardVertexFactor}
\end{aligned}\end{align}\end{minipage}
\vskip0.4cm
\noindent where in the last equality we substituted the expressions for $\xi$ and $\eta$ as derived in Appendix~\ref{app:Weyl_spinors}. The components $k_1$ and $k_2$ of the momentum $k$ correspond to the spatial directions perpendicular to the vector $\hat{v}$. In the following, $\Phi(k,k-p)$ will be referred to as ``hard-vertex factor'', because for the collinear kinematics in Eq.~\eqref{eq:collinear_kinematics}, all the momenta of the legs attached to this vertex are hard. The reduced self-energy $\hat\Pi$ satisfies the integral equation
\begin{align}\label{eq:LPM_Equation}
    \hat{\Pi}(p,\vec{k}) &= \frac{1}{\epsilon(\vec{k}_{\perp})}\Bigg\{{-} \left[ f_{+}(k_{\parallel}) - f_{+}(k_{\parallel}-p_{\parallel}) \right] \Phi^{\dagger}(k,k-p)\nonumber \\
    & \quad + i \sum_i C_2(\mathcal{R}_i) \, g_i^2 \, T \int \frac{d^2 \vec{q}_{\perp}}{(2\pi)^2} \, \mathcal{K}_i(\vec{q}_{\perp}) \left[ \hat{\Pi}(p,\vec{k}) - \hat{\Pi}(p,k_{\parallel},\vec{k}_{\perp}-\vec{q}_{\perp}) \right] \Bigg\}.
\end{align}
The first and second terms in the r.h.s originate, respectively, from the first and second diagrams in the last line of Fig.~\ref{fig:LPM_Equation}\,$c)$. The sum over $i$ accounts for the different gauge bosons that may mediate interactions, depending on the specific realization of the DM model. 
The quantity $C_2(\mathcal{R}_i)$ denotes the quadratic Casimir invariant in the representation $\mathcal{R}_i$ associated with the fields in the main loop, and $\epsilon(\vec{k}_{\perp})$ represents the energy difference between the pole locations of the two right-most internal fermionic propagators in any of the ladder diagrams, as indicated in red in Fig.~\ref{fig:LPM_Equation}\,$b)$. 
In our case, these are fermionic propagators, and the expression reads
\begin{equation}
\label{eq:polelocations}
\epsilon(\vec{k}_{\perp}) = \frac{m_{\mathrm{DM}}^2}{2 p_{\parallel}} - \frac{\vec{k}_{\perp}^2 + m_F^2}{2 k_{\parallel}} + \frac{(\vec{k}_{\perp} - \vec{p}_{\perp})^2  + m_{f}^2}{2 \left(k_{\parallel} - p_{\parallel} \right)}.
\end{equation}
Here, the quantities $m_F$ and $m_f$ denote the effective masses of the mediator and SM particles, respectively, incorporating both their vacuum and asymptotic contributions. The integral kernel $\mathcal{K}_i(\vec{q}_{\perp})$, shown as a modified vertex indicated with an orange dot in Fig.~\ref{fig:LPM_Equation}, represents the effect of a soft gauge boson propagator (indicated in red in Fig.~\ref{fig:LPM_Equation}\,$b)$). It is defined as
\begin{equation}
\mathcal{K}_i(\vec{q}_{\perp}) \equiv \frac{1}{\vec{q}_{\perp}^2} - \frac{1}{\vec{q}_{\perp}^2 + m_{D_i}^2},
\end{equation}
where $m_{D_i}$ is the Debye mass associated with the corresponding gauge boson.

For convenience, we redefine the reduced self-energy $\hat{\Pi}(p,\vec{k})$ as
\begin{equation}
\label{eq:fieldRed}
    \hat{\Pi}(p,\vec{k}) = -i y_{\rm DM}\frac{f_{+}(k_{\parallel}) - f_{+}(k_{\parallel}-p_{\parallel})}{2} 
\left( \frac{1}{2k_{\parallel}} - \frac{1}{2(k_{\parallel}-p_{\parallel})} \right)\chi(\vec{k}_{\perp}),
\end{equation}
where we have introduced the function $\chi(\vec{k}_{\perp})$. This redefinition allows us to express the spectral self-energy corresponding to the resummation of scalar ladder diagrams from Eq.~\eqref{eq:Pi_Pihat} as\footnotemark{}
\begin{align}
\label{eq:LPMScalarSelfEnergy}
\Pi^{\mathcal{A},\mathrm{LPM}}_s (p_{\parallel}) 
=y_{\rm DM}^2 \int \frac{d^3k}{(2\pi)^3} \frac{f_{+}(k_{\parallel}) - f_{+}(k_{\parallel}-p_{\parallel})}{2} 
\left( \frac{1}{2k_{\parallel}} - \frac{1}{2(k_{\parallel}-p_{\parallel})} \right)^2 
\mathrm{Re} \left[ (k_1 + i k_2)\, \chi(\vec{k}_{\perp}) \right].
\end{align}
Applying the field redefinition from Eq.~\eqref{eq:fieldRed} to the recursive relation for the reduced self-energy $\hat{\Pi}(p,\vec{k})$ in Eq.~\eqref{eq:LPM_Equation}, we obtain the integral equation that determines the function $\chi(\vec{k}_{\perp})$:
\begin{equation}
2(k_1 - i k_2)
= i \epsilon(\vec{k}_{\perp})\, \chi(\vec{k}_{\perp}) 
+ \sum_i g_i^2 C_2(\mathcal{R}_i)\, T \int \frac{d^2\vec{q}_{\perp}}{(2\pi)^2} 
\mathcal{K}_i(\vec{q}_{\perp}) 
\left[ \chi(\vec{k}_{\perp}) - \chi(\vec{k}_{\perp} - \vec{q}_{\perp}) \right].
\label{eq:LPMScalarDifferential}
\end{equation}
This equation represents the scalar analog of the LPM resummation integral equation and must be solved numerically for each set of values of momenta $p_{\parallel}$, $k_{\parallel}$, and the temperature $T$.

\footnotetext{
In Refs.~\cite{Arnold_20012, Arnold, Jeon_1995}, the LPM equations are commonly reformulated in transverse coordinate space via a Fourier transform $\vec{k}_{\perp} \rightarrow \vec{y}_{\perp}$. In this representation, the scalar ladder resummation takes the form
\begin{align} 
\label{eq:LPMScalarSelfEnergy2} 
\Pi^{\mathcal{A},\mathrm{LPM}}_s (p_0) = y_\text{DM}^2\int dk_0\, \frac{f_{+}(k_0) - f_{+}(k_0 - p_0)}{8 \pi} \lim_{y_{\perp} \to 0} \mathbb{P} \left\{ \frac{p_0^2}{k_0^2(k_0 - p_0)^2} \, \mathrm{Im} \left[ \vec{\nabla}_{\perp} \cdot \vec{f}(\vec{y}_{\perp}) \right] \right\},
\end{align}
where $\mathbb{P}$ denotes the Cauchy principal value. We have expressed the integral in terms of energy-like variables. In the ultra-relativistic regime relevant for the LPM effect, energy and longitudinal momentum scale similarly, while thermal and vacuum masses are parametrically suppressed: $k^0 \sim k_{\parallel} \sim T$ and $m^2 \sim m_{\infty}^2 \sim g^2 T^2 \ll T^2$. 

The function $\vec{f}(\vec{y}_{\perp})$ satisfies the inhomogeneous differential equation
\begin{equation}
\left( \hat{H} - i 0^+ \right) \vec{f}(\vec{y}_{\perp}) = - \vec{\nabla}_{\perp} \delta^{(2)}(\vec{y}_{\perp}),
\label{eq:LPMScalarDifferential2}
\end{equation}
with the effective Hamiltonian given by
\begin{equation}
\label{eq:polelocations2}
\hat{H} = -\frac{m_{\mathrm{DM}}^2}{2 p_0} + \frac{m_F^2 - \nabla_{\perp}^2}{2 k_0} + \frac{m_f^2 - \nabla_{\perp}^2}{2(k_0 - p_0)} - i \sum_i g_i^2 C_2(\mathcal{R}_i) \, \phi(m_{D_i}, \vec{y}_{\perp}).
\end{equation}
The function $\phi(m_{D_i}, \vec{y}_{\perp})$ is defined as the Fourier transform of the transverse momentum kernel $\mathcal{K}_i(\vec{q}_{\perp})$, and has the explicit form
\begin{equation}
\phi(m_{D_i}, \vec{y}_{\perp}) = \frac{1}{2\pi} \left[ \gamma_E + \ln \left( \frac{m_{D_i} y_{\perp}}{2} \right) + K_0(m_{D_i} y_{\perp}) \right],
\end{equation}
where $\gamma_E$ is the Euler–Mascheroni constant and $K_0$ denotes the modified Bessel function of the second kind.
}

\subsection{The limit of vanishing soft scatterings}
It is possible to perform a non-trivial check of the LPM rate obtained in Eq.~\eqref{eq:LPMScalarSelfEnergy}. For this, it is common to calculate the so-called \textit{Born limit} of the LPM \cite{Ghiglieri_2014,  Biondini, Ghisoiu_2014}. This limit corresponds to having no exchanged soft gauge bosons and it is obtained by simply taking the limit of vanishing soft scatterings to recover a collinear Born rate. Physically, the Born limit captures DM production from decays of the parent particle in the collinear limit and without any gauge boson absorption or emission.

To obtain this contribution, we solve Eq.~(\ref{eq:LPMScalarDifferential}) without the integral term. The solution satisfies a purely algebraic equation
\begin{equation}
\chi(p,\vec{k}) = \frac{-i}{\epsilon(p,\vec{k})} 2 (k_1-i k_2)\,,
\end{equation}
with $\mathbb{P}$ denoting the Cauchy principal value.
This solution is then inserted into the LPM rate from Eq.~\eqref{eq:LPMScalarSelfEnergy}. To perform the integral over $\vec{k}_{\perp}$, we set $\vec{p}_{\perp}=0$ and make use of the algebraic property
\begin{equation}
    \frac{1}{x\pm i \varepsilon} = \mathbb{P} \left( \frac{1}{x} \right) \mp i\pi \delta(x).
\end{equation}
We then obtain the expression for the Born limit of the LPM rate
\begin{align}
\Pi_s^{\mathcal{A},\mathrm{LPM} \, \mathrm{Born}}(p_{\parallel}) &= - \int_{k_-}^{k_+} d k_{\parallel} \frac{-f_{+}(k_{\parallel}) + f_{+}(k_{\parallel}-p_{\parallel})}{8\pi} \left( \frac{m_{\mathrm{DM}}^2}{2 p_{\parallel}} + \frac{m_f^2}{2(k_{\parallel}-p_{\parallel})} - \frac{m_F^2}{2k_{\parallel}} \right). \label{eq:LPMBornRate}
\end{align}
These integration limits arise from the requirement that the equation 
 $\epsilon(p,\vec{k})=0$ admits a positive solution for $\vec{k}_{\perp}^2$, leading to
\begin{equation}
k_- = \frac{\Delta-\sqrt{\lambda}}{2 m_{\mathrm{DM}}^2} p_{\parallel}, \quad k_+=\frac{ \Delta + \sqrt{\lambda}}{2 m_{\mathrm{DM}}^2 }p_{\parallel},
\label{eq:kBoundaries}
\end{equation}
where
\begin{gather}
\Delta \equiv m_{F}^2 - m_{\mathrm{DM}}^2 - m_f^2, \\
\lambda \equiv \big(m_{\mathrm{DM}}^2 - (m_F + m_f)^2\big)\big(m_{\mathrm{DM}}^2 - (m_F - m_f)^2\big).
\end{gather}
It is straightforward to verify that $\lambda \geq 0$ if $ m_F \geq m_f + m_{\mathrm{DM}}$, which corresponds to the kinematic threshold required for DM production via the decay of the mediator.

The LPM Born rate matches the full LPM rate in the non-relativistic regime. 
However, both exhibit an unphysical growth at low temperatures, where the rate is expected to vanish due to phase space suppression. 
This behavior arises from applying the collinear kinematics of Eq.~\eqref{eq:collinear_kinematics} beyond its domain of validity, for instance in the non-relativistic regime, where the DM momentum satisfies $p^2 = m_{\rm DM}^2 \gg g^2 T^2$, in contradiction with Eq.~\eqref{eq:collinear_kinematics}. 
In the next section, we introduce a prescription based on the LPM Born rate that regulates this unphysical behavior and ensures a physically consistent suppression at low temperatures.

\subsection{Switch-off of the LPM rate}
\label{sec:switchoff0}
As previously discussed, the LPM effect relies on the collinearity between emitted and emitting particles—a characteristic feature of ultra-relativistic regimes, where thermal masses satisfy $m_i \sim gT \ll T$.
However, when the temperature is no longer much higher than the vacuum masses, this collinear approximation breaks down. 
In such cases, Eq.~\eqref{eq:LPMScalarSelfEnergy} no longer holds. 
To address this, we must introduce a prescription that smoothly suppresses the LPM rate as the temperature approaches the largest vacuum mass in the system.
Our approach follows the phenomenological method proposed in Ref.~\cite{Biondini, Ghiglieri_2016}.
To this end, we introduce the total DM rate as 
\begin{equation}
\gamma_{\mathrm{DM}} = \gamma_{\mathrm{DM}}^{\operatorname{1PI}}+ \left( \gamma_{\mathrm{DM}}^{\mathrm{LPM}} - \gamma_{\mathrm{DM}}^{\operatorname{LPM \, Born}} \right) f (m_{F,0}/T) .
\label{eq:Prescription}
\end{equation}
We refer to $\gamma_{\mathrm{DM}}^{\mathrm{LPM}}$ as the LPM interaction rate density and it is given by Eq.~\eqref{eq:RateDensity} evaluated for the LPM self-energy given in Eq.~\eqref{eq:LPMScalarSelfEnergy}. 
Similarly, $\gamma_{\mathrm{DM}}^{\operatorname{LPM \, Born}}$ is the Born limit of the LPM interaction rate density, corresponding to the $1\leftrightarrow 2$ body decays of the mediator $F$ in the collinear limit, which is obtained by evaluating Eq.~\eqref{eq:RateDensity} with the self-energy given in Eq.~\eqref{eq:LPMBornRate}.
Lastly, $\gamma_{\mathrm{DM}}^{\operatorname{1PI}}$ is the production rate corresponding to the leading order process, calculated from the 1-loop self-energy calculated with 1PI-resummed propagators.
This rate was previously calculated in Ref.~\cite{Copello} and we use those results throughout the rest of the paper. 
The 1PI rate includes not only the $1\leftrightarrow 2$ body decay with general kinematics, but also the matrix element squared corresponding to the $2\leftrightarrow 2$ scatterings, which are present in the ultra-relativistic regime. 
Lastly, $f(m_{F,0}F/T)$ is a phenomenological switch-off function that efficiently suppresses the LPM contribution at low temperatures. 
It is required to satisfy  $f(0) = 1$, $ f(m_{F,0}/T \rightarrow \infty) = 0$, and to be monotonically decreasing. The reason is that in the non-relativistic limit, $m_{F,0}/T\gg 1$, in which the LPM effect is suppressed, one simply recovers the 1PI rate $\gamma_{\mathrm{DM}}^{\operatorname{1PI}}$ accounting for $1\leftrightarrow 2$ decays and $2\leftrightarrow2$ scatterings. On the other hand, in the relativistic limit $m_{F,0}/T\ll 1$, one has to account for the LPM effect captured by $ \gamma_{\mathrm{DM}}^{\mathrm{LPM}}$. However, the latter includes the rate from decays in the collinear limit, which are more accurately captured by $\gamma_{\mathrm{DM}}^{\operatorname{1PI}}$. Hence, to avoid  double counting one has to remove the $1\leftrightarrow 2$ decay contribution from $ \gamma_{\mathrm{DM}}^{\mathrm{LPM}}$, which is given by the Born limit $\gamma_{\mathrm{DM}}^{\operatorname{LPM \, Born}}$. This explains the subtraction in Eq.~\eqref{eq:Prescription}.

\begin{figure}[ht]
     \centering
         \includegraphics[width=0.6 \textwidth]{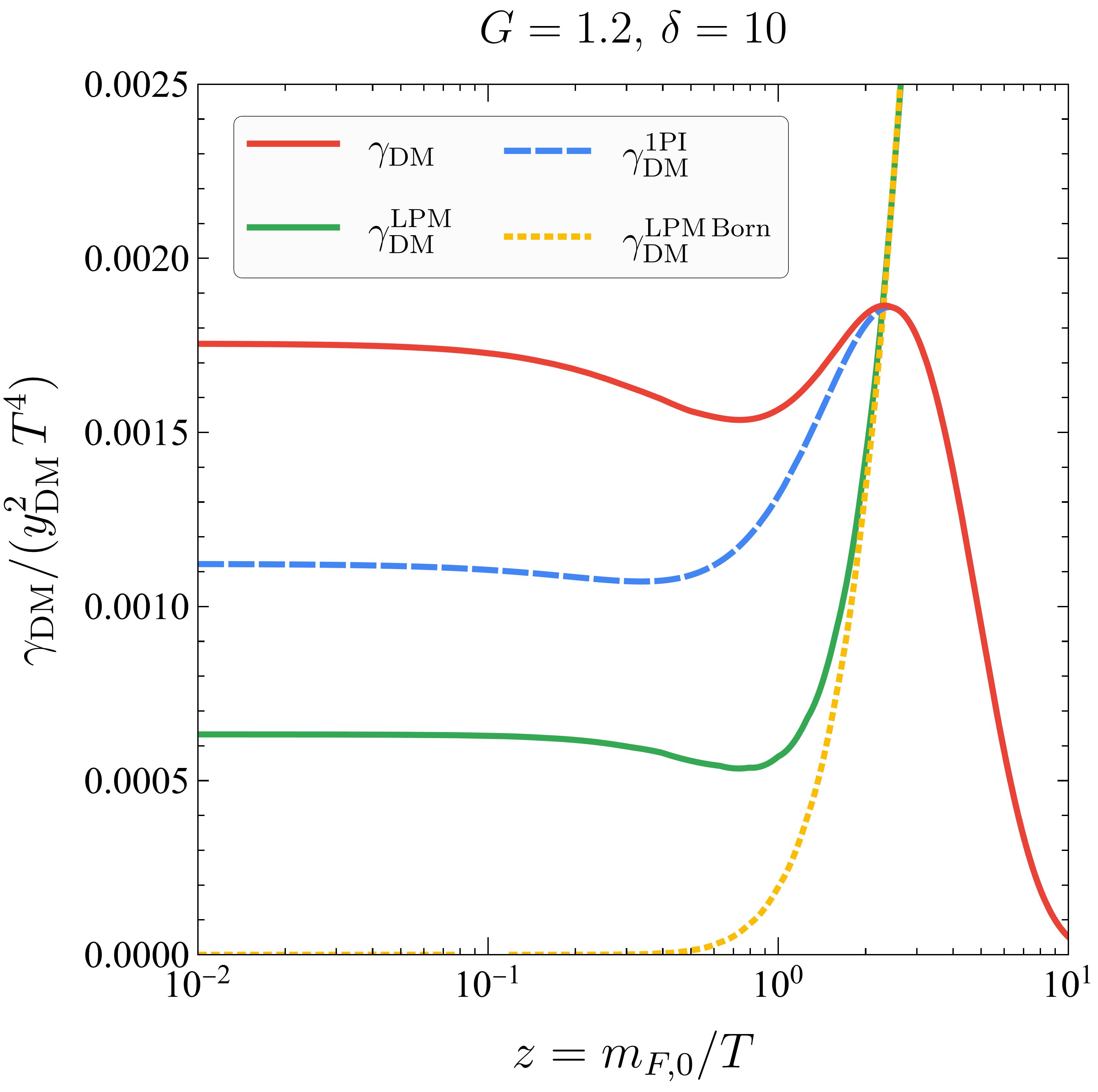} 
        \caption{Breakdown of the different components involved in the switch-off prescription that defines the total dark matter production rate $\gamma_{\mathrm{DM}}$. In this plot, the thermal function $f_{\zeta}$ is used to implement the switch-off. The gauge coupling is fixed to $G = 1.2$, and the mass splitting is set to $\delta = 10$. The DM mediator is taken to be the $d_R$ realization.}
        \label{fig:Prescription}
\end{figure}

We show how the prescription in Eq.~\eqref{eq:Prescription} works in Fig.~\ref{fig:Prescription}. 
Firstly, we take the 1PI rate $\gamma_{\mathrm{DM}}^{\mathrm{1PI}}$, i.e., the one-loop self-energy calculated with 1PI-resummed propagators from Ref.~\cite{Copello}. 
This term, indicated by a blue line, includes the decay and the $s$- and $t$-channel scattering elements squared. 
Secondly, we add the LPM rate $\gamma_{\mathrm{DM}}^{\mathrm{LPM}}$, given by the green line. 
This rate shows an unphysical growth in the non-relativistic regime $z\gtrsim1$, where collinearity breaks down. 
However, we know that in this region the physical rate runs out of phase space and approaches zero. 
To remove this growth, we substract the Born limit of the LPM $\gamma_{\mathrm{DM}}^{\mathrm{LPM \, Born}}$, indicated by the yellow line. 
The LPM and its Born limit coincide in the non-relativistic regime, which serves as a consistency check of our calculations. 
The final result of the prescription, $\gamma_{\mathrm{DM}}$, is given by the red line. 
We obtain a rate which includes the LPM and scatterings in the ultra-relativistic regime, while keeping the 1PI rate in the non-relativistic regime.

There are many valid options for defining a function $f(m_{F,0})$ that suppresses the LPM contribution in the relativistic regime. In the next subsection, we introduce and qualitatively compare these switch-off functions, which allow us to assess the associated uncertainty in the LPM rate.


\subsection{Theoretical uncertainity of the LPM Contribution}\label{sec:switchoff}

In the following, we outline three different approaches for switching off the LPM contribution and explain the rationale behind the prescription chosen in this work. 
The first two build upon Eq.~\eqref{eq:Prescription} and essentially correspond to different choices for the switch-off function $f(m_{F,0} / T)$. 
These approaches aim to suppress the LPM effect in the relativistic and non-relativistic regimes. 
The third approach introduces an interpolation at the integrand level between the tree-level rate and the LPM rate.

The first method, previously used in Refs.~\cite{Biondini, Ghiglieri_2016}, is based on the \textit{susceptibility function}, while the second method—which we will refer to as the \textit{thermal function}—is the one introduced and employed throughout this work. The motivation for this choice arises from both theoretical and practical considerations. On the theoretical side, the thermal function appears naturally in thermal field theory calculations. On the practical side, it yields a more conservative estimate of the relic abundance, as it suppresses the LPM contribution more strongly in the relativistic regime $T \sim m$. Moreover, if the $2 \leftrightarrow 2$ scatterings were considered in their ultrarelativistic limit and subsequently switched off, as done in Ref.~\cite{Biondini}, this function offers a more accurate representation of the rate at which such scatterings should be suppressed. We will return to this point in a later section.

It is important to emphasize that none of the three methods provides an accurate treatment of the LPM effect in the relativistic regime; rather, they should be regarded as interpolation schemes. A precise calculation that properly tracks the vacuum mass dependence of the LPM effect in this regime is highly nontrivial and is left for future work. Furthermore, comparing the results obtained from all three methods allows us to estimate the current theoretical uncertainty associated with the LPM contribution. Below, we summarize the three interpolation methods considered in this work:

\begin{enumerate}

    \item \textbf{Susceptibility Function.} \\
    This method uses the normalized susceptibility function $f_{\kappa}(m_{F,0}/T)$, which characterizes particle number fluctuations and is defined as the second derivative of the free energy density with respect to the chemical potential, evaluated at zero chemical potential. Its expression, normalized to its ultra-relativistic value, reads
    \begin{equation}
    f_{\kappa}(m_{F,0}/T) = \frac{6}{\pi^2 T^3} \int_0^{\infty} dp \, p^2 f_{+}(E_F) \left[ 1 - f_{+}(E_F) \right],
    \label{eq:realSuscept}
    \end{equation}
    where $E_F=\sqrt{m_{F,0}^2+p^2}$ is the energy of the mediator. This function vanishes in the non-relativistic regime and approaches one in the massless limit. 
    It has previously been used—particularly its bosonic analogue—in Refs.~\cite{Biondini, Ghiglieri_2016} to model the suppression of the LPM effect. 
    In our work, it serves as a benchmark for comparison with alternative suppression schemes.

    \item \textbf{Thermal Function.} \\
    This is the approach we adopt throughout this work. It is based on the thermal loop function
    \begin{equation} 
    J_F(z^2 = m_{F,0}^2/T^2) = \int_0^{\infty} dy \, y^2 \log \left( 1 + e^{-\sqrt{z^2 + y^2}} \right) \, , 
    \end{equation}
    and defines the suppression factor as
    \begin{equation} 
    f_{\zeta}(z = m_{F,0}/T) = \frac{\frac{d}{dz^2} J_F(z^2)}{\left.\frac{d}{dz^2}J_F(z^2)\right|_{z=0}} \, . \label{eq:SusceptibilityFactor} 
    \end{equation}
    The function $J_F(z)$ characterizes the 1-loop finite temperature contribution of a fermion (with mass $m_{F,0}$) to a scalar potential.
    Its derivative with respect to $z = m_{F,0}/T$
    \begin{equation} 
    \frac{d}{dz^2}J_F(z^2) = \int_0^{\infty} dy  \frac{y^2}{2 \sqrt{z^2 + y^2}} \frac{1}{e^{\sqrt{z^2 + y^2}} + 1} \, , \end{equation} 
    is directly related to the fermionic contribution to the scalar thermal mass in the static limit.
    Thus, it is directly related to the real part of the scalar self-energy and describes how its finite temperature contribution becomes suppressed when leaving the ultra-relativistic regime. The fact that $dJ_F/dz^2$ appears in scalar self-energies and appropriately captures the decoupling of degrees of freedom in the non-relativistic limit has been noted in Ref.~\cite{Ringwald:2020vei} and used to define an improved Daisy resummation in the scalar effective potential, which remains valid beyond the high-temperature, relativistic limit. Here, we use the suppression of the 1-loop contribution to the real part of the scalar self-energy as a proxy for the suppression of its imaginary part when the temperature decreases.  
    The switch-off function defined above in Eq.~\eqref{eq:SusceptibilityFactor} satisfies $f_{\zeta} (0) = 1$, decreases monotonically with $z$, and approaches zero at large $z$, i.e., $f_{\zeta}(z) \rightarrow 0$ as $z\rightarrow \infty$.

    \item \textbf{Smooth Interpolation} \\
    This approach, developed in Ref.~\cite{Ghisoiu_2014}, differs from the previous methods by implementing the interpolation directly at the integrand level. Specifically, it modifies the difference in the pole locations of the fermionic propagators in the integral equation for the LPM effect. In this framework, the function $\epsilon(\vec{k}_{\perp})$ from Eq.~\eqref{eq:polelocations} becomes
    \begin{align}
    \label{eq:SmoothOperator}
    \epsilon(\vec{k}_{\perp}) = \frac{p^0 m_{\mathrm{DM}}^2}{2 p_{\parallel}^2} 
    - &\frac{\vec{k}_{\perp}^2 + m_{F}^2 + \frac{\left( m_{f,0}^2 - m_{F,0}^2 - m_{\mathrm{DM}}^2 \right)^2}{4 p_{\parallel}^2}}{2 k_0} \nonumber \\
    &+ \frac{(\vec{k}_{\perp} - \vec{p}_{\perp})^2 + m_{f}^2 + \frac{\left( m_{F,0}^2 - m_{f,0}^2 - m_{\mathrm{DM}}^2 \right)^2}{4 p_{\parallel}^2}}{2 \left( k_0 - p_0 \right)} \, .
    \end{align}
    In addition to this modification, the expression in Eq.~\eqref{eq:LPMScalarSelfEnergy} must be adjusted so that the resulting rate reproduces both the LPM behavior in the ultra-relativistic regime and the tree-level decay rate in the non-relativistic limit. We provide this adjusted expression in Appendix~\ref{appendixB}. However, since the interpolation is applied directly to the tree-level rate, we want to stress that this method does not capture the full physics of the 1PI rate. In particular, it does not include the proper decay contribution nor the $2\leftrightarrow 2$ scattering processes.
\end{enumerate}

\begin{figure*}[ht]
    \centering
    \begin{subfigure}[b]{0.49\textwidth}
         \centering
         \includegraphics[width=\textwidth]{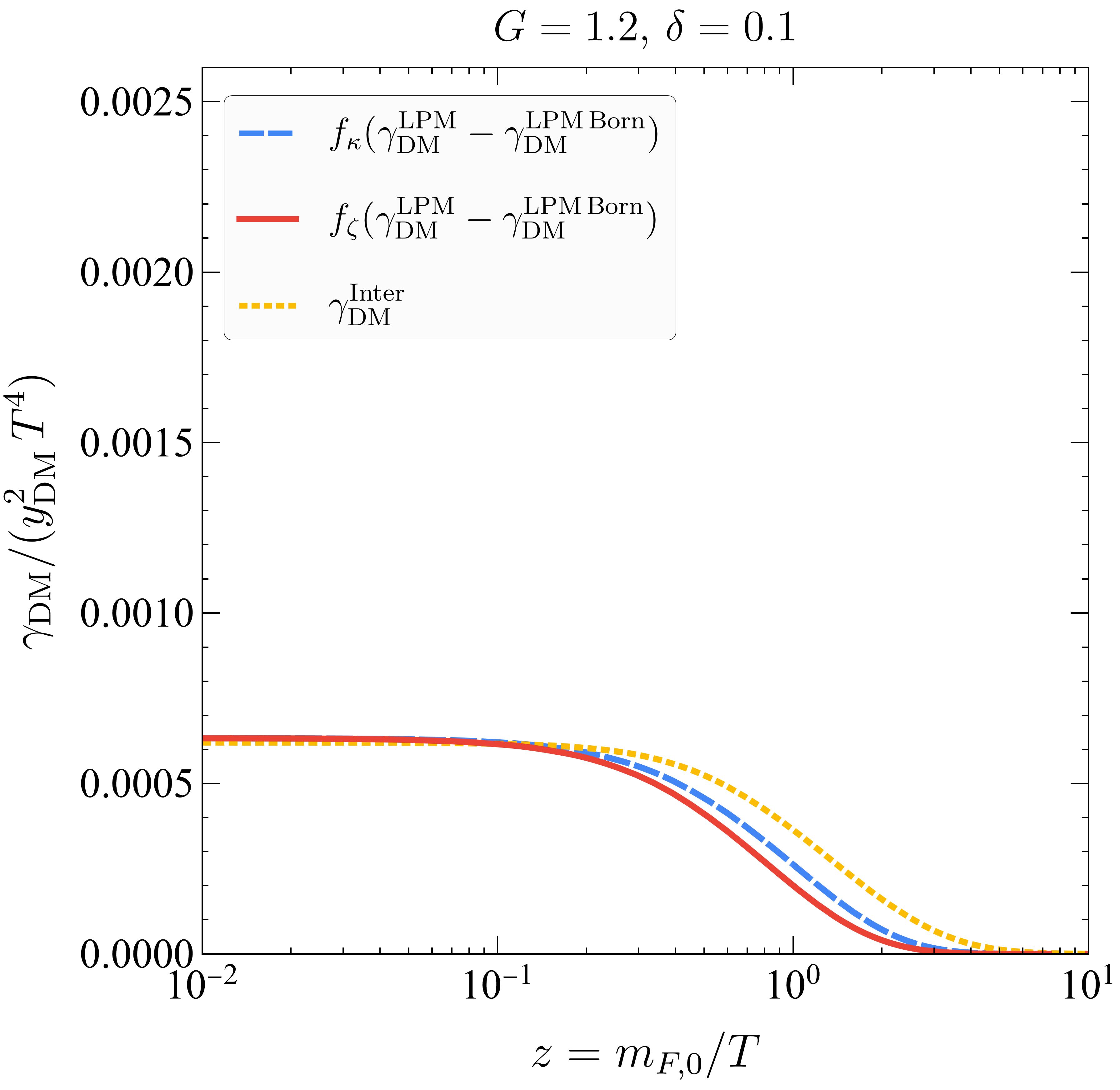}
    \end{subfigure}
    \begin{subfigure}[b]{0.49\textwidth}
         \centering
         \includegraphics[width=\textwidth]{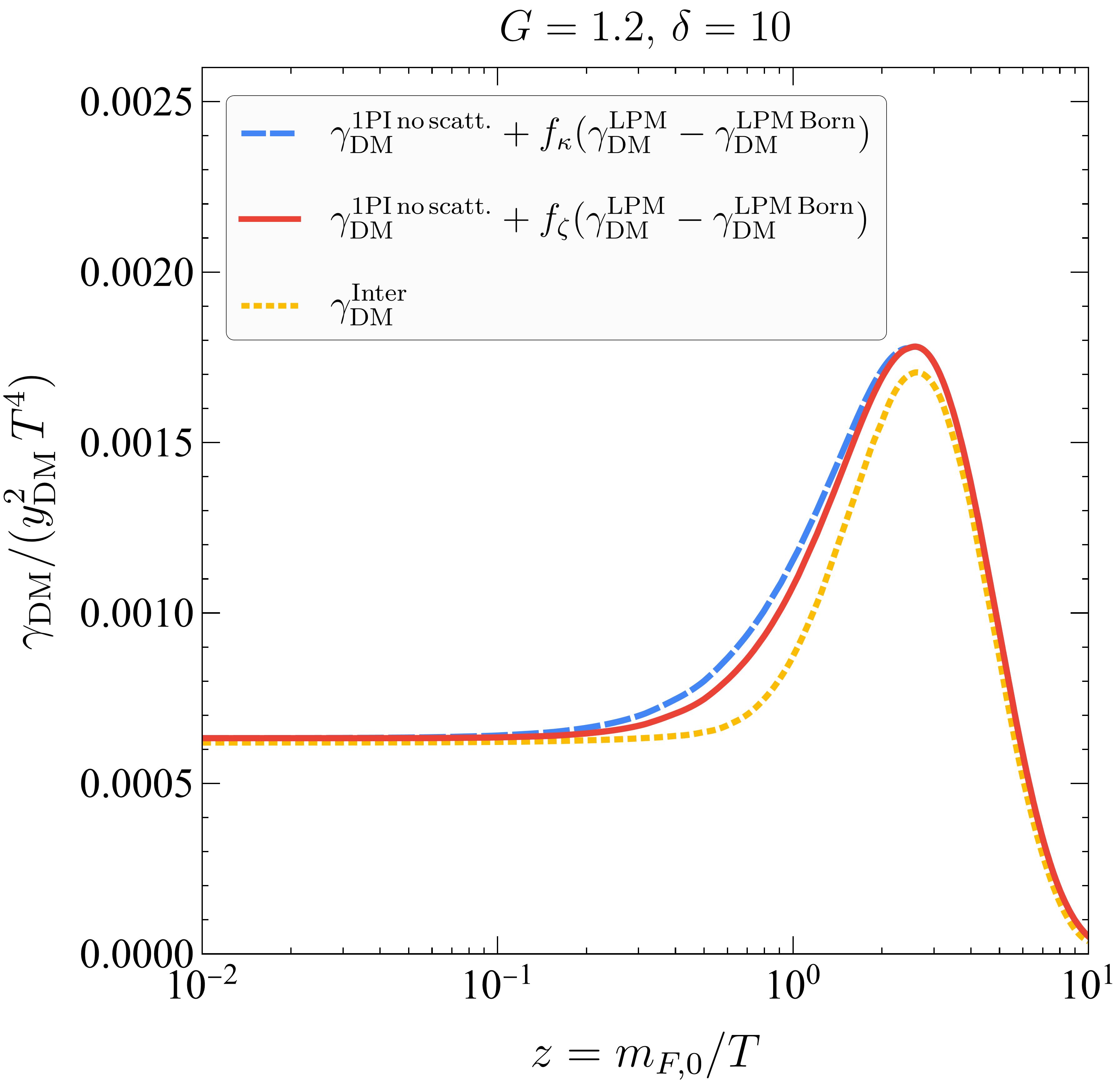}
    \end{subfigure} 
    \caption{Comparison of three different switch-off methods for the LPM rate. On the left, the 1PI rate is not included since decays are suppressed for small mass splittings. On the right, the scattering contributions have been removed from the 1PI rate. 
    This allows a fair comparison with the smooth interpolation rate $\gamma_{\mathrm{DM}}^{\mathrm{Inter}}$, which does not include $2\leftrightarrow 2$ scatterings. The plots consider a mediator of type $d_R$ with a gauge coupling constant $G=1.2$ and a mass splitting $\delta=10$.}
    \label{fig:Uncertainities1}
\end{figure*}

We compare the three approaches used to switch off the LPM contribution in Fig.~\ref{fig:Uncertainities1}. To illustrate the differences, we plot only the LPM contribution in the left panel, where the decay peak is negligible. In the right panel, we include only the decay contribution from the 1PI rate by subtracting the $2\leftrightarrow 2$ scatterings\footnote{The $2 \leftrightarrow 2$ scatterings are subtracted using a fit to the 1PI one-loop DM rate from Ref.~\cite{Copello} (see Eq.~(4.2) therein), given by $\gamma_\text{DM} \sim A z^3 K_1 (a z) + B z K_1 (b z)$. The first term resembles a decay contribution, the second a scattering. To isolate decays, we set $B = 0$.} to allow a fair comparison with the interpolation rate, which also does not include these scattering processes. 
The difference between the two switch-off functions (i) and (ii) translates into a discrepancy in the DM abundance of 21\% and 1\% for mass splittings $\delta = 0.1$ and $\delta = 10$, respectively, corresponding to the rates shown in Fig.~\ref{fig:Uncertainities1}. The procedure for computing the relic abundance from the DM production rate will be detailed in Section~\ref{sec:relicabundance}. At this point, the main advantage of using the thermal function (ii) over the previously suggested susceptibility function (i) is that it provides a more conservative estimate by suppressing the LPM rate more strongly. This is desirable because we expect the LPM effect to become irrelevant in the relativistic regime, where collinear kinematics assumed in the power counting break down. 
The smooth interpolation method (iii) shows a larger deviation from the other two approaches. This is expected, as it relates not to the 1PI rate but instead to the tree-level calculation, introducing an additional source of error. It turns off the LPM contribution too late for small mass splittings (left panel), leading to an overestimate of DM production, and underestimates the peak of the decay contribution for large mass splittings (right panel). This results in a 76\% overestimation for $\delta = 0.1$ and a 9\% underestimation for $\delta = 10$ compared to the thermal function approach.

\begin{figure*}[ht]
    \centering
    \begin{subfigure}[b]{0.49\textwidth}
         \centering
         \includegraphics[width=\textwidth]{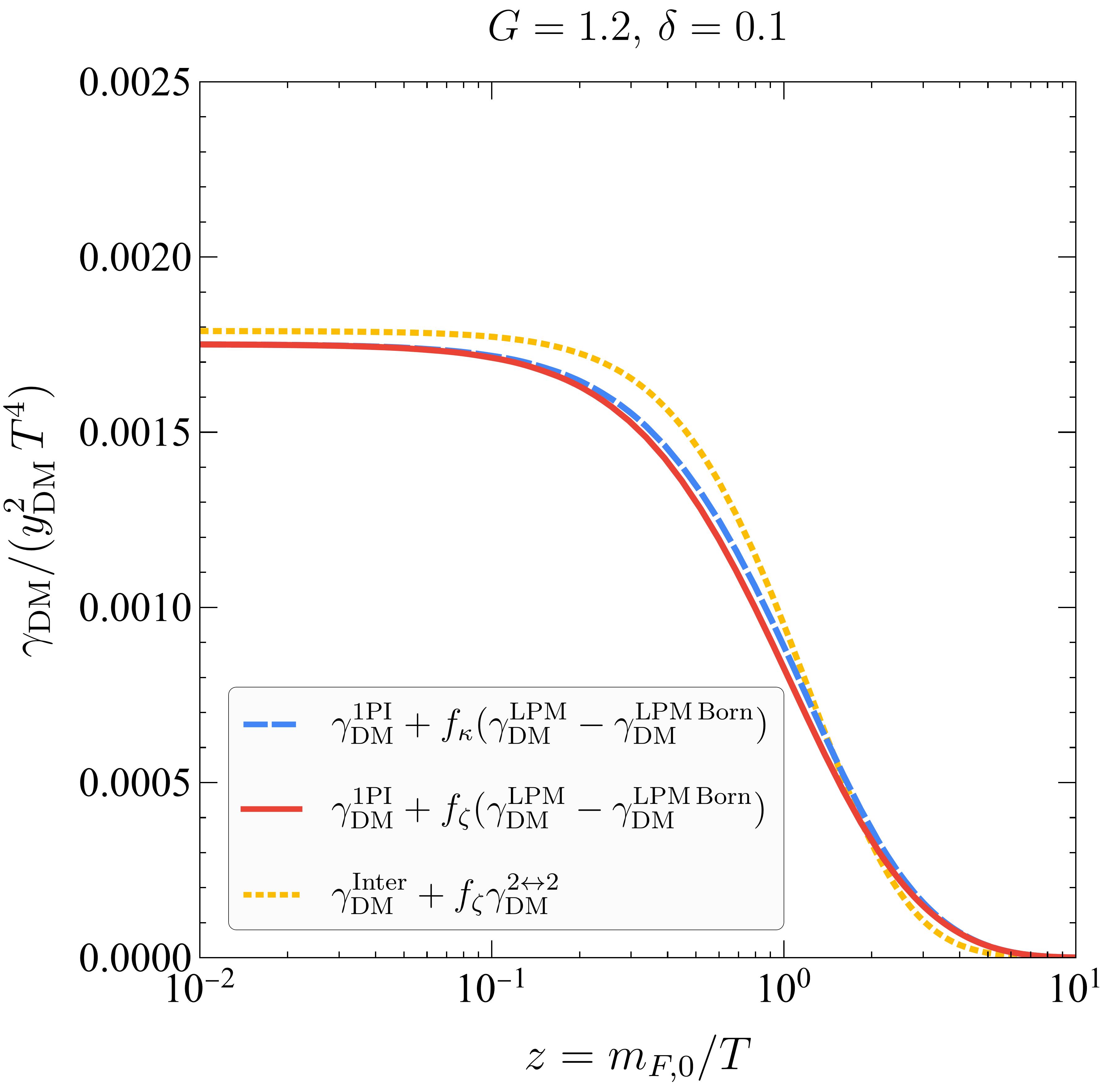}
    \end{subfigure}
    \begin{subfigure}[b]{0.49\textwidth}
         \centering
         \includegraphics[width=\textwidth]{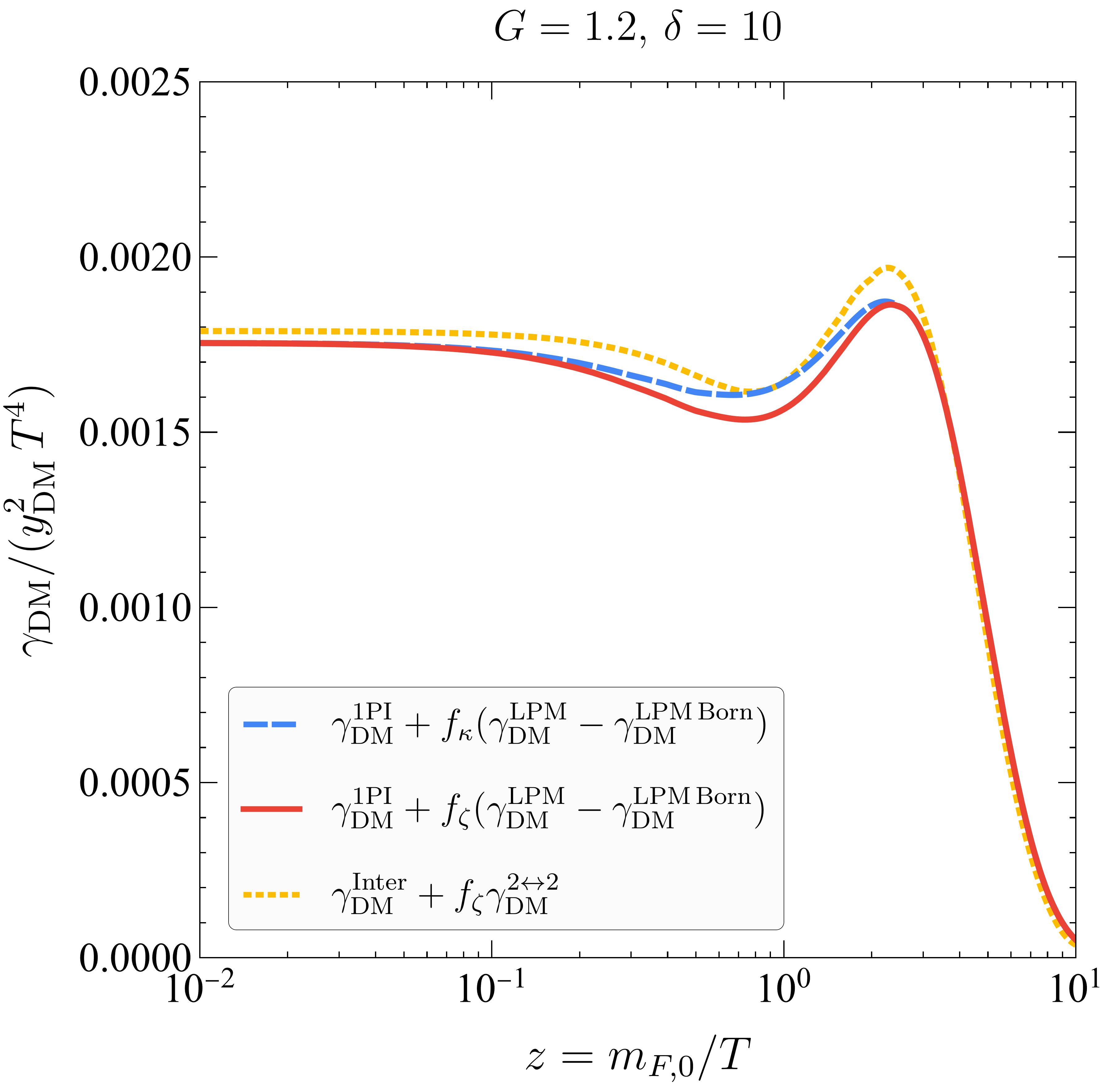}
    \end{subfigure} 
    \caption{The three different prescriptions regarding how to switch-off the LPM contribution. The dark mediator is of the type $d_R$.}
    \label{fig:Methods}
\end{figure*}

Next, we compare the total rates obtained from three different approaches in Fig.~\ref{fig:Methods}, this time including as well the $2\leftrightarrow 2$ scatterings. The blue and red line follow the prescription given in Eq.~\eqref{eq:Prescription}. 
The matrix element squared of the $2\leftrightarrow 2$ scatterings are encoded in $\gamma_{\mathrm{DM}}^{\mathrm{1PI}}$.
The third approach, based on the smooth interpolation, includes only decays and the LPM effect, but excludes scatterings.
To compensate for the missing scatterings, we compute them separately using the method described in Refs.~\cite{Besak_2012, Moore_2001, Arnold_2001, Biondini}. 
These calculations are performed in the ultra-relativistic regime, assume massless particles, and incorporate Fermi-Dirac and Bose-Einstein distributions. 
Due to the massless assumption, the scattering contributions must be manually suppressed at low temperatures. 
For this purpose, we apply the same thermal switch-off function in Eq.~\eqref{eq:SusceptibilityFactor} used for the LPM contribution. 

We recall that the 1PI rate does not include interference effects between the \( s\)- and \( t\)-channel scattering processes. However, these interference terms are known to be subleading~\cite{Copello}, and we have verified that they contribute only at the level of \( \mathcal{O}(1\%) \) relative to the squared matrix elements of the ultra-relativistic \( 2 \leftrightarrow 2 \) scattering rate, \( \gamma_\text{DM}^{2 \leftrightarrow 2} \). For this reason, we remove the interference terms between matrix elements from the \( 2 \leftrightarrow 2 \) scattering contributions, which are not included in the 1PI calculation, in order to allow for a consistent comparison. The explicit expressions for the scattering contributions are provided in Appendix~\ref{appendix}. As shown in Fig.~\ref{fig:Methods}, including the scattering terms in the smooth interpolation rate leads to an overestimation of the bulk peak. Unlike the 1PI rate, the scattering contribution $\gamma_{\mathrm{DM}}^{2\leftrightarrow 2}$ does not track the vacuum mass dependence and is only valid in the ultra-relativistic regime. The phenomenological switch-off is meant to overcome this, but we find that it reduces the $2 \leftrightarrow 2$ scattering rate slower than what is found with the 1PI treatment.

We are now in a position to qualitatively compare the different switch-off methods for the complete rate. 
The difference between using the thermal function $f_{\zeta}$ and the susceptibility function $f_{\kappa}$ amounts to approximately 5\% and 1\% for mass splittings of $\delta = 0.1$ and $\delta = 10$, respectively, in terms of the resulting relic density.
In contrast, the difference between the smooth interpolation with scatterings and the 1PI method using the thermal function $f_{\zeta}$ is around 30\% for $\delta = 0.1$ and 0.4\% for $\delta = 10$. 
The reason for the smaller discrepancies at larger mass splittings is straightforward: in this regime, the decay contribution dominates over both LPM and scattering effects.
In all cases, the rate computed using the 1PI method with the thermal function $f_{\zeta}$ as switch-off for the LPM contribution yields the most conservative result, as it consistently predicts a smaller DM abundance from the multiple soft scatterings.
Based on these results, we will later choose the thermal function $f_{\zeta}$ (ii) as more conservative choice for the switch-off of the LPM contribution in the remainder of the paper. 

While our results based on the 1PI-framework consistently include the mass scales of the problem, which are relevant for $T \sim M$, other approaches in the literature rely on an additional switch-off function for the scattering processes to obtain a physical result. Therefore, we compare in the following our DM production rate with the HTL-based approach used in Ref.~\cite{Biondini}, where the LPM effect was incorporated for the first time in the context of freeze-in production of fermionic DM. 
Their rate consists of three components: the decay rate, computed by using the UV limit of the HTL-resummed fermionic propagators and relying on momentum-dependent thermal masses derived taking into account the vacuum mass of the parent particle; the $2 \leftrightarrow 2$ scattering processes, evaluated in the ultra-relativistic regime with proper quantum statistics and neglecting vacuum masses; and the LPM contribution. 
Both the LPM and scattering terms were modulated using the susceptibility-based switch-off (see Eq.~\eqref{eq:realSuscept}) function discussed earlier.
Note that in contrast to Ref.~\cite{Biondini}, we (i) apply the thermal switch-off function (based on the thermal loop function defined in Eq.~\eqref{eq:SusceptibilityFactor}) to ensure a consistent comparison with our approach (the 1PI method including the LPM contribution) and (ii) evaluate the decay contributions relying on momentum-independent thermal masses.
With these modifications, the prescription used in Ref.~\cite{Biondini} is then given by
\begin{equation}
\gamma_{\mathrm{DM}} = \gamma_{\mathrm{DM}}^{\operatorname{HTL*}} + \left( \gamma_{\mathrm{DM}}^{\mathrm{LPM}} - \gamma_{\mathrm{DM}}^{\operatorname{LPM \, Born}} \right) f_{\zeta} (m_{F,0}) +  \gamma_{\mathrm{DM}}^{2 \leftrightarrow 2}f_{\zeta} (m_{F,0}),
\label{eq:PrescriptionGhiglieri}
\end{equation}
where $\gamma_{\mathrm{DM}}^{\operatorname{HTL*}}$ denotes the decay rate, $\gamma_{\mathrm{DM}}^{\mathrm{LPM}}$ the full LPM-resummed contribution, and $\gamma_{\mathrm{DM}}^{2 \leftrightarrow 2}$ the scattering contribution. Since these propagators do not correspond to the full HTL expressions and fail to capture the $2\leftrightarrow 2$ scatterings, we denote them with an asterisk as $\mathrm{HTL}^*$. The explicit expressions for $\gamma_{\mathrm{DM}}^{2 \leftrightarrow 2}$ and $\gamma_{\mathrm{DM}}^{\mathrm{HTL^*}}$ are provided in Appendix~\ref{appendix}.

\begin{figure*}[ht]
    \centering
    \begin{subfigure}[b]{0.49\textwidth}
        \centering
        \includegraphics[width=\textwidth]{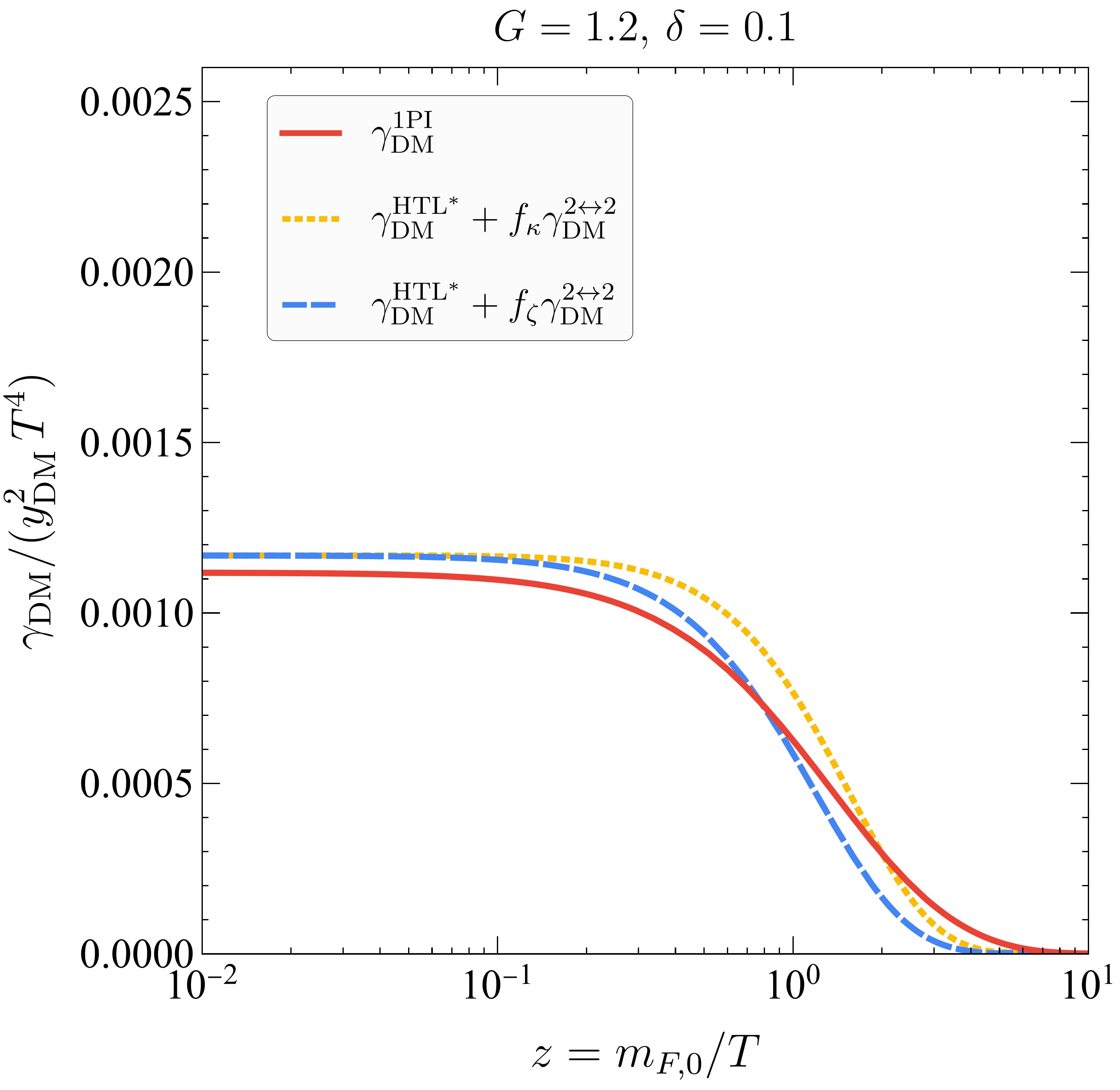}
    \end{subfigure}
    \begin{subfigure}[b]{0.49\textwidth}
        \centering
        \includegraphics[width=\textwidth]{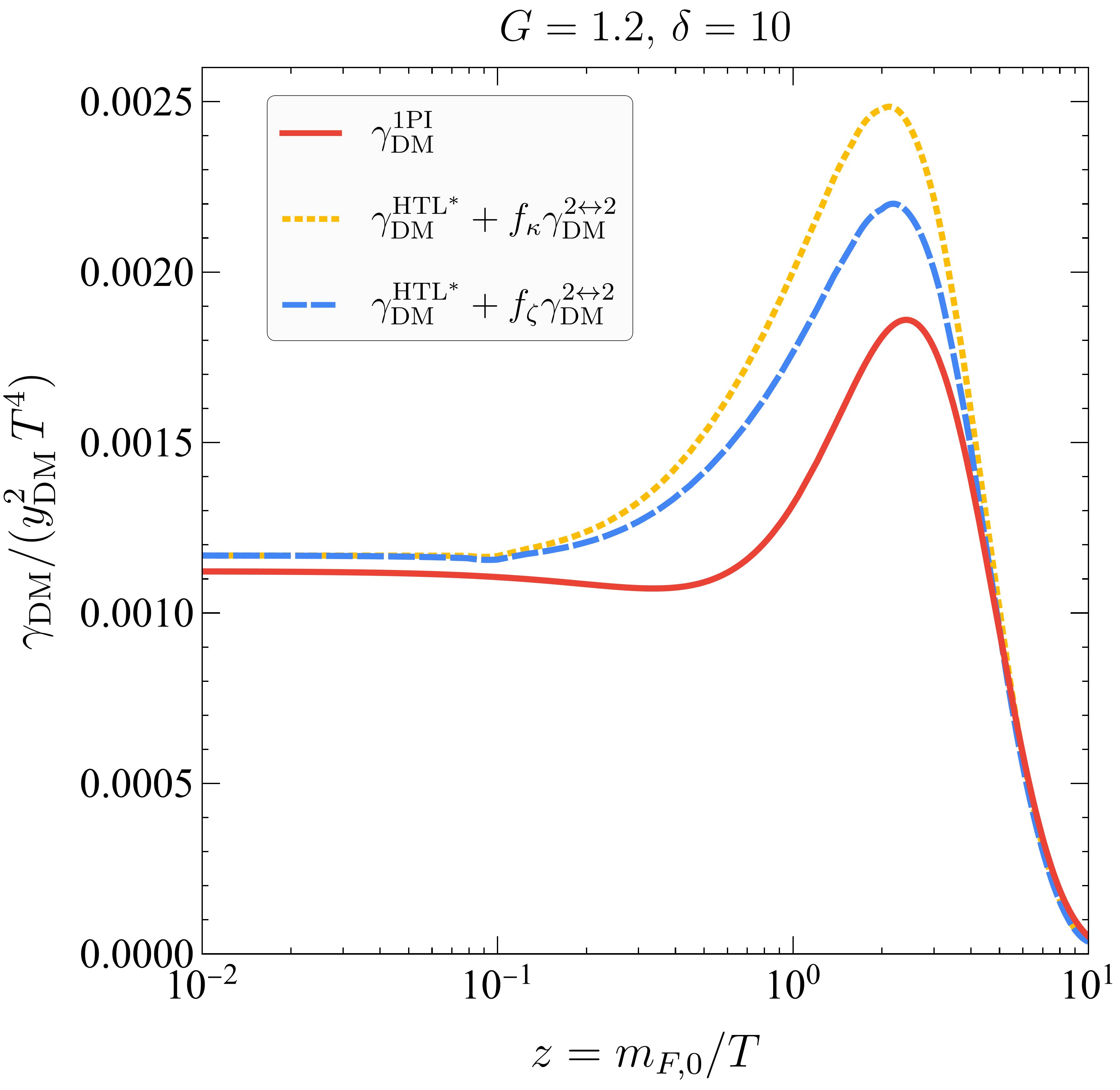}
    \end{subfigure} 
    \caption{Impact of different switch-off functions on the scattering contribution to the rate calculated in the ultra-relativistic regime that needs to be switched-off. For a consistent comparison, the interference terms have been removed from the $\gamma_{\mathrm{DM}}^{2\leftrightarrow2}$ rate. The LPM contribution is not considered for this comparison.}
    \label{fig:Uncertainities2}
\end{figure*}

In the following, we discuss the differences between the different switch-off functions (yellow and blue lines) with respect to the scattering processes within the $\mathrm{HTL}^*$ scheme and argue why we believe that our proposed thermal function is in particular for larger mass splittings better suitable than the susceptibility function. This comparison is shown in Fig.~\ref{fig:Uncertainities2}, where we compare the $\mathrm{HTL}^*$ results to the 1PI rate (red), which serves as a benchmark. 
For small mass splittings, see Fig.~\ref{fig:Uncertainities2} (left), where decays are suppressed, the thermal switch-off function $f_{\zeta}$ captures more accurately the onset the suppression around $z \lesssim 1$, while the susceptibility function $f_\kappa$ captures the behavior more accurately for $z \gtrsim 1$. For larger mass splittings, see Fig.~\ref{fig:Uncertainities2} (right), the largest difference appears around the decay peak. In this case, the slower switch-off of scatterings leads to an enhanced peak. Between the two functions, the thermal function switches off the scatterings earlier and therefore provides a more accurate representation.

Additionally, the figure highlights the size of the uncertainty introduced by computing scatterings solely in the ultra-relativistic regime, compared to the 1PI rate, which naturally includes the squared matrix elements for both $s$- and $t$- channel processes.  Quantitatively, using the susceptibility function to switch off the scatterings according to Eq.~\eqref{eq:realSuscept} leads to an overestimation of the DM relic abundance by approximately 2\% for a mass splitting of $\delta = 0.1$ and 22\% for $\delta = 10$ for $G=1.2$.
In contrast, the rate from Eq.~\eqref{eq:PrescriptionGhiglieri}, which employs the thermal switch-off function in Eq.~\eqref{eq:SusceptibilityFactor}, underestimates the relic abundance by about $20\%$ for $\delta = 0.1$ and overestimates by $11\%$ it for $\delta = 10$.
These discrepancies highlight the advantage of the 1PI rate, which already incorporates the relevant scattering processes in a manner that consistently accounts for the presence of vacuum mass scales at all stages of the computation. 

However, we note that at this point neither of the currently available LPM switch-off prescriptions are strictly valid in the relativistic or non-relativistic regimes.
Nonetheless, for the rest of this work, we adopt the thermal switch-off function  $f_{\zeta}$. 
This choice is motivated by the fact that the LPM rate itself corresponds to the imaginary part of a resummed self-energy, while $f_{\zeta}$ encodes the suppression of the scalar thermal mass in the static limit as the system exits the ultra-relativistic regime. 
Moreover, among the available options, the thermal function $f_{\zeta}$ offers the most conservative estimate of the LPM contribution, making it a well-justified choice.
\FloatBarrier
\section{Impact of the LPM effect} \label{sec:results}

In the following, we quantify the LPM contribution by directly comparing it to the 1PI rate obtained from the one-loop DM self-energy calculated in Ref.~\cite{Copello}, which did not account for the LPM rate.
We perform a parameter scan over $G\in \left[0.4, 1.6 \right]$ in steps of 0.1 and $\log_{10} \delta \in \left[-1, 1\right]$ in steps of $2/9$, the same grid as used in Ref.~\cite{Copello}. 
At leading order in the expansion of the DM self-energy, the 1PI DM production rate depends non-trivially only on the mass splitting $\delta$ and the effective gauge coupling $G$. 
This simplification arises because contributions from different gauge groups to the 1PI-resummed fermion propagators add linearly, as shown in Eq.~\eqref{eq:effective_gauge_coupling}, and the DM self-energy is truncated at one-loop level. 
In contrast, this linearity no longer holds for the LPM contribution. 
As seen in Eq.~\eqref{eq:LPMScalarDifferential}, interference effects between multiple gauge interactions break the simple additive structure. 
Consequently, the LPM rate becomes model-dependent: the interaction rate is sensitive to the specific gauge charges of the parent particles, not just to the effective gauge coupling $G$. 
As such we present results for the five possible model realizations corresponding to the charge assignment of the SM fermions and accordingly labeled $q_L$, $d_R$, $u_R$, $e_L$, $e_R$. 
To obtain the LPM contribution, we reformulate the integral equation in Eq.~(\ref{eq:LPMScalarDifferential}) as a system of four coupled differential equations, which we then solve dynamically using Mathematica. 
The latter formalism has been extensively developed and refined over the past decades \cite{Aurenche_2002newnew, Anisimov_2011, Besak, Hutig:2013oka, Depta}.

We now illustrate the impact of the LPM effect at the level of the DM interaction rate, followed by its effect on the relic abundance relative to the 1PI one-loop result across our five model realizations.

\subsection{The DM production rate}
\begin{figure*}[h!]
     \centering
     \begin{subfigure}[b]{0.49\textwidth}
         \centering
         \includegraphics[width=\textwidth]{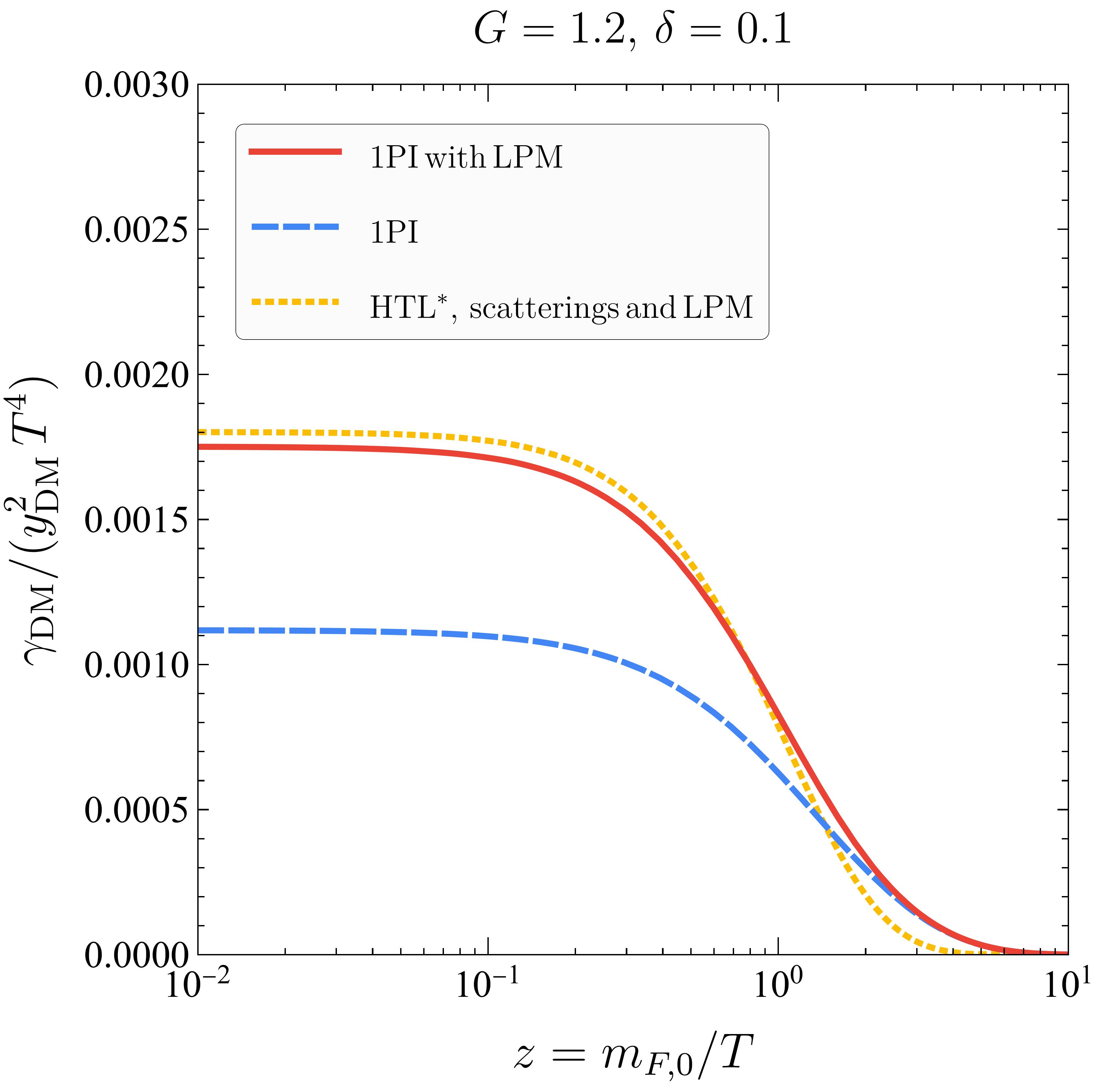}
     \end{subfigure}
     \begin{subfigure}[b]{0.49\textwidth}
         \centering
         \includegraphics[width=\textwidth]{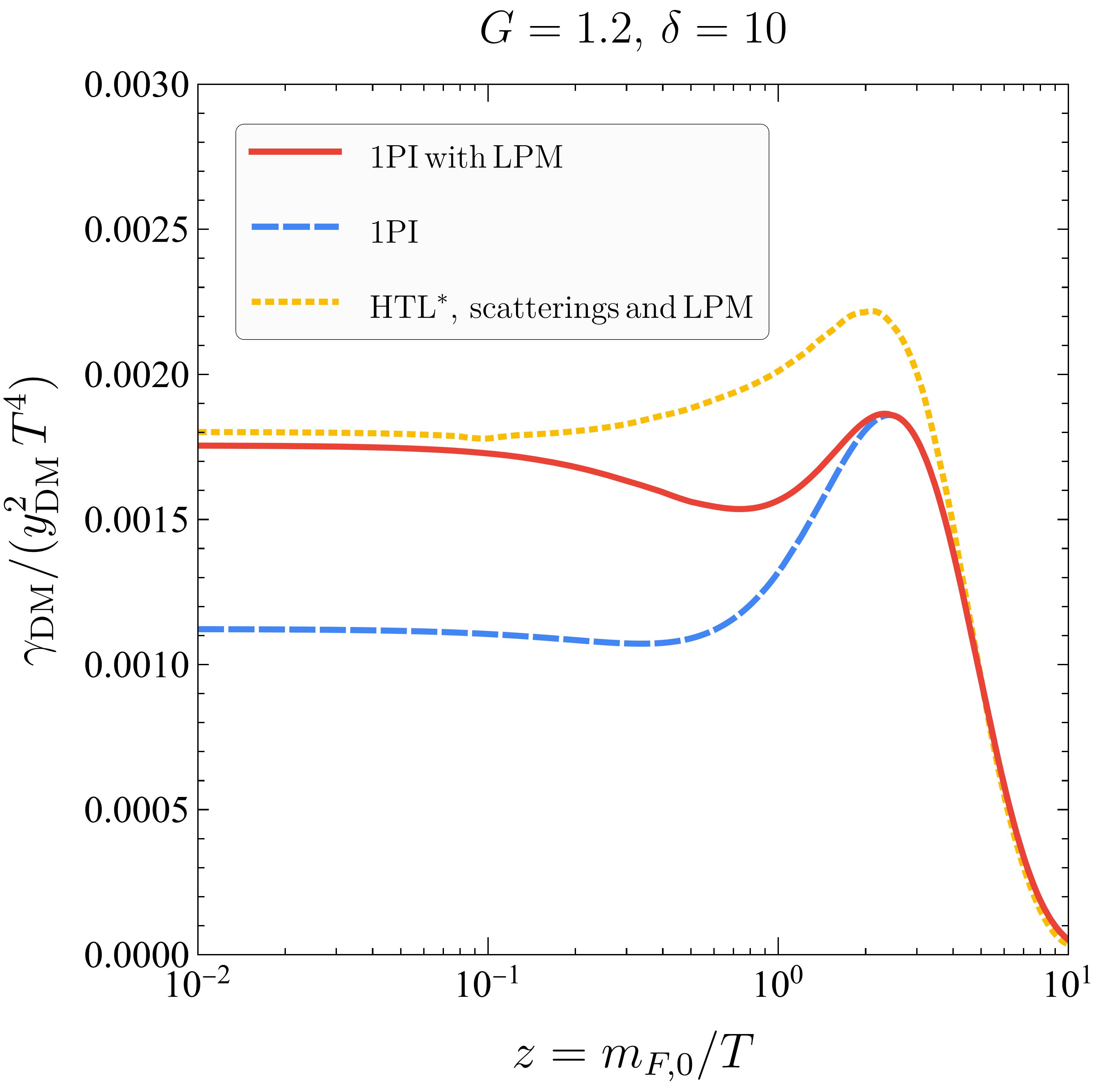}
     \end{subfigure} \\
     \vspace{0.3cm}
     \begin{subfigure}[b]{0.49\textwidth}
         \centering
         \includegraphics[width=\textwidth]{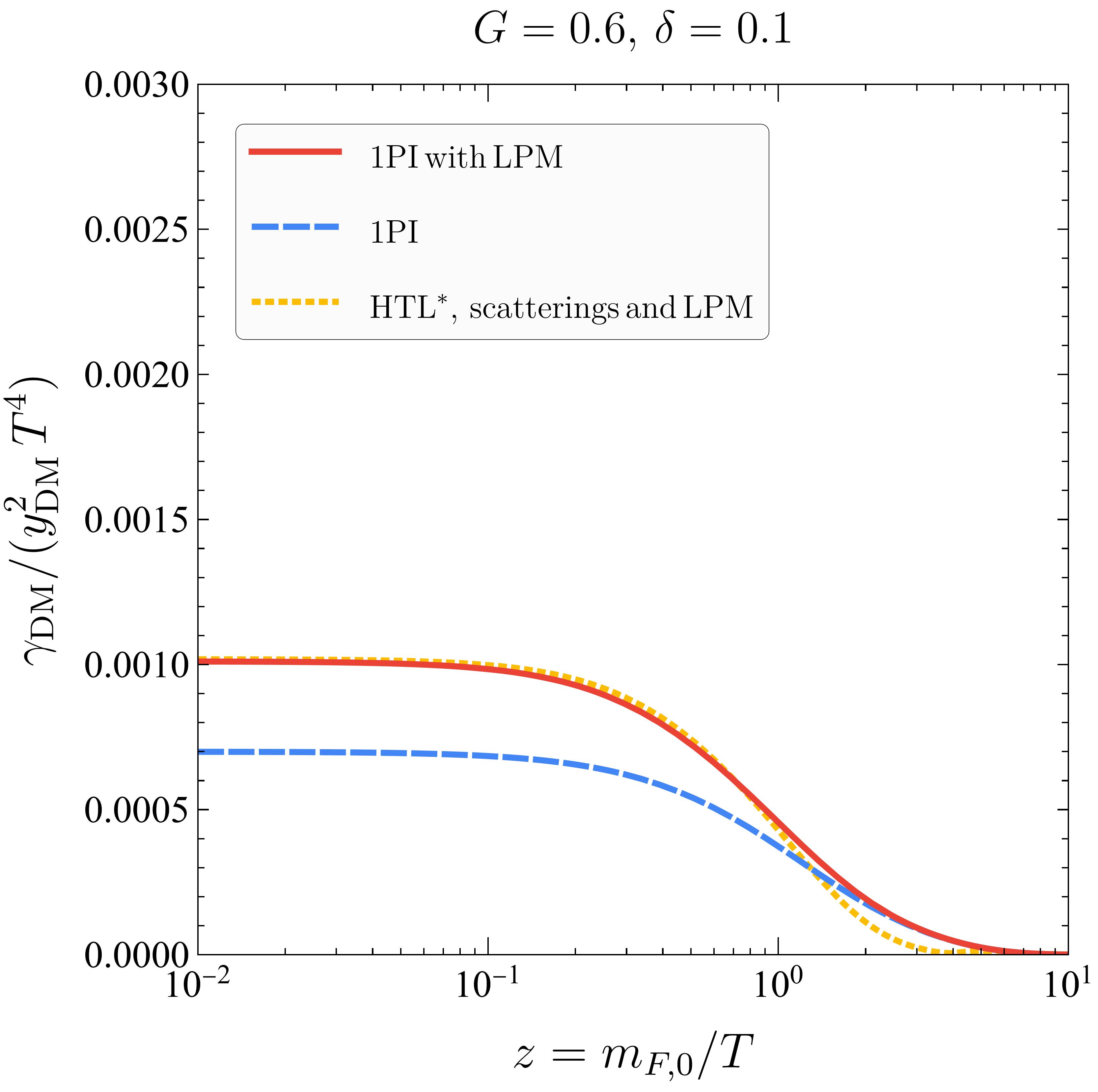}
     \end{subfigure}     
     \begin{subfigure}[b]{0.49\textwidth}
         \centering
         \includegraphics[width=\textwidth]{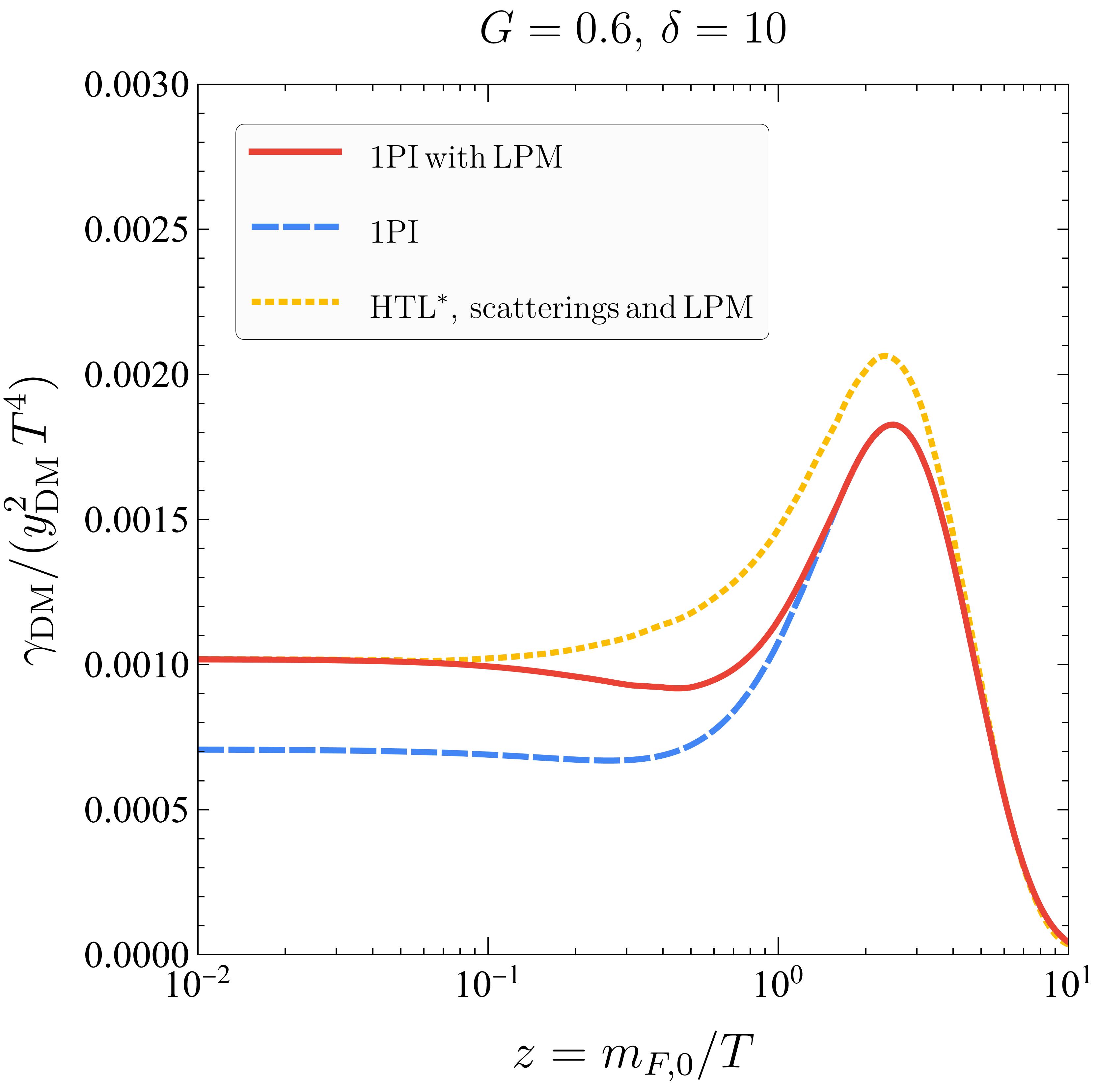}
     \end{subfigure}
        \caption{Time evolution of the DM rate as a function of the time-like variable $z=m_{F,0}/T$. These plots have been calculated for the $d_R$ representation of the mediator particle for different values of the gauge coupling $G$ and the mass splitting $\delta$.}
        \label{fig:1PILPMRate}
\end{figure*}

In Fig.~\ref{fig:1PILPMRate}, we plot the DM interaction rate for four data points, $G = \left \lbrace 0.6 , 1.2 \right \rbrace$ and $\delta = \left \lbrace 0.1, 10 \right \rbrace$, obtained in three different ways: 
\begin{enumerate}
    \item In blue, labeled \emph{1PI}, we show the interaction rate defined in Eq.~\eqref{eq:RateDensity}, derived from the one-loop DM self-energy using 1PI-resummed fermion propagators. This result, taken from Ref.~\cite{Copello}, does not include LPM contributions or interference terms between $s$- and $t$-channel $2 \leftrightarrow 2$ scatterings. Consequently, the 1PI rate exhibits a residual gauge dependence. 
    The latter is expected to be of the order of the theoretical uncertainty associated with omitted higher contributions~\cite{Nishiyama:2013dca}. 
    While a comprehensive analysis of gauge dependence in the 1PI framework remains to be completed, we have estimated its magnitude by comparing non-interfering contributions to thermal mass regulated scatterings in both Feynman and axial gauges, which vary at the percent level.
    \item In red, labeled \emph{1PI with LPM}, we present the same 1PI rate as above, supplemented by the LPM contribution according to Eq.~\eqref{eq:Prescription}, modulated by the switch-off function defined in Eq.~\eqref{eq:SusceptibilityFactor}.
    \item In yellow, labeled \emph{$\textit{HTL}^*$, scatterings and LPM}, we show the rate obtained in the spirit of Refs.~\cite{Ghiglieri_2016, Biondini}, as given by Eq.~\eqref{eq:PrescriptionGhiglieri}. The decay contributions are computed from the one-loop DM self-energy evaluated with HTL-resummed fermionic propagators at time-like momentum, taken in their ultra-relativistic limit, and, in contrast to Refs.~\cite{Biondini}, using momentum-independent thermal masses. The $2 \leftrightarrow 2$ scattering contribution is obtained from massless scattering diagrams with hard momentum exchange, while for soft momentum exchange in the $t$-channel, the corresponding propagator is HTL-resummed, as done in Ref.~\cite{Ghiglieri_2016}. We omit interference terms between the $s$- and $t$-channel scatterings to allow a more direct comparison with the 1PI rate. Both the $2 \leftrightarrow 2$ scattering contribution and the LPM rate are switched off using the thermal function defined in Eq.~\eqref{eq:SusceptibilityFactor}. 
   
\end{enumerate}

There are two distinct regimes in Fig.~\ref{fig:1PILPMRate}. 
In the ultra-relativistic regime ($z \ll 1$), the LPM rate and $2\leftrightarrow 2$ scatterings dominate.
For smaller temperatures ($z \gtrsim 1$), decays become the main contribution. 
The decay contribution is characterized by a peak, which is absent for smaller mass splittings due to phase space suppression. 

We can find a fit for the LPM rate in the ultra-relativistic regime given by
\begin{equation} 
\label{eq:linearfit}
\gamma_{\mathrm{DM}}^{\mathrm{LPM}}/(y_{\mathrm{DM}}^2T^4) \bigg|_{z=0.01} = a + b G \, , 
\end{equation}
where the fitted values of $a$ and $b$ are listed in Table~\ref{tab:fit_params}. 
The agreement of this fit with the data is presented in Appendix~\ref{appendixC}. Comparing this result with the fit for the $2 \leftrightarrow 2$ scattering rate 
\begin{align}
    \gamma_\text{DM}^{2 \leftrightarrow 2} /(y_{\mathrm{DM}}^2T^4) \bigg|_{z=0.01}  = 10^{-3} G - 3.32 \cdot 10^{-4} G \log G \, ,
\end{align} found in Ref.~\cite{Copello}, indicates that a larger $G$ enhances the LPM contribution relative to the 1PI rate, which is visible in Fig.~\ref{fig:1PILPMRate} in the offset of the blue and red lines for $z \ll 1$.

\begin{table}[h]
    \centering
    \begin{tabular}{c ccccc}
        \rowcolor{gray!20} 
         & $e_L$ & $q_L$ & $e_R$ & $u_R$ & $d_R$ \\
        \hline
        $a$ & -0.000020 & -0.000040 & $-8.1 \times 10^{-8}$ & -0.000038 & -0.000010 \\
        $b$ & 0.00067 & 0.00047 & 0.00061 & 0.00053 & 0.00054 \\

    \end{tabular}
    \caption{Fitted parameters $a$ and 
$b$ for the five different realizations of the DM mediator.}
    \label{tab:fit_params}
\end{table}

A direct comparison between the \textit{1PI} DM production rate and the \textit{HTL$^*$, scatterings and LPM} rate (similar to the HTL-based approach proposed in Ref.~\cite{Biondini}) is now possible. This comparison was not performed in Ref.~\cite{Copello}, as the analysis did not yet account for the LPM contribution. 
The comparison is illustrated by the red and yellow curves in Fig.~\ref{fig:1PILPMRate}.
First, we note a mild discrepancy at high temperatures, which becomes more pronounced with increasing gauge coupling $G$. 
This deviation arises from the fact that the yellow curve neglects both in-vacuum and thermal masses in the $2 \leftrightarrow 2$ scattering rates. 
Second, the HTL-based result (yellow) features a more pronounced peak for larger mass splittings (right), which is also slightly shifted towards higher temperatures. 
This stems from two key differences: (i) the decay rate in the HTL approach is computed using HTL-resummed propagators in the ultra-relativistic limit, whereas the 1PI result does not rely on such a limit; and (ii) more significantly, the 1PI calculation consistently tracks the in-vacuum mass scale $m_{F,0}$ throughout the computation. 
This affects the evaluation of $2 \leftrightarrow 2$ scatterings at $z \not\ll 1$, where the HTL approach---based on a UV-limit rate---relies on a phenomenological switch-off function. 
As a result, the HTL-based method tends to overestimate the scattering contribution near the decay peak, particularly for large mass splittings, resulting in the enhanced peak.
Third, especially for small mass splittings, the switched-off UV $2 \leftrightarrow 2$ scattering rate suffers from a stronger suppression compared to the 1PI result at small temperatures $z \gtrsim 1$.
We observe these behaviors across all switch-off functions discussed in Sec.~\ref{sec:switchoff}.
These observations motivate a calculation based on the 1PI approach for obtaining the most accurate calculation of the relic abundance prediction and hence an error estimate of other semi-classical approaches used in the literature.

\subsection{The relic abundance}
\label{sec:relicabundance}
\begin{figure*}[t]
     \centering
     \begin{subfigure}[b]{0.49\textwidth}
         \centering
         \includegraphics[width=\textwidth]{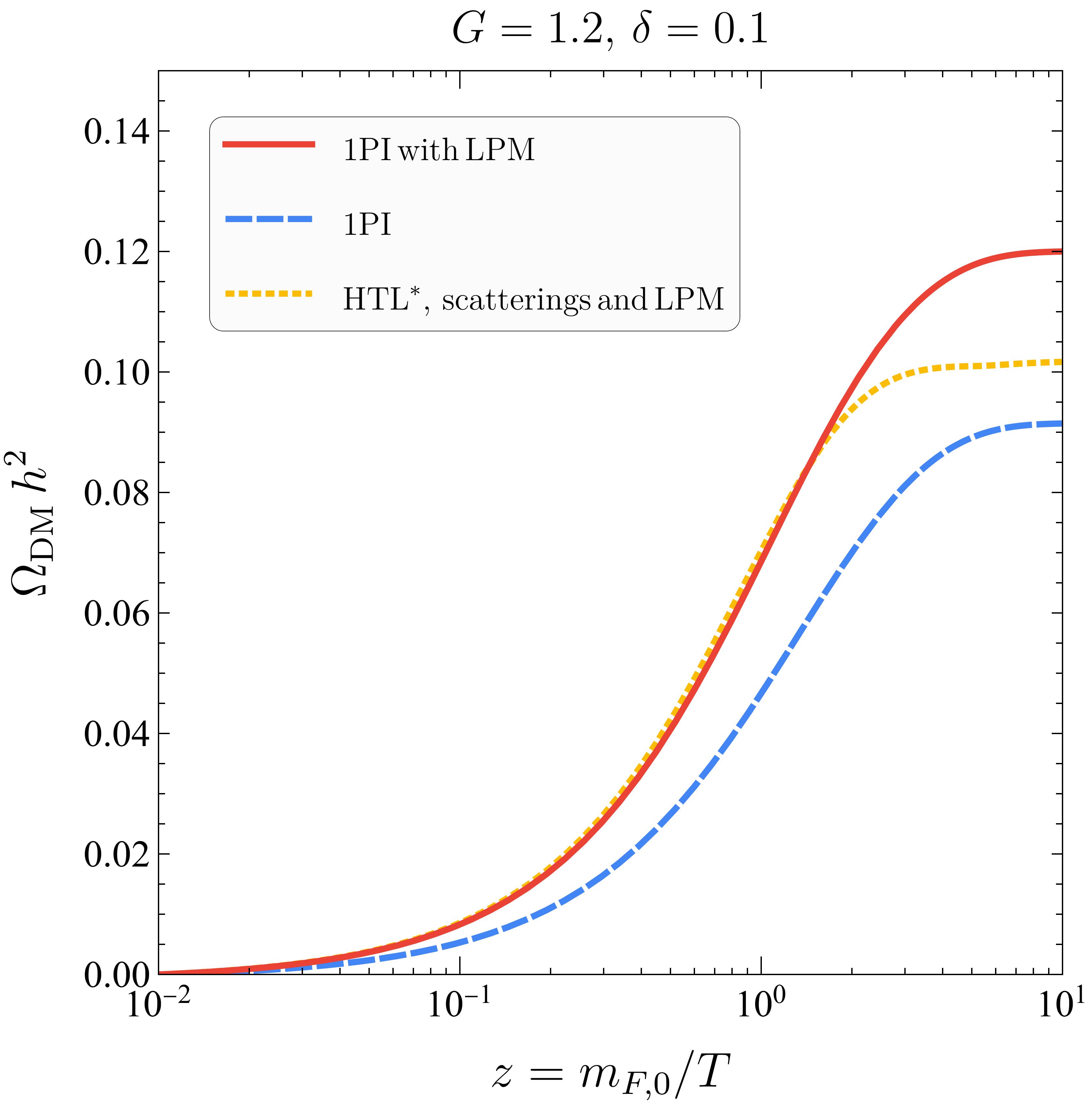}
     \end{subfigure}
     \begin{subfigure}[b]{0.49\textwidth}
         \centering
         \includegraphics[width=\textwidth]{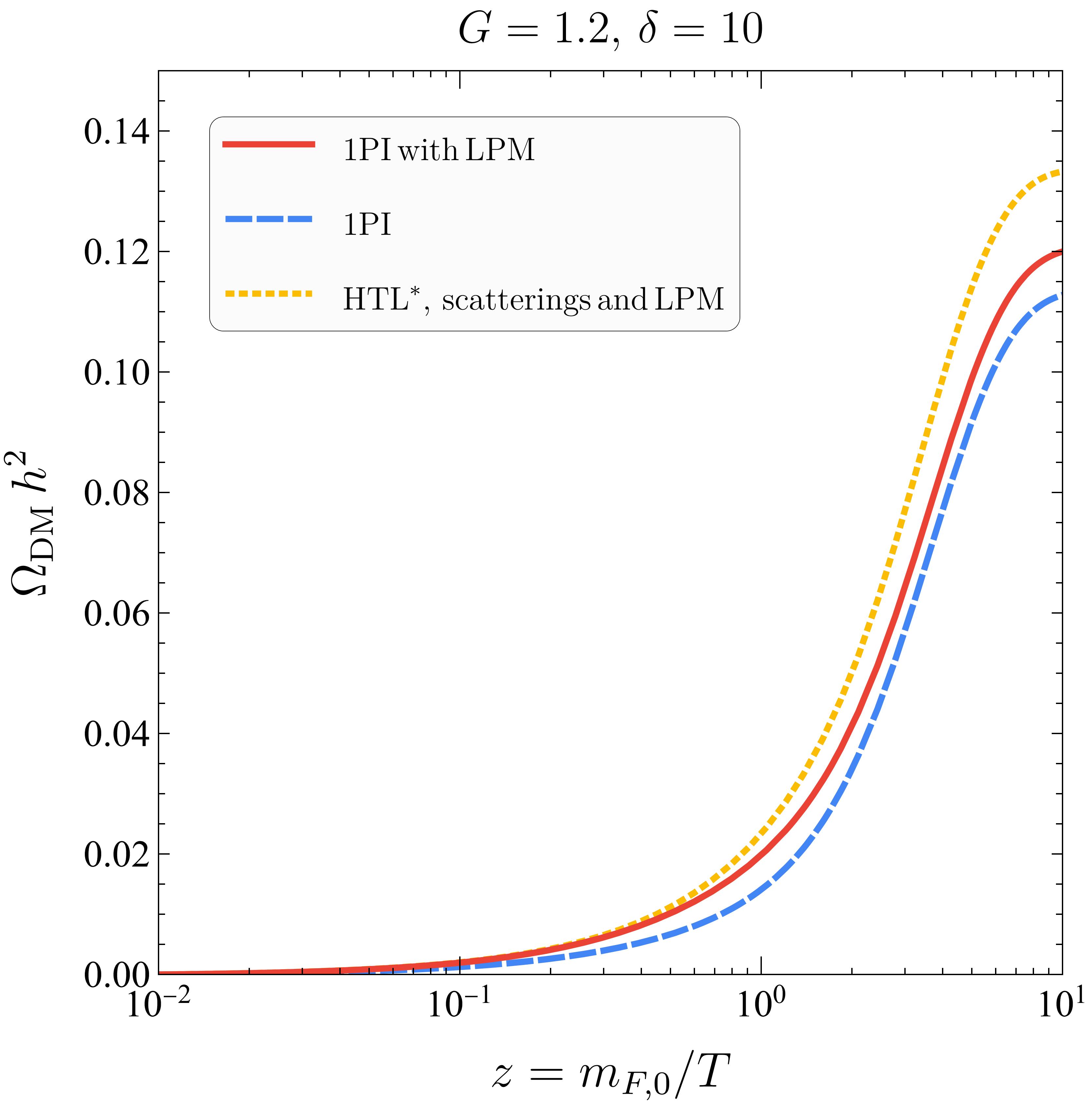}
     \end{subfigure} \\
     \vspace{0.3cm}
     \begin{subfigure}[b]{0.49\textwidth}
         \centering
         \includegraphics[width=\textwidth]{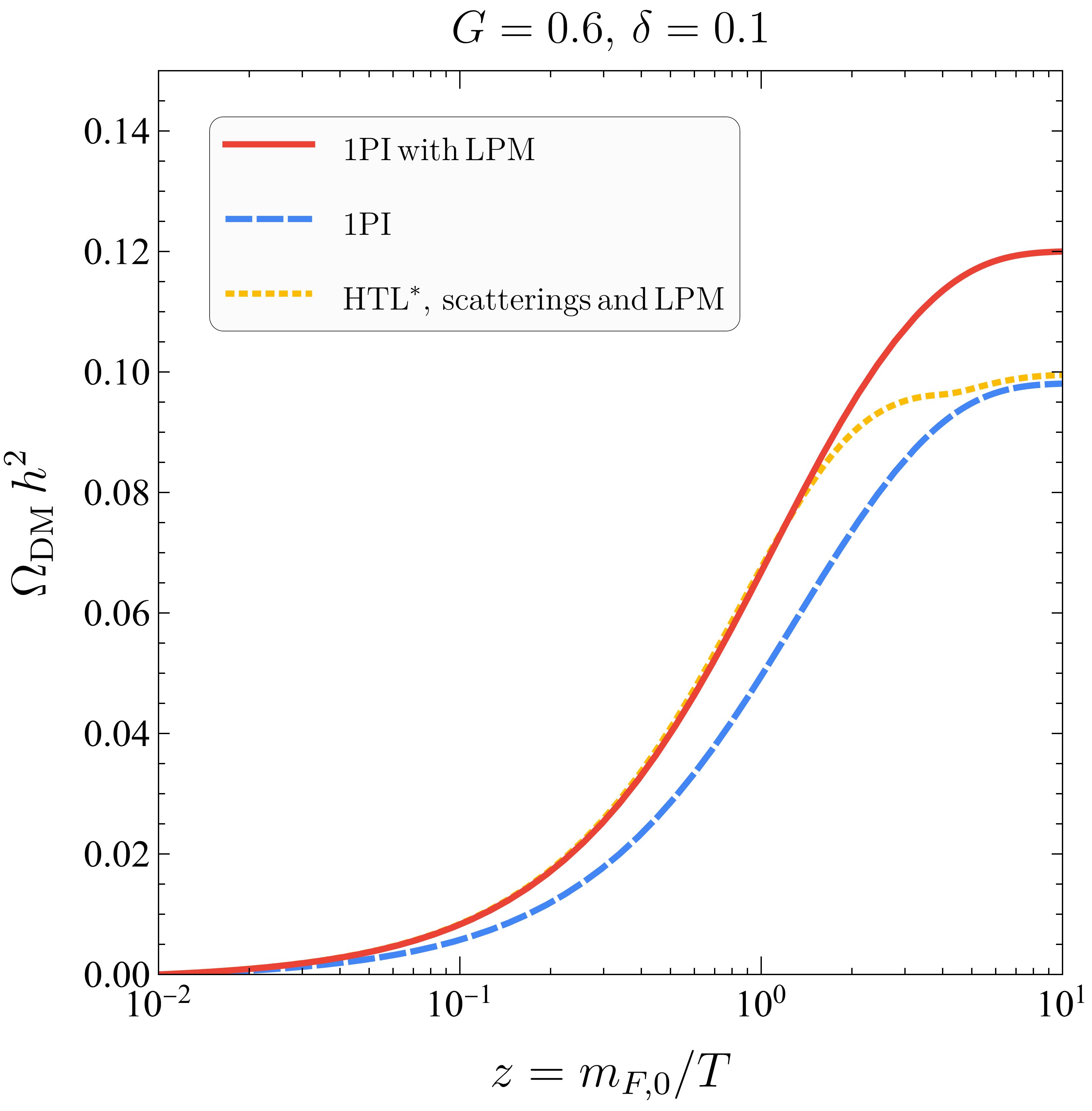}
     \end{subfigure}
     \begin{subfigure}[b]{0.49\textwidth}
         \centering
         \includegraphics[width=\textwidth]{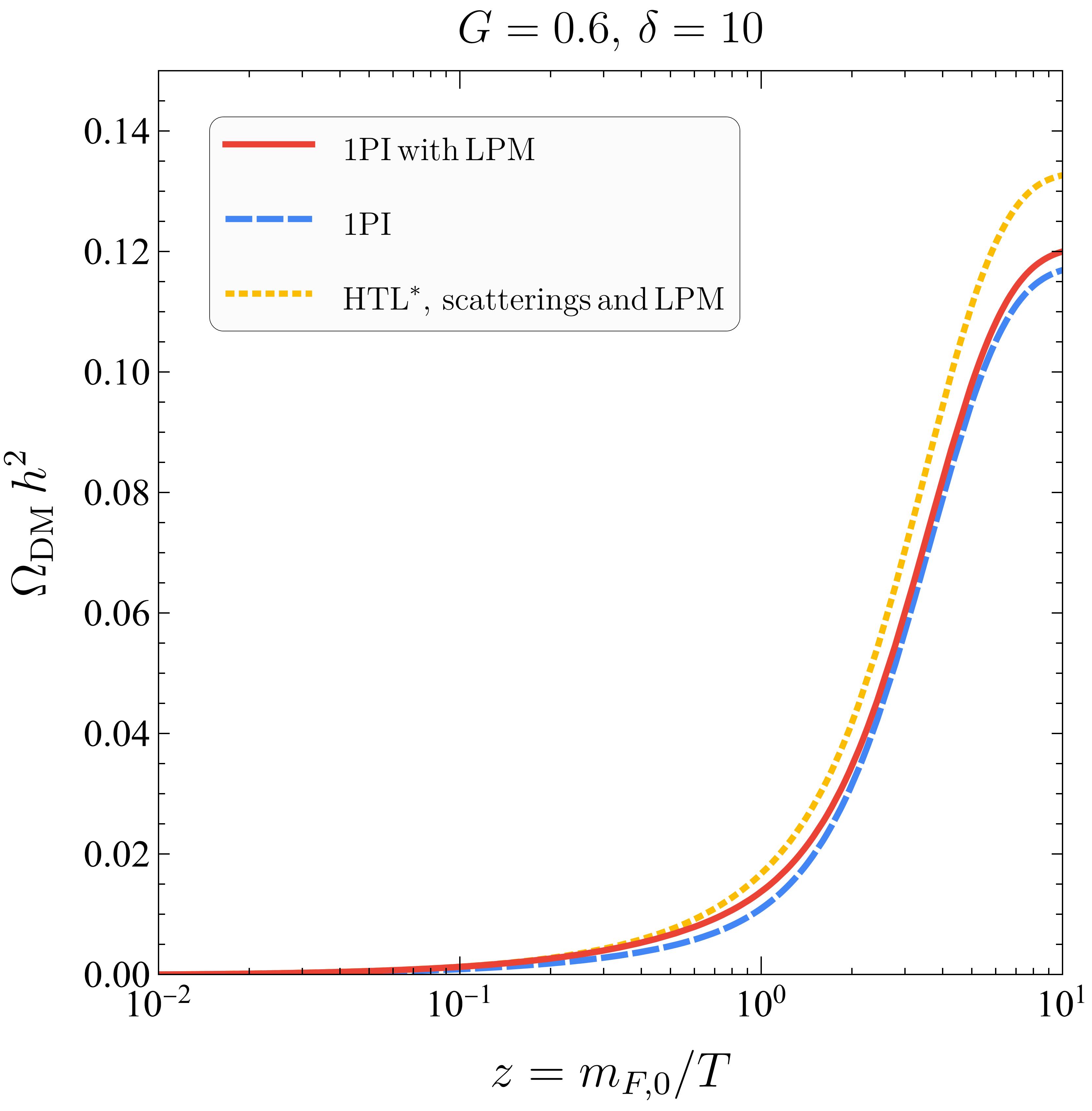}
     \end{subfigure}
        \caption{Time evolution of the DM abundance. The particle mediator is of the type $d_R$.}
        \label{fig:RelicDensityProgression}
\end{figure*}

The relic abundance at time $z$ can be written as
\begin{align}
\Omega_{\mathrm{DM}}h^2(z) &= 0.12 \left( \frac{y_{\mathrm{DM}} }{2.46 \times 10^{-13} } \right)^2 \frac{ \mathcal{I}(G,\delta;z)}{1+\delta}\, , \label{eq:RelicDensityZ}
\end{align}
where $\mathcal{I}(G,\delta;z)=\int_0^z dz' \gamma_{\mathrm{DM}} /(T^4 y_{\mathrm{DM}}^2)$ is the integrated DM rate. 
The time evolution of the DM relic density is shown in Fig.~\ref{fig:RelicDensityProgression}, calculated with the rates also shown previously in Fig.~\ref{fig:1PILPMRate}.

Fig.~\ref{fig:RelicDensityProgression} demonstrates that the key features identified in our discussion of the interaction rate are directly reflected in the DM relic density.
In particular, as the mass splitting $\delta$ decreases, the phase space available for decays becomes increasingly suppressed. 
Consequently, scattering processes and the LPM contribution dominate the dynamics. 
This is evident from the comparison between the red and blue curves, where neglecting the LPM effect leads to a systematic underestimation of the rate---a discrepancy that grows for smaller $\delta$ and larger gauge coupling $G$.

\begin{figure*}[ht]
     \centering
     \begin{subfigure}[c]{0.49\textwidth}
         \centering
         \includegraphics[width=\textwidth]{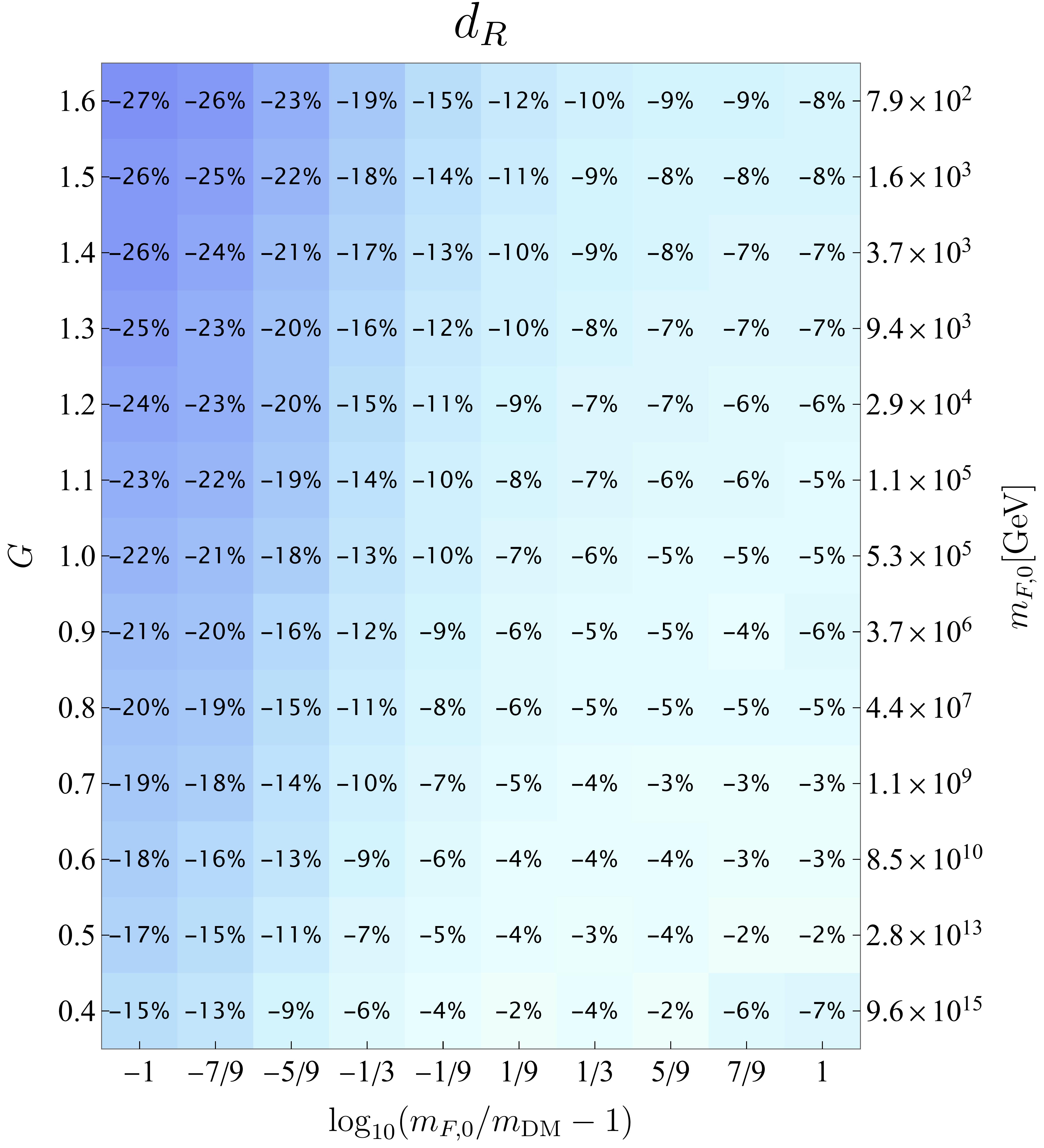}
     \end{subfigure}
     \begin{subfigure}[c]{0.49\textwidth}
         \centering
         \includegraphics[width=\textwidth]{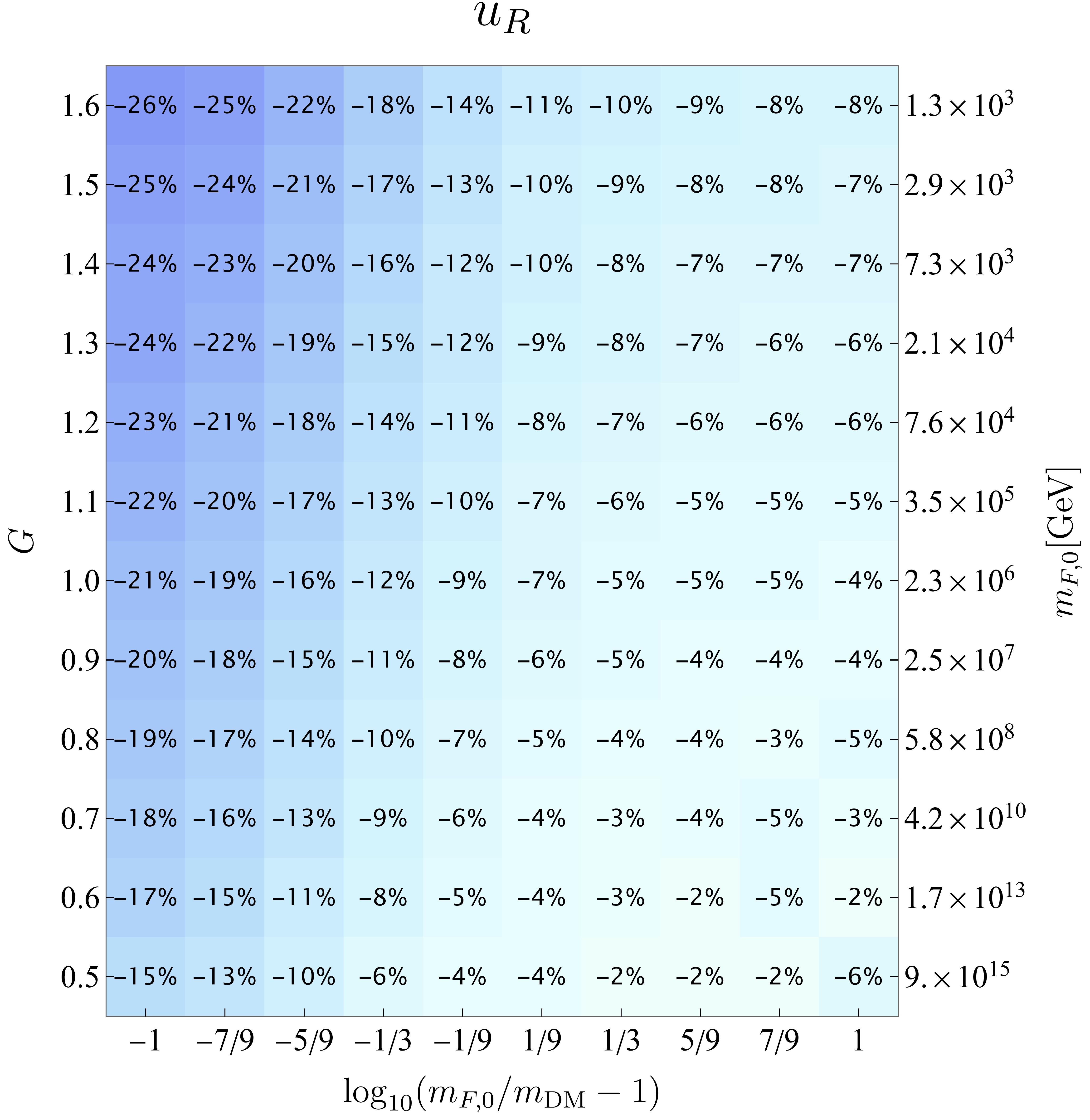}
     \end{subfigure} \\
     \vspace{0.3cm}
     \begin{subfigure}[c]{0.49\textwidth}
         \centering
         \includegraphics[width=\textwidth]{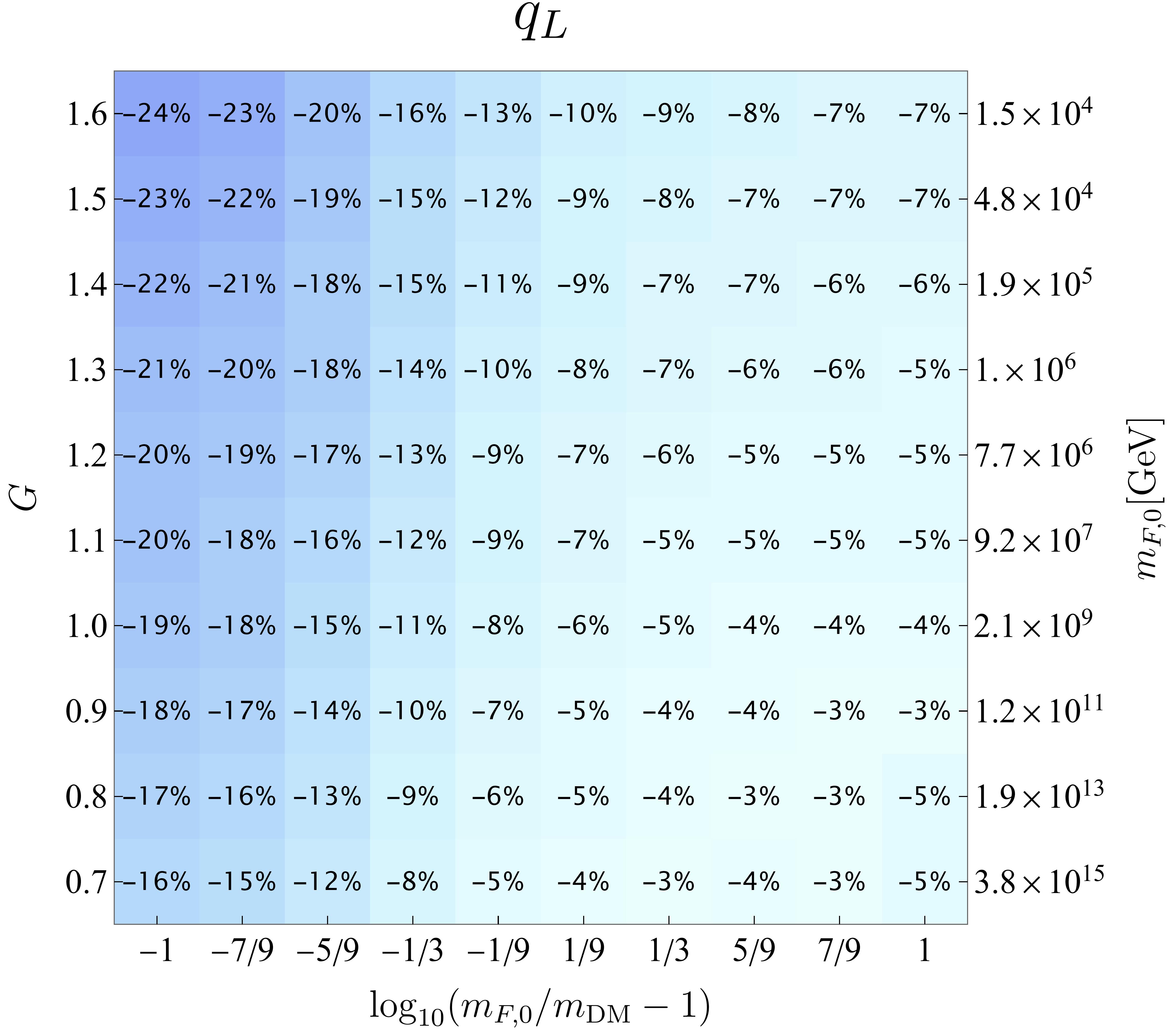}
     \end{subfigure}
     \begin{subfigure}[t]{0.49\textwidth}
         \centering
         \includegraphics[width=\textwidth]{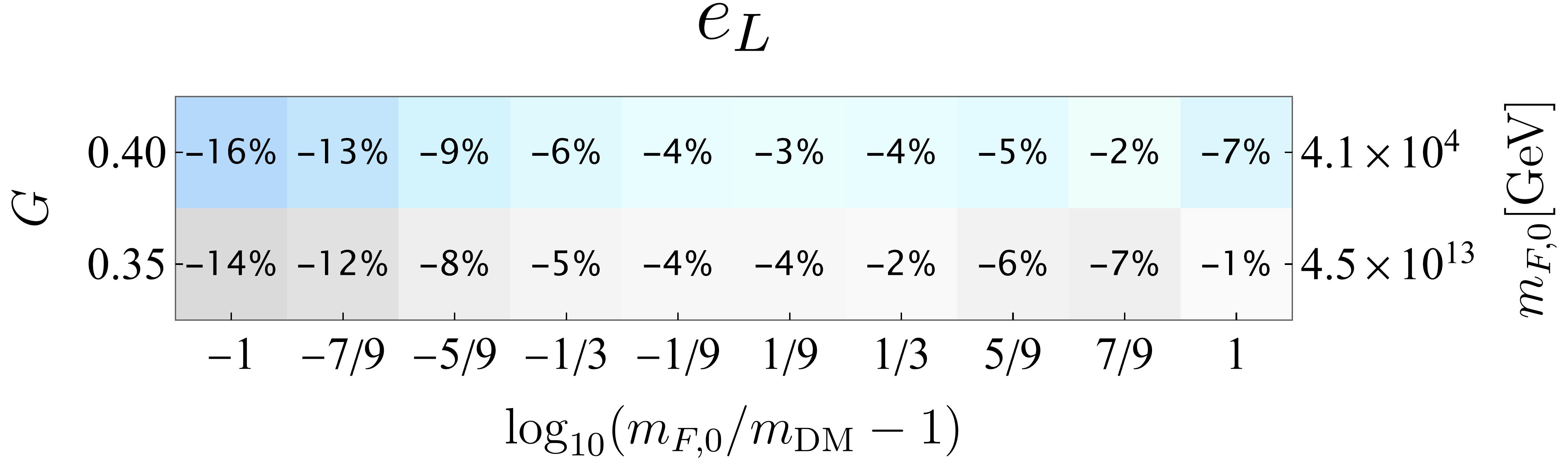} \vspace{1.5cm}
         \includegraphics[width=\textwidth]{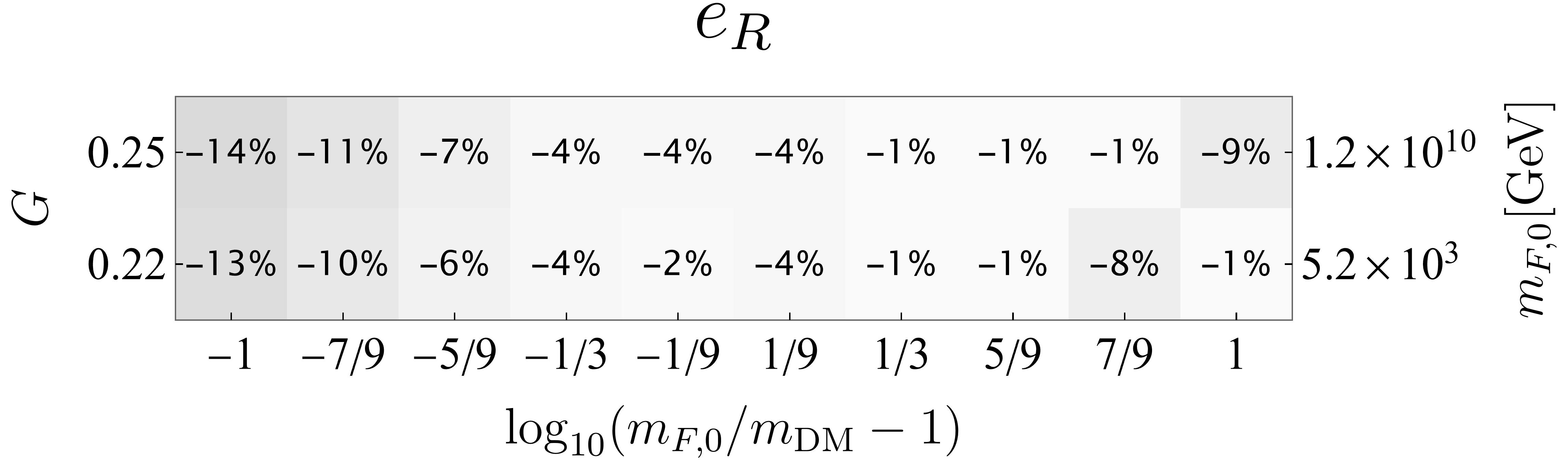}
     \end{subfigure}
    \caption{Relative impact of the LPM effect on the total relic abundance, based on the DM rate calculated within the 1PI approach, i.e., $\left(\Omega_{\mathrm{DM}}h^2\right)^{\mathrm{1PI}} / \left(\Omega_{\mathrm{DM}}h^2\right)^{\mathrm{1PI \, with \, LPM}} - 1$. The gauge coupling $G$ is evaluated at the energy of the mass of the parent particle $m_{F,0}$. For values of $G < 0.4$, indicated in gray colors, our results rely on a extrapolation of the 1PI results taken from Ref.~\cite{Copello}.}
        \label{fig:Beach1}
\end{figure*}

To quantify this deviation in terms of the relic density, Fig.~\ref{fig:Beach1} presents results across a the $\delta$--$G$ parameter space for our five model realizations. 
The gauge couplings were chosen such that the corresponding fermion mass $m_{F,0}$ remains within $m_{F,0} \in \left[ M_Z, 10^{16}~\mathrm{GeV} \right]$\footnote{The upper limit is an order of magnitude estimate set to ensure that DM production dominantly occurs during the radiation-dominated era after inflationary reheating without conflicting with CMB constraints~\cite{BICEP:2021xfz,Haque_2020} on the tensor-to-scalar ratio. Throughout our calculation the gauge coupling is evaluated at the scale $\mu=m_{F,0}$.}. 
Our analysis confirms  the increasing importance of the LPM contribution for smaller mass splittings  across all model realizations: for a color-driven scenario, the LPM effect can contribute as much as $27\%$ of the scalar DM relic abundance when $\delta = 0.1$, and around $8\%$ when $\delta = 10$. In contrast, colorless models, which are characterized by a smaller gauge coupling, exhibit a reduced impact from the LPM process.

Similarly, the previously discussed differences between the 1PI rate (including the LPM effect) and the HTL-based approach, is reflected in the evolution of the DM abundance, comparing the red and yellow curves in Fig.~\ref{fig:RelicDensityProgression}. 
For large mass splittings, the HTL-based method predicts a higher relic density than the 1PI approach, while for small mass splittings it predicts a lower one. 
We attribute this behavior to the phenomenological switch-off applied to the scattering contributions in the HTL-based scheme, which delays the onset of UV scattering rate suppression but enhances its strength relative to the 1PI calculation that accounts for the vacuum mass scale $m_{F,0}$ at all stages of the calculation. 
We have quantified these deviations for our five model realizations across the $\delta$--$G$ grid in Fig.~\ref{fig:Beach1PIvsHTL} (see Appendix~\ref{appendix}). 
Overall, the HTL-based approach deviates from the 1PI results by approximately $-20\%$ for small mass splittings and $+10\%$ for large mass splittings. 

\begin{figure*}[ht]
    \centering
    \begin{subfigure}[c]{0.49\textwidth}
        \centering
        \includegraphics[width=\textwidth]{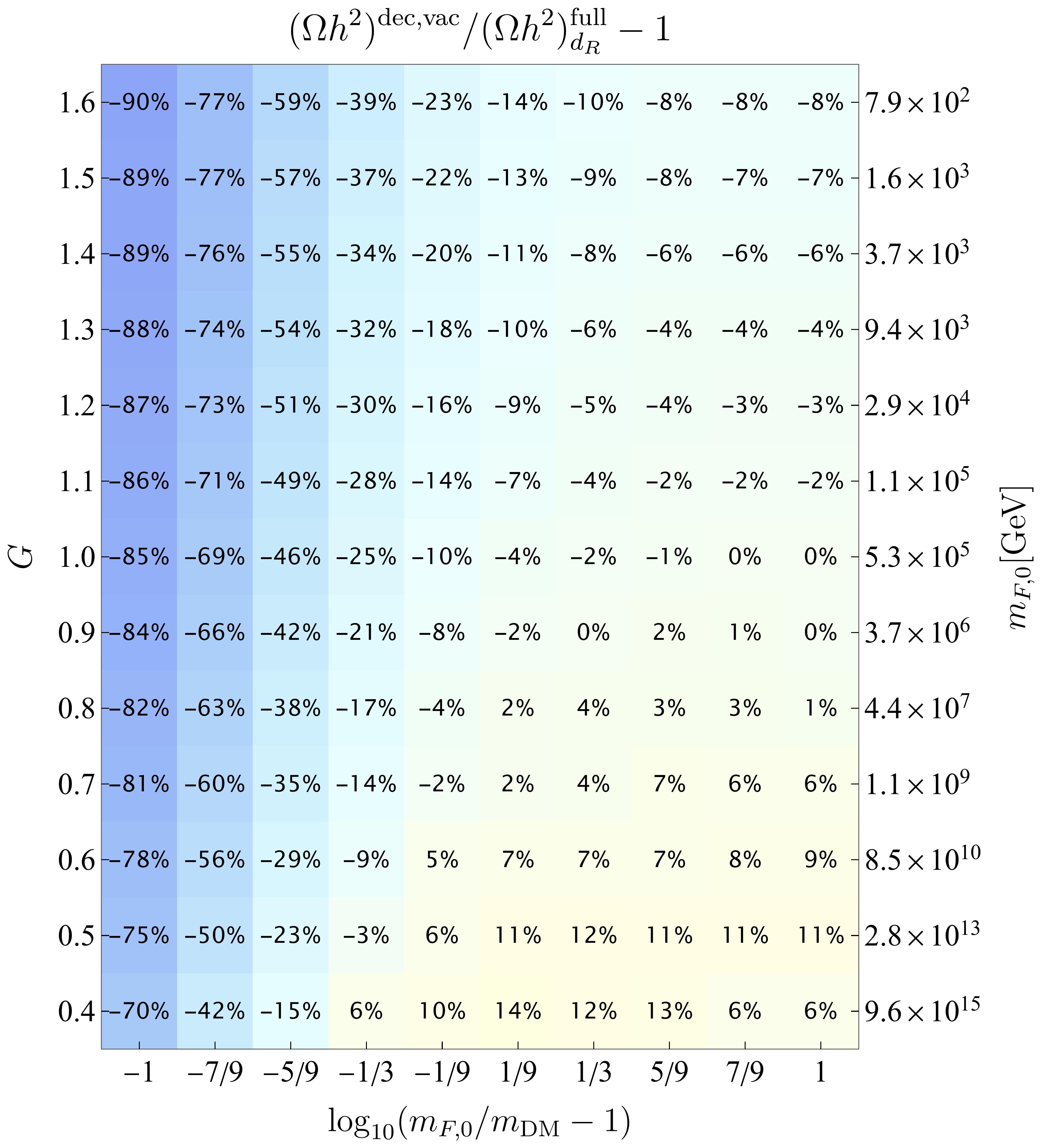}
        \caption*{(i) Vacuum‐mass decays}
    \end{subfigure}
    \begin{subfigure}[c]{0.49\textwidth}
        \centering
        \includegraphics[width=\textwidth]{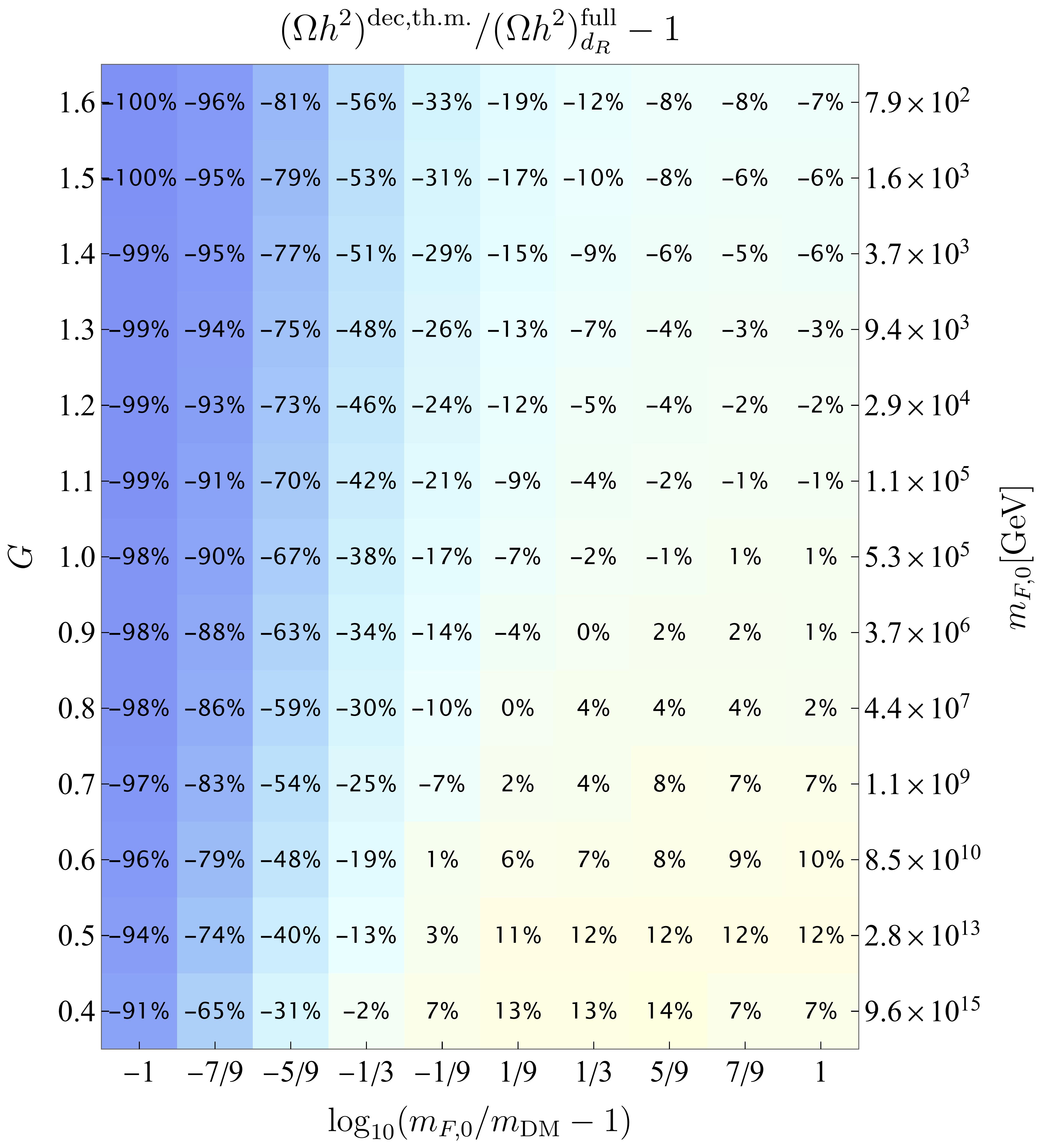}
        \caption*{(ii) Thermal‐mass decays}
    \end{subfigure} \\[0.3cm]
    \begin{subfigure}[c]{0.49\textwidth}
        \centering
        \includegraphics[width=\textwidth]{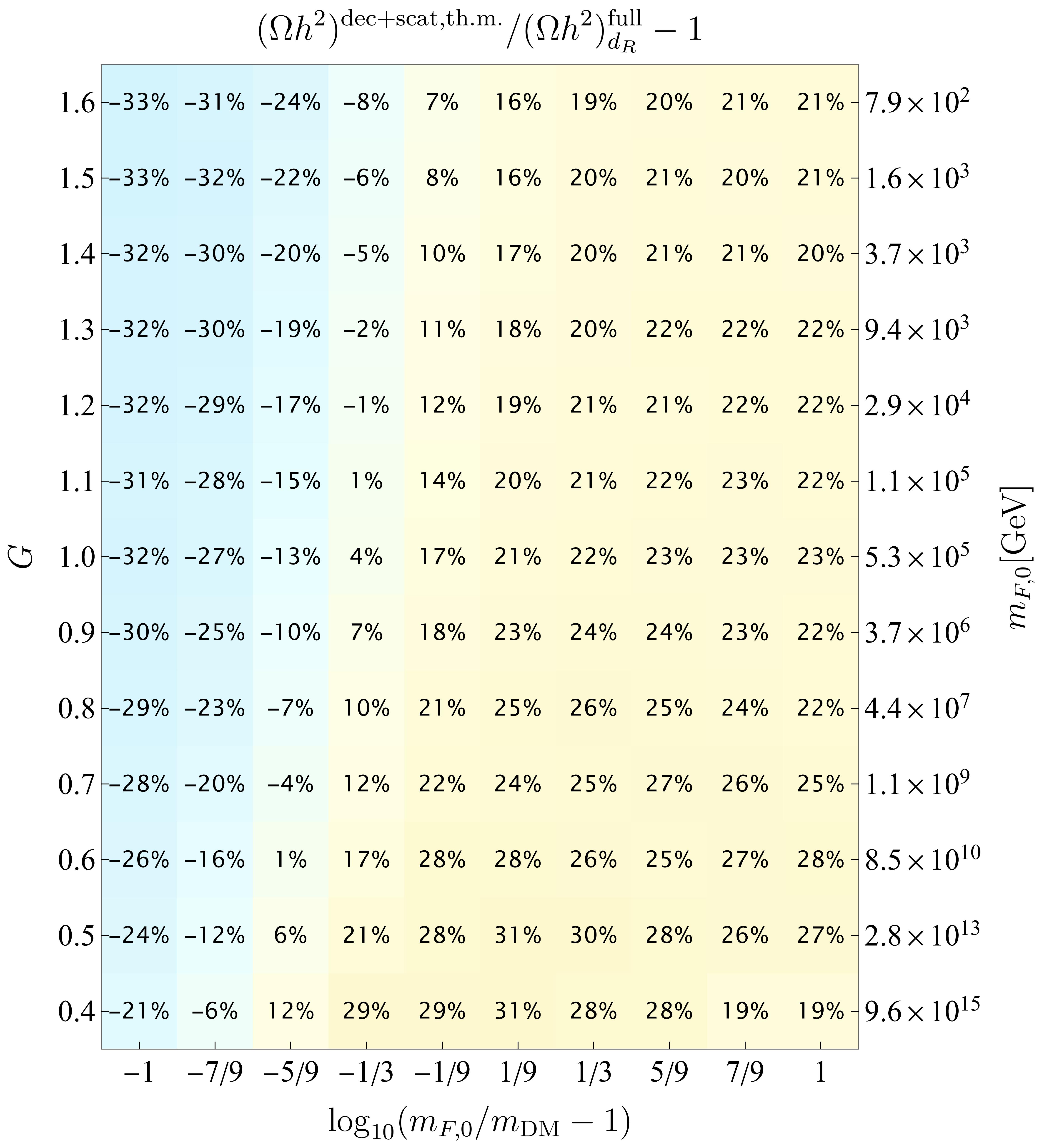}
        \caption*{(iii) Thermal‐mass decays + scatterings}
    \end{subfigure}
    \begin{subfigure}[c]{0.49\textwidth}
        \centering
        \includegraphics[width=\textwidth]{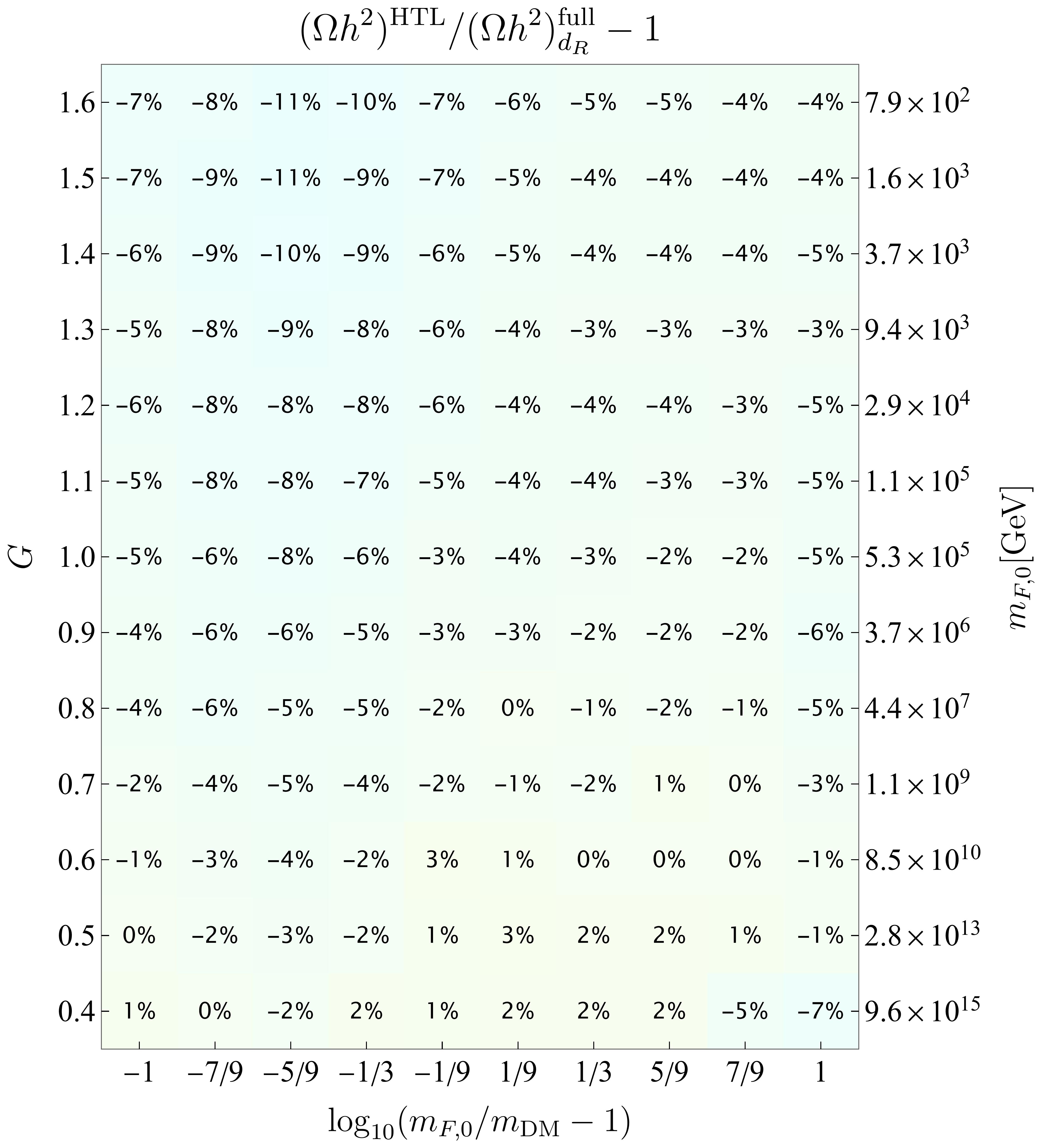}
        \caption*{(iv) HTL‐resummed 1PI rate}
    \end{subfigure}
    \caption{Relative discrepancy between each approximation scheme (i–iv) and our full 1PI based DM production rate including the LPM effect.  Panels (i)–(iv) correspond, respectively, to vacuum‐mass decays; thermal‐mass decays; thermal‐mass decays plus t-channel–regulated scatterings; and the HTL-resummed 1PI interaction rate.}
    \label{fig:BeachAltMethods}
\end{figure*}

We conclude this section by presenting the updated comparison—now including the LPM effect via the prescription in Eq.~\eqref{eq:Prescription}—between our 1PI-based production rate and the four approximation schemes previously considered in Ref.~\cite{Copello}. 
Specifically, we compare:
\begin{enumerate}[label=(\roman*)]
  \item \textbf{Vacuum-mass decays}: semi-classical Boltzmann equations including only decays with zero-temperature masses;  
  \item \textbf{Thermal-mass decays}: semi-classical Boltzmann equations including only decays but using momentum-independent thermal masses;  
  \item \textbf{Thermal-mass decays + scatterings}: as in (ii), plus $2\to2$ scatterings whose $t$-channel divergences are regulated by thermal masses;  
  \item \textbf{HTL-approximated 1PI rate}: the interaction rate derived from the DM self-energy at one-loop level using HTL–approximated, 1PI-resummed propagators\footnote{Note that this method is distinct from the earlier discussed "$\text{HTL}^*$, scatterings and LPM" rate.}
\end{enumerate}
We want to emphasize that methods (i)–(iii) are commonly used in phenomenological freeze-in DM studies.
For full definitions of these four schemes we refer the reader to the end of Sec.~2 in Ref.~\cite{Copello}. 
The resulting relative corrections for the relic density calculated from the methods (i)-(iv) with respect to the relic abundance calculated from the \textit{1PI with LPM} scheme are displayed in Fig.~\ref{fig:BeachAltMethods}.

The two decay-only schemes (methods (i) and (ii)) perform well in the regime of large mass splittings ($\delta \gtrsim 1$). In this case, they differ from the 1PI+LPM result by less than approximately $10\%$, typically underestimating the relic abundance at small gauge couplings and overestimating it at larger ones. 
However, as expected, both approaches substantially underestimate the relic density in the regime of small mass splittings and large gauge couplings, with deviations reaching up to $100\%$. 
Introducing momentum-independent thermal masses without including scatterings, as in method (ii), further worsens the prediction in this regime.

Method (iii), which incorporates thermal-mass-regulated $2 \to 2$ scatterings, improves upon the decay-only approaches for small mass splittings. 
The deviations from the full result range from an underestimate of approximately $30\%$ at $\delta = 0.1$ to an overestimate of around $25\%$ as $\delta$ increases to 10. 
Method (iv), based on HTL-resummed propagators but excluding LPM contributions, achieves the best overall agreement. 
Across the full parameter space considered, it underestimates the 1PI+LPM result by no more than roughly $10\%$. Interestingly, this level of accuracy appears somewhat accidental. 
As previously observed in Ref.~\cite{Copello}, the pure HTL rate tends to overestimate the full 1PI result. 
This overestimate is (over)compensated by complementing the 1PI rate with LPM rate---an effect not included in method (iv).

We find that the best choice of approximation depends on the size of the mass splitting. 
For large to moderate splittings ($\delta \gtrsim 1$), including only vacuum decays in the semi-classical Boltzmann equation yields relic abundances that are already close to the 1PI+LPM result. 
In this regime, even the HTL-resummed rate does not offer a significant improvement over the much simpler decay-only treatments. 
For small splittings ($\delta \lesssim 1$), however, thermal scatterings must be included to capture the relevant dynamics: 
The semi-classical Boltzmann based-approach (iii) still results in an underestimate of up to $30\%$ relative to the full result. The HTL-approximated 1PI rate (iv), however, deviates from the 1PI+LPM rate by no more than $10\%$ across all parameters.

\FloatBarrier
\section{Conclusion} \label{sec:conclusion}

In this work, we considered a scenario in which dark matter is a Standard Model (SM) singlet, feebly interacting with a SM fermion and a vector-like fermion beyond the SM that shares the same gauge charges.
Since freeze-in dynamics occur at temperatures $T \gtrsim M$, where $M$ is the largest mass scale in the model, it is crucial to incorporate finite temperature corrections to the DM interaction rate.
Previous work~\cite{Copello} calculated the interaction rate via the DM self-energy using 1PI-resummed propagators, ensuring that all vacuum mass scales were accounted for---a key point as freeze-in dynamics are most relevant when $T \sim M$, a regime where neither ultra-relativistic nor non-relativistic approximations hold. 
However, that approach did not yet include all leading-order ultra-relativistic contributions, such as the LPM effect.

Here, we have incorporated the LPM contribution into the DM production rate computed using 1PI-resummed propagators. 
To this end, we derive for the first time an equation for the LPM rate of scalar.
We then use this result to extend the 1PI-resummed analysis~\cite{Copello} to include the LPM contribution. 
We note that the derivation of the LPM rate is so far based on collinear dynamics—strictly valid when $T \gg M$—and therefore requires a phenomenological switch-off as the temperature approaches the largest vacuum mass scale. Until today there exists no calculation of the LPM effect accounting for vacuum mass scales, which will be part of a follow-up work.
To quantify the uncertainty from this ad hoc switch-off, we considered three distinct implementations: (i) based on the fermionic susceptibility as in Ref.~\cite{Biondini}, (ii) via the derivative of a fermionic thermal loop function with respect to the vacuum mass squared, and (iii) through an interpolation between the ultra-relativistic LPM rate and the Born-level decay rate based on~\cite{Ghiglieri_2022}.
We furthermore assess how the phenomenological switch-off applied to the ultra-relativistic $2 \leftrightarrow 2$ scattering rate compares to the 1PI-resummed result that includes the vacuum mass at all stages of the calculation.

We find that different phenomenological switch-off prescriptions introduce an uncertainty of circa $30 \%$ ($5 \%$) in the final relic density obtained from the LPM-extended 1PI approach for small (large) mass splittings.
Among these, we identified the fermionic-loop function (ii) as the most conservative choice and adopted it throughout the remainder of this work.
Moreover, we compare the LPM-extended 1PI approach with an HTL-based method that also switches off the $2\leftrightarrow2$ scattering rates by hand ($\mathrm{HTL}^*$+scatterings+LPM).
Since the LPM contribution is treated identically in both frameworks, this comparison allows to investigate the differences between the 1PI and $\mathrm{HTL}^*$+scatterings approaches.
We observe discrepancies ranging from approximately $+10\%$ for large mass splittings to $-20\%$ for small mass splittings.  

We have calculated the LPM contribution over a wide range of model parameters, specifically scanning the effective gauge coupling $G \in [0.2, 1.6]$ and the mass splitting $\delta \in [0.1, 10]$. 
Unlike the 1PI-resummed one-loop result, the LPM rate depends explicitly on the charge assignments of the particles involved; hence, we have provided results for five specific realizations: DM coupling to $q_L$, $e_L$, $u_R$, $d_R$, and $e_R$. 
Our findings demonstrate that including the LPM effect in the calculation of the dark matter production rate leads to a substantial enhancement of the predicted relic density—by as much as \(8\%-27\%\) for color-charged mediators—with a milder but still relevant effect for electroweak mediators. 
This enhancement is reduced for smaller effective gauge couplings and larger mass splittings.
This highlights the importance of consistently incorporating all leading-order ultra-relativistic processes. 
Besides providing a new state-of-the-art computation, which for the first time combines the rate derived using 1PI-resummed propagators with the full LPM effect, we have comprehensively and quantitatively compared our results to several phenomenological approaches commonly used in the literature (e.g. to various semi-classical Boltzmann approaches).

Remarkably, we find that Boltzmann equations including only in-vacuum decay contributions already achieve percent-level accuracy for large mass splittings.
At smaller mass splitting, the inclusion of scatterings is required and integrating the resulting Boltzmann equations results in an underestimate of up to $30\%$.
Relying on a HTL-approximated 1PI-rate without the LPM contribution is accidentally at least $\sim 10 \%$ accurate across all the parameter space analyzed.

Finally, a consistent approach for calculating the LPM rate, fully accounting for all vacuum mass scales (both for the parent particle and the DM) remains desirable. 
The observed differences introduced by the different manual switch-off procedures highlight the need for a more systematic treatment, in particular to obtain an accurate calculation in the for freeze-in relevant regime $T \sim M$.  
We will address this task in a follow-up work.


\FloatBarrier
\acknowledgments
All authors acknowledge support by the Cluster of Excellence “Precision Physics, Fundamental Interactions, and Structure of Matter” (PRISMA$^+$ EXC 2118/1) funded by the Deutsche Forschungsgemeinschaft (DFG, German Research Foundation) within the German Excellence Strategy (Project No. 390831469).
M.~B. and J.~H. acknowledge support from the Emmy Noether grant "Baryogenesis, Dark Matter and Neutrinos: Comprehensive analyses and accurate methods in particle cosmology" (HA 8555/1-1, Project No. 400234416) funded by the Deutsche Forschungsgemeinschaft (DFG, German Research Foundation).
Furthermore, this work is supported in part by the Italian MUR Department of Excellence grant 2023-2027 “Quantum Frontiers”.
M.~B. is supported by Istituto Nazionale di Fisica Nucleare (INFN) through the Theoretical Astroparticle Physics (TAsP) project.

Parts of this research were conducted using the supercomputer MOGON NHR and advisory services offered by Johannes Gutenberg University Mainz (hpc.uni-mainz.de), which is a member of the AHRP (Alliance for High Performance Computing in Rhineland Palatinate,  www.ahrp.info) and the Gauss Alliance e.V. The authors gratefully acknowledge the computing time granted on the supercomputer MOGON NHR at Johannes Gutenberg University Mainz (hpc.uni-mainz.de).
\vfill

\pagebreak
\appendix

\FloatBarrier
\section{Fermionic propagators in the on-shell collinear limit}
\label{app:Weyl_spinors}

For a hard momentum $k={\cal O}(T)$, the resummed fermionic propagator in the imaginary-time formalism can be written in momentum space as~\cite{Besak_2012}
\begin{align}
    \slashed{S}(k)=-\frac{\slashed{k}-\frac{m^2_\infty}{2k_\parallel}\gamma^0}{k^2-m^2_\infty},
\end{align}
with $k^0=2\pi i\left(n+\frac{1}{2}
\right)\pi T,n\in\mathbb{Z}$, and where $m^2_\infty$ is the asymptotic thermal mass. In terms of the $2\times 2$ matrices $\sigma^\mu=(\mathbb{I},\vec{\sigma}),\bar\sigma^\mu=(\mathbb{I},-\vec{\sigma})$, where $\vec{\sigma}$ are the ordinary Pauli matrices, one has
\begin{align}
    \slashed{k}=\gamma^\mu k_\mu=\left[\begin{array}{cc}0 & \sigma^\mu k_\mu\\
    \bar\sigma^\mu k_\mu & 0
    \end{array}\right].
\end{align}
The matrices $\sigma^\mu k_\mu$, $\bar\sigma^\mu k_\mu$ admit a spectral decomposition in terms of projectors constructed from  normalized spinors $\eta(k), \xi(k)$. These spinors are defined as eigenstates of the helicity operator $\vec{\sigma}\cdot \hat{k}$, 
\begin{align}\label{eq:etaxi}\begin{aligned}
\vec{\sigma} \cdot \hat{k}\, \eta(k) &= -\eta(k), \\
\vec{\sigma} \cdot \hat{k}\, \xi(k) &= +\xi(k),
\end{aligned}\end{align}
where \(\hat{k} = \vec{k}/|\vec{k}|\) is the unit vector in the direction of the momentum \(\vec{k}\). From Eq.~\eqref{eq:etaxi}, the following spectral decompositions follow:
\begin{align}\begin{aligned}
    \sigma^\mu k_\mu =&\,\sigma^0 k^0-\vec{\sigma}\cdot\vec{k}=|\vec{k}|\left(\frac{k^0}{|\vec{k}|}\,\mathbb{I}-\vec{\sigma}\cdot\hat{k}\right)\\
   = &\,|\vec{k}|\left(\frac{k^0}{|\vec{k}|}\,+1\right)\eta(k)\eta(k)^\dagger+|\vec{k}|\left(\frac{k^0}{|\vec{k}|}-1\right)\xi(k)\xi(k)^\dagger,\\
    \bar\sigma^\mu k_\mu =&\,\sigma^0 k^0+\vec{\sigma}\cdot\vec{k}=|\vec{k}|\left(\frac{k^0}{|\vec{k}|}\,\mathbb{I}+\vec{\sigma}\cdot\hat{k}\right)\\
   = &\,|\vec{k}|\left(\frac{k^0}{|\vec{k}|}\,-1\right)\eta(k)\eta(k)^\dagger+|\vec{k}|\left(\frac{k^0}{|\vec{k}|}+1\right)\xi(k)\xi(k)^\dagger.
\end{aligned}\end{align}
Specializing now into the kinematic region relevant for the LPM effect, i.e. nearly on-shell and collinear momenta with $k^2={\cal O}(g^2T^2)$ and $|\vec{k}|=k_\parallel(1+{\cal O}(g^2))$, which imply   $k^0=|k_\parallel|(1+{\cal O}(g^2))$, the previous results can be approximated as
\begin{align}\begin{aligned}
    \sigma^\mu k_\mu\approx &\,2k_\parallel\eta(k)\eta(k)^\dagger,\\
    \bar\sigma^\mu k_\mu \approx &\,2k_\parallel\xi(k)\xi(k)^\dagger,
\end{aligned}\end{align}
resulting in a propagator 
\begin{align}\label{eq:DiracProp}
    \slashed{S}(k)\approx -\frac{2k_\parallel}{k^2-m^2_\infty}\left[\begin{array}{cc}0 & \eta(k)\eta(k)^\dagger\\
    \xi(k)\xi(k)^\dagger & 0
    \end{array}\right]\equiv D(k) \left[\begin{array}{cc}0 & \eta(k)\eta(k)^\dagger\\
    \xi(k)\xi(k)^\dagger & 0
    \end{array}\right].
\end{align}
Finally, in this kinematic region one can also obtain simple solutions for the spinors $\eta(k),\xi(k)$. Choosing for simplicity the momentum vector \(\vec{k}\) to lie along the \(z\)-axis, i.e., \(k_3 = k_\parallel\),
inserting the explicit form of the Pauli matrices and solving the eigenvalue equations, we obtain the following normalized Weyl spinors:
\begin{align}\label{eq:WeylSpinors}\begin{aligned}
\eta(k) &= \begin{pmatrix} -\dfrac{k_1 - i k_2}{2 k_\parallel} \\ 1 \end{pmatrix},  \\
\xi(k) &= \begin{pmatrix} 1 \\ \dfrac{k_1 + i k_2}{2 k_\parallel} \end{pmatrix}. 
\end{aligned}\end{align}
These spinors enter the hard vertex factor $\Phi(k,k-p)$ of Eq.~\eqref{eq:hardVertexFactor}, which appears in the integral equation \eqref{eq:LPM_Equation} for the self-energy that captures the LPM effect. The vertex factor is defined by writing the one-loop self-energy diagram  as\\
\vskip0.1cm
\begin{minipage}{0.28\textwidth}
        \includegraphics[width=\textwidth]{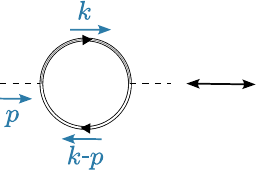}
\end{minipage}\begin{minipage}{0.72\textwidth}
\begin{align}\label{eq:PiTree}\begin{aligned}
    &-\Pi(p)=-y_{\rm DM}^2T\,{\rm tr}\SumInt\,\frac{d^4k}{(2\pi^4)} \slashed{S}(k)P_L\slashed{S}(k-p)\\
   & \equiv -T\SumInt\,\frac{d^4k}{(2\pi^4)} \,\Phi^\dagger(k,k-p)D(k)D(k-p)\Phi(k,k-p),
\end{aligned}\end{align}
\end{minipage}
\vskip0.2cm
\noindent where we considered the case of a DM coupling to a left-handed SM fermion. Above, $\SumInt$ denotes the integration over the 3-momentum $\vec{k}$, together with the sum over the discretized values of $k^0$ in the imaginary time formalism. $D(k)$ are scalar quantities introduced in Eq.~\eqref{eq:DiracProp}. It follows that the vertex factors are absorbing the Yukawa couplings from the ordinary Feynman rules, plus the spinor structure from the numerators of the fermionic propagators. Substituting Eq.~\eqref{eq:DiracProp} in the first line of Eq.~\eqref{eq:PiTree}, one finds
\begin{align}
    \Phi(k,k-p)=y_{\rm DM}\eta^\dagger(k-p)\xi(k).
\end{align}
Substituting Eq.~\eqref{eq:WeylSpinors} and choosing the DM momentum aligned with the $z$ direction, such that $p_1=p_2=0$, one obtains the result of Eq.~\eqref{eq:hardVertexFactor}.
For a DM coupling to a right-handed fermion, corresponding to a projector $P_R$ instead of $P_L$ in the first line of Eq.~\eqref{eq:PiTree}, one finds instead
\begin{align}
     \Phi(k,k-p)=y_{\rm DM}\eta^\dagger(k)\xi(k-p),
\end{align}
and upon substitution of Eq.~\eqref{eq:WeylSpinors} the result is minus the one in Eq.~\eqref{eq:hardVertexFactor}. An overall change in $\Phi$ does not affect the DM production rate, which can be seen as follows. First, note that if the reduced self-energy $\hat{\Pi}$ solves the recursion relation of Eq.~\eqref{eq:LPM_Equation} with a given hard vertex function $\Phi$, then $-\hat{\Pi}$ solves the same equation with a hard vertex function $-\Phi$. Hence a change in sign in ${\Phi}$ leads to a change in sign in $\hat{\Pi}$. But the DM production rate  is proportional to $\Pi^{\mathcal A}_s$ which involves $\Phi \,\hat\Pi$, c.f.~Eq.~\eqref{eq:Pi_Pihat}. The latter  does not change under $\Phi\rightarrow-\Phi$, from which one concludes that one can use the result of Eq.~\eqref{eq:hardVertexFactor} for DM coupling to either left or right-handed fermions in the SM.

\FloatBarrier
\section{Expression for the interpolation rate}
\label{appendixB}
In this section, we present the explicit expression for the smoothly interpolated LPM rate, denoted by $\gamma_{\mathrm{DM}}^{\mathrm{Int}}$. This rate interpolates between the LPM result in the ultra-relativistic regime and the tree-level decay rate in the non-relativistic regime. The construction and justification of this interpolation have been thoroughly discussed in Section~\ref{sec:switchoff}; here, we simply provide the resulting expression.
The interpolated LPM contribution to the spectral function is given by
\begin{align}
\Pi^{\mathcal{A},\mathrm{LPM}}_s (p_0) = \int \frac{dk_{0}}{4\pi} \left[f_{+}(k_{0}) - f_{+}(k_0 - p_0)\right] 
\lim_{y_{\perp} \to 0} \mathbb{P} \left\{ \frac{\tilde{\Theta}(\epsilon_a, y_{\perp}^2)}{\epsilon_a (\omega - \epsilon_a)} \right\},
\end{align}
where energy-like variables $\epsilon_a$ and $\omega$ are introduced for consistency with the notation of Ref.~\cite{Ghiglieri_2022}, as in Eq.~\ref{eq:LPMScalarDifferential2}. The function $\tilde{\Theta}$ denotes an interpolated splitting kernel that reproduces the LPM behavior in the ultra-relativistic limit and matches the tree-level decay rate in the non-relativistic regime.

We define the interpolated splitting function $\tilde{\Theta}$ as
\begin{align}
    \tilde{\Theta} &= \frac{p_0^2}{2k_0(k_0 - p_0)} \, \mathrm{Im} \left[\nabla_{\perp} \cdot \vec{f}(\vec{y}_{\perp}) \right] \nonumber \\
    &\quad + \Bigg[ \frac{p_0^2}{2 p^2} m_{\mathrm{DM}}^2 
    - \frac{(p_0 - k_0)^2 + k_0^2}{2 k_0 p_0} \left(m_{F,0}^2 +  \frac{\left(m_f^2 - m_F^2 - m_{\mathrm{DM}}^2\right)^2}{4 p^2} \right) \nonumber \\
    &\qquad + \frac{(p_0 - k_0)^2 + k_0^2}{2 p_0 (k_0 - p_0)} \left(m_{q,0}^2 + \frac{\left(m_F^2 - m_f^2 - m_{\mathrm{DM}}^2\right)^2}{4 p^2} \right) 
    - \frac{\left(m_F^2 - m_f^2\right)^2 - m_{\mathrm{DM}}^4}{4 p^2} \nonumber \\
    &\qquad - \frac{k_0 p_0}{p^2} \left(m_f^2 - m_F^2\right) 
    + \frac{p_0^2}{2 p^2} \left(m_f^2 - m_F^2 - m_{\mathrm{DM}}^2 \right) \Bigg] \mathrm{Im} \left[g(\vec{y}_{\perp}) \right],
\end{align}
where the term $\mathrm{Im} \left[\nabla_{\perp} \cdot \vec{f}(\vec{y}_{\perp}) \right]$ is obtained by solving the differential equation in Eq.~\eqref{eq:LPMScalarDifferential2}, with the modified dispersion relation $\epsilon(\vec{k}_{\perp})$ introduced in Eq.~\eqref{eq:SmoothOperator}.

An interesting feature of this expression is the appearance of the function $\mathrm{Im}\left[g(\vec{y}_{\perp})\right]$, which does not arise in the scalar LPM self-energy but does appear in the LPM expressions for photon and fermion self-energies. To compute $\mathrm{Im}[g(\vec{y}_{\perp})]$, one must solve the inhomogeneous equation
\begin{equation}
    \left( \hat{H} - i 0^+ \right) g(\vec{y}_{\perp}) = \delta^{(2)}(\vec{y}_{\perp}),
    \label{eq:LPMScalarDifferential3}
\end{equation}
where the Hamiltonian $\hat{H}$ is defined in Eq.~\eqref{eq:polelocations2}.

\FloatBarrier
\section{Expressions for the HTL-based approach}
\label{appendix}
In this appendix, we present the expressions for the decay and scattering rates relevant for our comparison with the results of Ref.~\cite{Biondini}, where fermionic dark matter production was studied. The decay rate is computed using the UV limit of the HTL propagators, while the scattering rate is evaluated in the ultra-relativistic limit, assuming all particles are massless. Both rates contribute to the total production rate as given in Eq.~\eqref{eq:PrescriptionGhiglieri}.

\subsection{Decay contribution}

The decay term is obtained by evaluating the self-energy using the UV limit of the HTL propagators. This limit reproduces the correct dispersion relation in the regime $p \gg gT$. We denote the resulting spectral self-energy as $\Pi_s^{\mathcal{A},\mathrm{HTL*}}$, and its expression is given by
\begin{align}
\Pi_s^{\mathcal{A},\mathrm{HTL*}} = \frac{1}{16 \pi p} \int_{k_{\min}}^{k_{\max}} dk & \Bigl[ m_F^2 + m_f^2 - m_{\mathrm{DM}}^2 - 2 E_k \left( E_{ |\vec{k}-\vec{p}|} - |\vec{k}-\vec{p}| \right) \nonumber \\
& - 2 E_{ |\vec{k}-\vec{p}|} \left( E_k - |\vec{k}| \right) \Bigr] \left[ -f_+\left(E_{ |\vec{p}-\vec{k}|} \right) + f_+\left(E_k\right) \right], \label{eq:BornRate}
\end{align}
where $E_k = \sqrt{k^2 + m_F^2}$ and $E_{ |\vec{p}-\vec{k}|} = \sqrt{ |\vec{p}-\vec{k}|^2 + m_f^2}$ are the energies of the mediator and the SM particle, respectively. The integration boundaries $k_{\min}$ and $k_{\max}$ are derived using Gram determinants and are identical to those given in Eq.~\eqref{eq:kBoundaries}.

\subsection{Scattering processes}

\begin{figure*}[ht]
    \centering
    \includegraphics[width=0.9\textwidth]{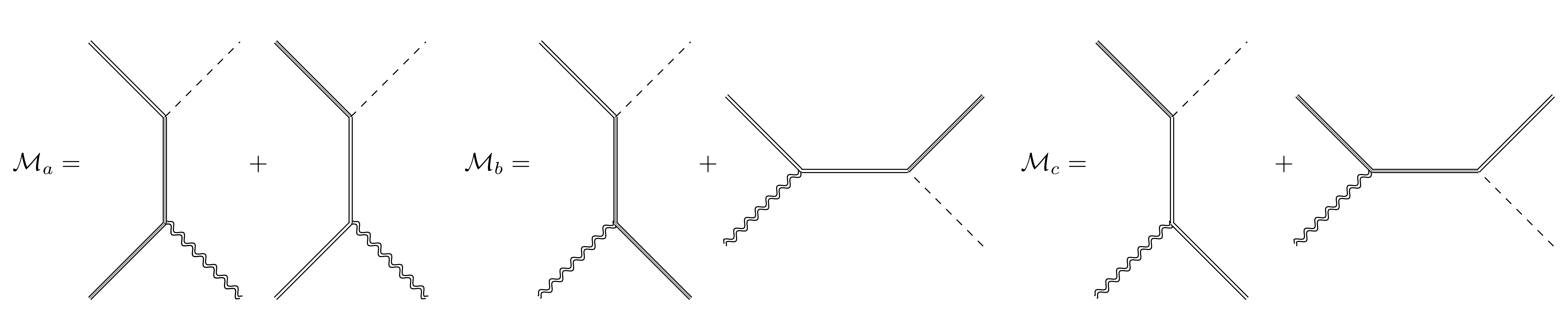}
    \caption{The $s$- and $t$- channel scattering diagrams that contribute to the production of the scalar singlet $s$. Fermion flow arrows are omitted.}
    \label{fig:scatterings}
\end{figure*}

The scattering processes from Fig.~\ref{fig:scatterings} are calculated following the approach of Refs.~\cite{Ghiglieri_2016, Biondini}. This approach uses the LO equivalence between the spectral self-energy $\Pi^{\mathcal{A}}(p)$ and the Boltzmann equation where all momenta are hard. This approach neglects both in-vacuum and thermal masses, and thus is only valid in the ultra-relativistic regime, requiring to be manually switched-off in the relativistic and non-relativistic regimes. The $t$-channel divergence is corrected through a HTL resummation in the $t$-channel term: one substracts the \textit{would-be} HTL contribution and subsequently adds the \textit{soft} contribution, as done in Ref.~\cite{Ghiglieri_2016}. For the phase-space integrals we use the parametrization given in Ref.~\cite{Besak_2012}. To the best of our knowledge, the expressions for the scattering processes producing a scalar particle using this approach has not been presented anywhere before.

We proceed to present the final results and direct the reader to the previous references for further explanation as to how to derive these results. The leading-order contribution to the $2\leftrightarrow 2$ scatterings for a hard momentum exchange is given by
\begin{align}
f_{-}(p^0) \Pi^{\mathcal{A}}\biggr\rvert^{\mathrm{hard}}_{2\leftrightarrow2} &=\frac{G}{2}\int d\Omega_{2\leftrightarrow2} \biggl\{ f_{+}(k_1^0) f_{+}(k_2^0)  [1+f_{-}(k_3^0)] \left(-\frac{2(s+t)}{t}-\frac{2t}{s+t}+4\right) \nonumber \\
&-2 f_{+}(k_1^0) f_{-}(k_2^0) [1-f_{+}(k_3^0)] \left(\frac{2s}{s+t}+\frac{4t}{s}+\frac{2(s+t)}{s} -\frac{4(s+t)}{s}\right) \biggr\},
\end{align}
where $d\Omega_{2\leftrightarrow 2}$ is the phase space measure together with 4-momentum conservation, as defined in Ref.~\cite{Ghiglieri_2016}. The momenta $k_1$, $k_2$ stand for the incoming particles, whereas $k_3$ stands for the outgoing particle and $p$ for our DM scalar $s$. In the massless limit, $s+t+u=0$, which allows us to write this contribution in just terms that depend on $s$ and $t$. Next, one introduces the $s$-channel parametrization, where the integration variable is chosen to be $q=k_3+p$ so that $s=q^2$. This applies to the terms where $\Sigma |\mathcal{M}|^2$ is proportional to $u/s$. For the $t$-channel parametrization, we choose $q=k_1-k_3$ so that $t=q^2$. After performing the subtraction prescription of the $t$-channel divergence introduced in Ref.~\cite{Ghiglieri_2016}, the expression for the scattering processes is given by
\begin{align}
\Pi^{\mathcal{A}}\biggr\rvert_{2\leftrightarrow 2} &=\Pi^{\mathcal{A}}\biggr\rvert_{2\leftrightarrow 2}^{\mathrm{hard}} -\Pi^{\mathcal{A}}\biggr\rvert_{2\leftrightarrow 2}^{\mathrm{HTL,expanded}} +\Pi^{\mathcal{A}}\biggr\rvert_{2\leftrightarrow 2}^{\mathrm{HTL,soft}}  \nonumber \\
&=\frac{2G}{(4\pi)^3p^0} \int_{p^0}^{\infty} dq_+ \int_0^{p^0} dq_- \left[-f_{+}\left(q^0\right) + f_{+}\left(q^0-p^0\right)\right] \left( 2 \Phi_{s1} + \Phi_{s2} \right) \nonumber \\
&+ \frac{2G}{(4\pi)^3p^0}  \int_0^{p^0} dq_+ \int_{-\infty}^{0} dq_-\Bigg\{\left[1-f_{+}\left(q^0\right) - f_{+}\left(p^0-q^0\right)\right] \left( \Phi_{t1} + \Phi_{t2} \right)  \nonumber \\
&  - \frac{3}{2} \left[\frac{1}{2}-f_{+}\left(p^0\right)\right] \frac{p^0 \pi^2 T^2}{q^2} \Bigg\} \nonumber \\
&+\frac{3}{2} \frac{m_f^2}{16\pi} \left[\frac{1}{2}-f_{+}\left(p^0\right)\right] \ln\left(1 + \frac{4\left(p^0\right)^2}{m_f^2} \right) + \mathcal{O}\left(\frac{m_f^4}{\left(p^0\right)^3} \right),\label{eq:ScatteringsPrescription}
\end{align}
where $q_{\pm} = \left(q^{0} \pm |\vec{q}|\right)/2$.
Here, the fermionic $s$-channel and fermionic $t$-channel exchange functions are
\begin{align}
\Phi_{s1} = & \frac{T}{q} \left[ \left(p^0-q_-\right) \left( \ln_f^-(q_+)-\ln_b^+(q_-)\right) + \left(p^0-q_+\right) \left( \ln_f^-(q_-) - \ln_b^+(q_+) \right) \right]  \nonumber \\
& +\frac{T^2}{q^2} \left(2p^0-q^0\right) \left[ \mathrm{li}_{b}^+(q_-)-\mathrm{li}_f^-(q_+) - \mathrm{li}_b^+(q_+) + \mathrm{li}_f^-(q_-) \right] ,\\
\Phi_{s2}=& \frac{T}{q} \left[ \left(p^0-q_-\right) \left( \ln_b^+(q_+)-\ln_f^-(q_-) \right) + \left(p^0-q_+\right)\left( \ln_b^+(q_-)- \ln_f^-(q_+) \right) \right]  \nonumber \\
& +\frac{T^2}{q^2} \left(2p^0-q^0\right) \left[ \mathrm{li}_b^+(q_+) - \mathrm{li}_f^-(q_-) -\mathrm{li}_b^+(q_-) + \mathrm{li}_f^-(q_+) \right] ,\\
\Phi_{t1}=& \frac{T}{q} \left( p^0 - q_- \right) \left[ \ln_f^-(q_+) - \ln_b^+(q_-) \right] + \frac{T^2}{q^2} \left(2p^0 - q^0\right) \left[ \mathrm{li}_b^+(q_-) - \mathrm{li}_f^-(q_+) \right], \\
\Phi_{t2}=& \frac{T}{q} \left( p^0 - q_- \right) \left[ \ln_b^-(q_+) - \ln_f^+(q_-) \right] + \frac{T^2}{q^2} \left(2p^0 - q^0\right) \left[ \mathrm{li}_f^+(q_-) - \mathrm{li}_b^-(q_+) \right],
\end{align}
where we have introduced the functions
\begin{align}
    & \ln_{f}^-(x) = \ln\left(1 + e^{-x/T}\right),  & \ln_f^{+}(x) = \ln\left(1 + e^{x/T}\right), \\
    & \ln_b^-(x) = \ln\left(1 - e^{-x/T}\right),  & \ln^+_{b}(x) = \ln\left(1 - e^{x/T}\right), \\
    & \mathrm{li}_{f}^-(x) = \mathrm{Li}_2\left(-e^{-x/T}\right),   &\mathrm{li}_{f}^+(x) =\mathrm{Li}_2\left(-e^{x/T}\right), \label{eq:LiLogarithm1}\\
    & \mathrm{li}_{b}^-(x) =\mathrm{Li}_2\left(e^{-x/T}\right),   &\mathrm{li}_{b}^+(x) =\mathrm{Li}_2\left(e^{x/T}\right). \label{eq:LiLogarithm2}
\end{align}

\subsection{Impact on the Relic Abundance}
We compare the total rate calculated based on the 1PI and LPM contributions, as defined in Eq.~\eqref{eq:PrescriptionGhiglieri}, with the $\mathrm{HTL}^*$-rate calculated presented in this section, given in Eq.~\eqref{eq:Prescription}. In the Figs.~\ref{fig:1PILPMRate} and \ref{fig:RelicDensityProgression}, both rates are shown for selected values of the gauge coupling $G$ and the mass splitting~$\delta$. In this section, we extend the comparison to the full $\delta$--$G$ parameter parameter space.

\begin{figure*}[ht]
     \centering
     \begin{subfigure}[c]{0.49\textwidth}
         \centering
         \includegraphics[width=\textwidth]{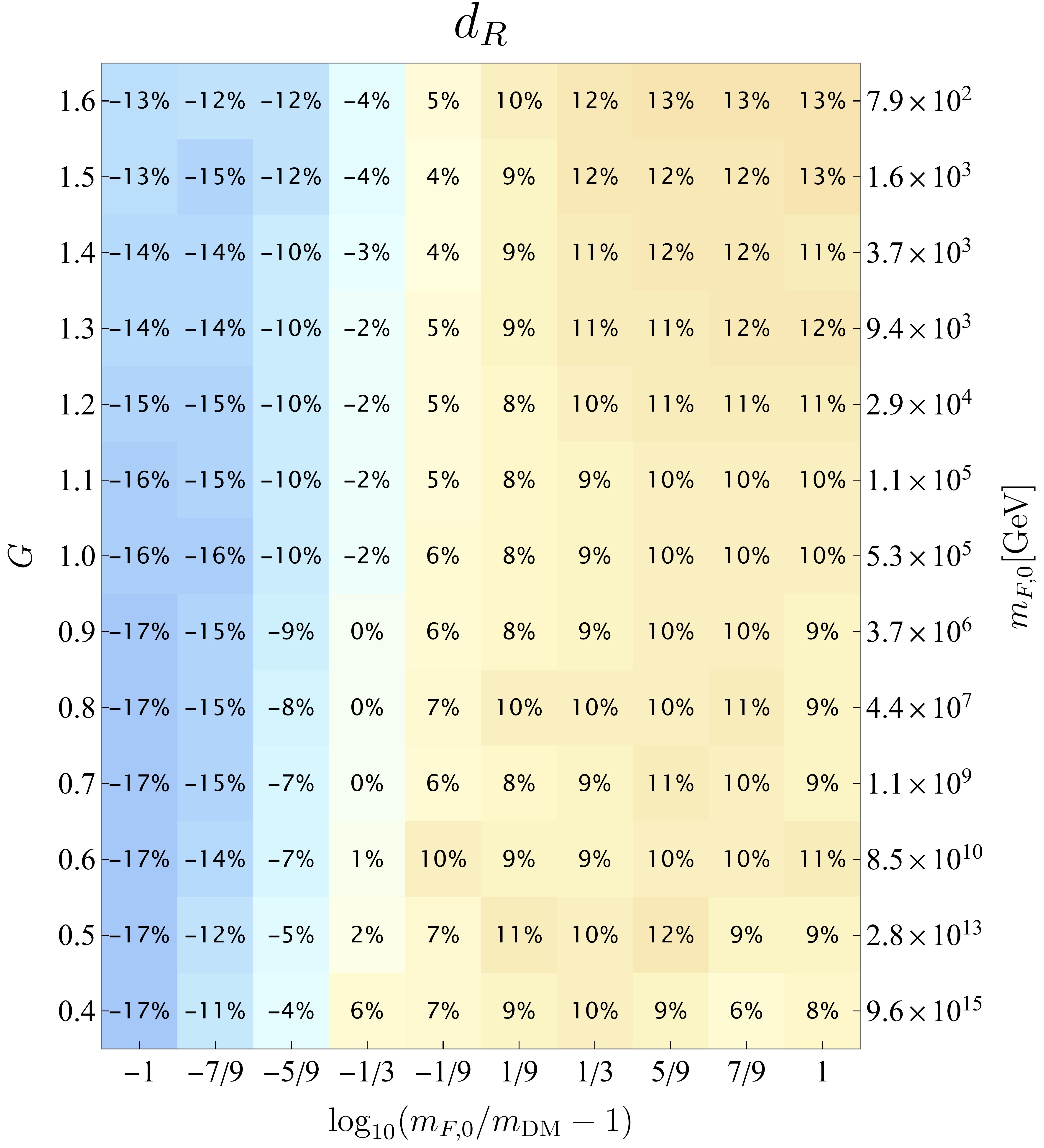}
     \end{subfigure}
     \begin{subfigure}[c]{0.49\textwidth}
         \centering
         \includegraphics[width=\textwidth]{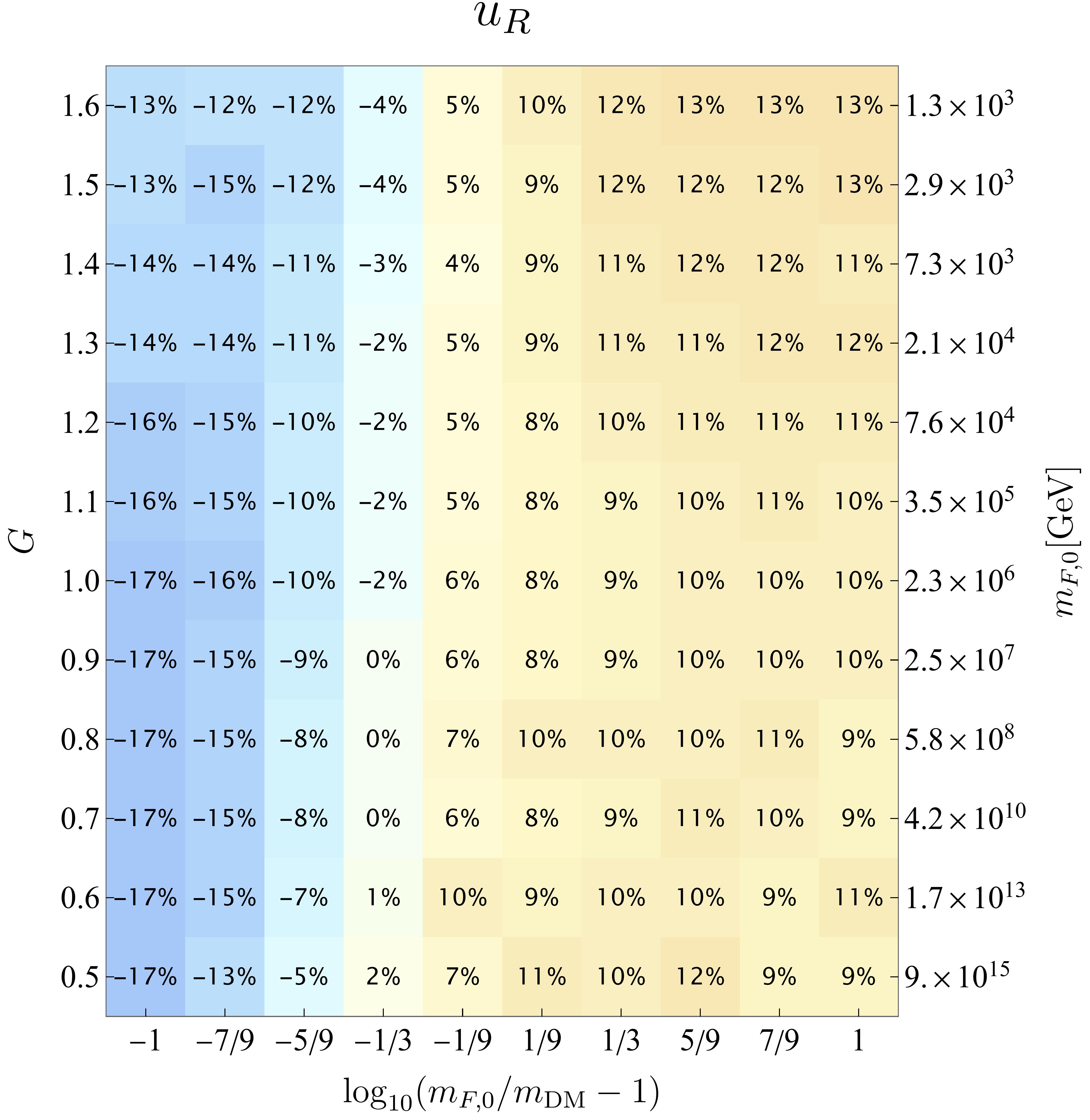}
     \end{subfigure} \\
     \vspace{0.3cm}
     \begin{subfigure}[c]{0.49\textwidth}
         \centering
         \includegraphics[width=\textwidth]{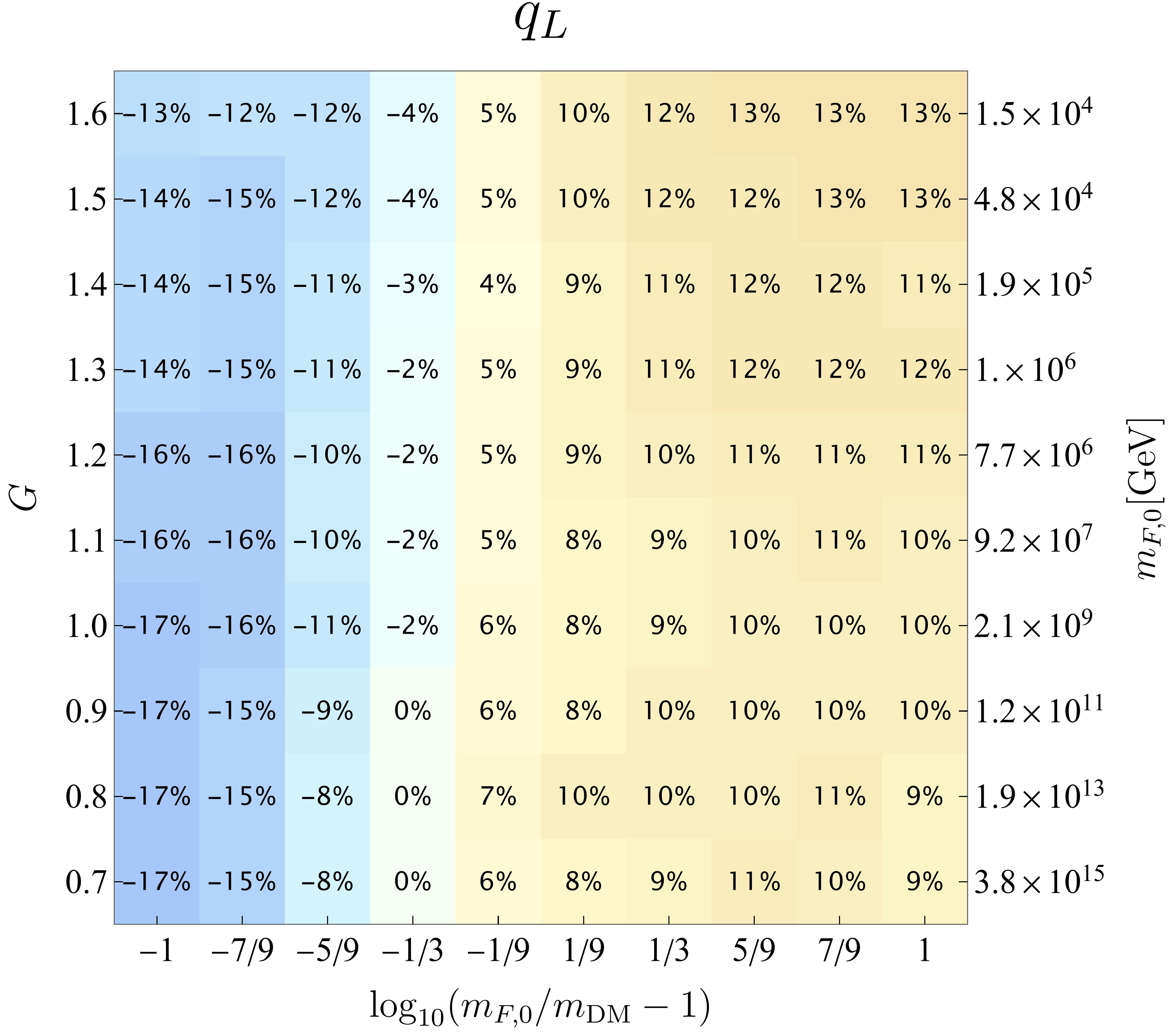}
     \end{subfigure}
     \begin{subfigure}[t]{0.49\textwidth}
         \centering
         \includegraphics[width=\textwidth]{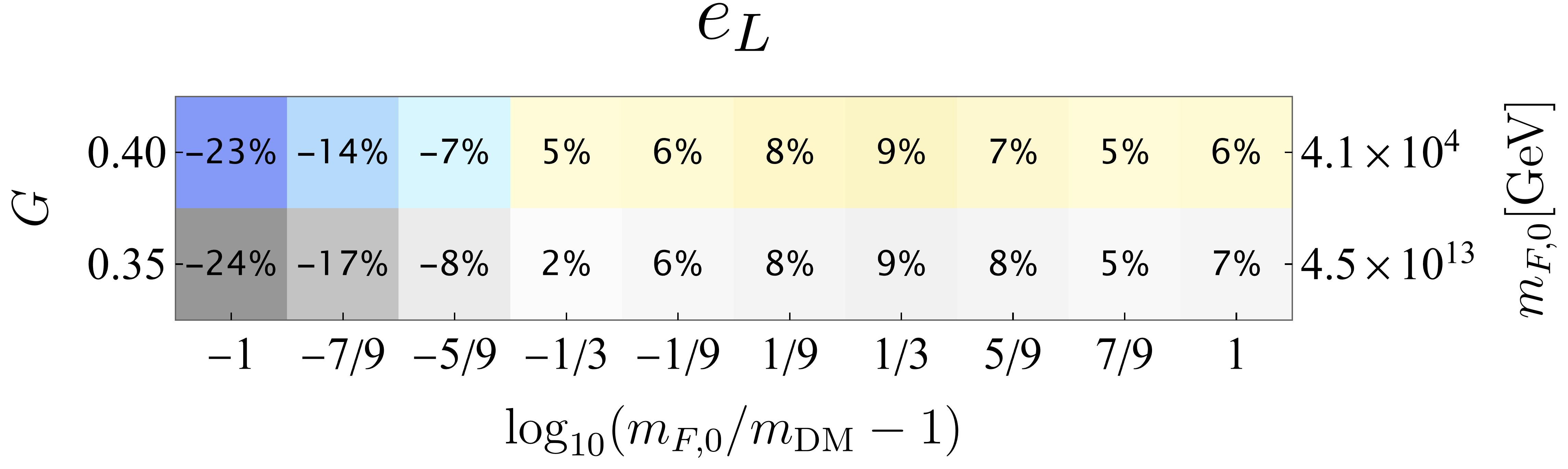} \vspace{1.5cm}
         \includegraphics[width=\textwidth]{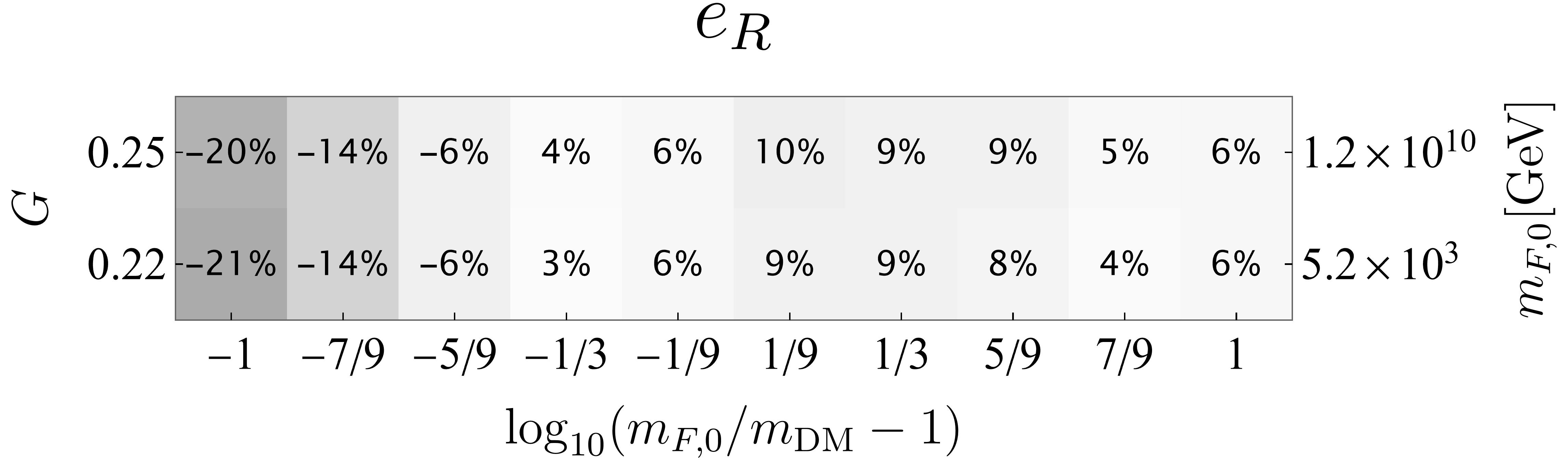}
     \end{subfigure}
    \caption{Relative difference in the final relic abundance between the rate labeled as \textit{$\textit{HTL}^*$, scatterings and LPM} and the rate labeled \textit{1PI with LPM}, i.e., $\left(\Omega_{\mathrm{DM}}h^2\right)^{\mathrm{HTL^*, \,scatterings\, and \, LPM}} / \left(\Omega_{\mathrm{DM}}h^2\right)^{\mathrm{1PI \, with \, LPM}} - 1$.}
        \label{fig:Beach1PIvsHTL}
\end{figure*}

From Fig.~\ref{fig:Beach1PIvsHTL}, we observe that the \textit{$\mathrm{\it HTL}^*$, scatterings and LPM} rate underestimates the relic abundance by approximately 20\% for small mass splittings, while it overestimates it by ca. 10\% for larger mass splittings. The dominant source of this discrepancy is the treatment of the scattering processes, which must be manually switched off. This behavior is also evident in Fig.~\ref{fig:Uncertainities2}. If, instead of the thermal function, the susceptibility function had been used to switch off the scatterings, the discrepancy at small mass splittings would be reduced. However, this comes at the cost of significantly larger deviations at larger mass splittings, where the decay channel dominates and the relic abundance peak becomes overly suppressed due to the scatterings not being turned off quickly enough.

\FloatBarrier
\section{The fit of the LPM rate}
\label{appendixC}
\begin{figure*}[ht]
    \centering
    \includegraphics[width=0.5\textwidth]{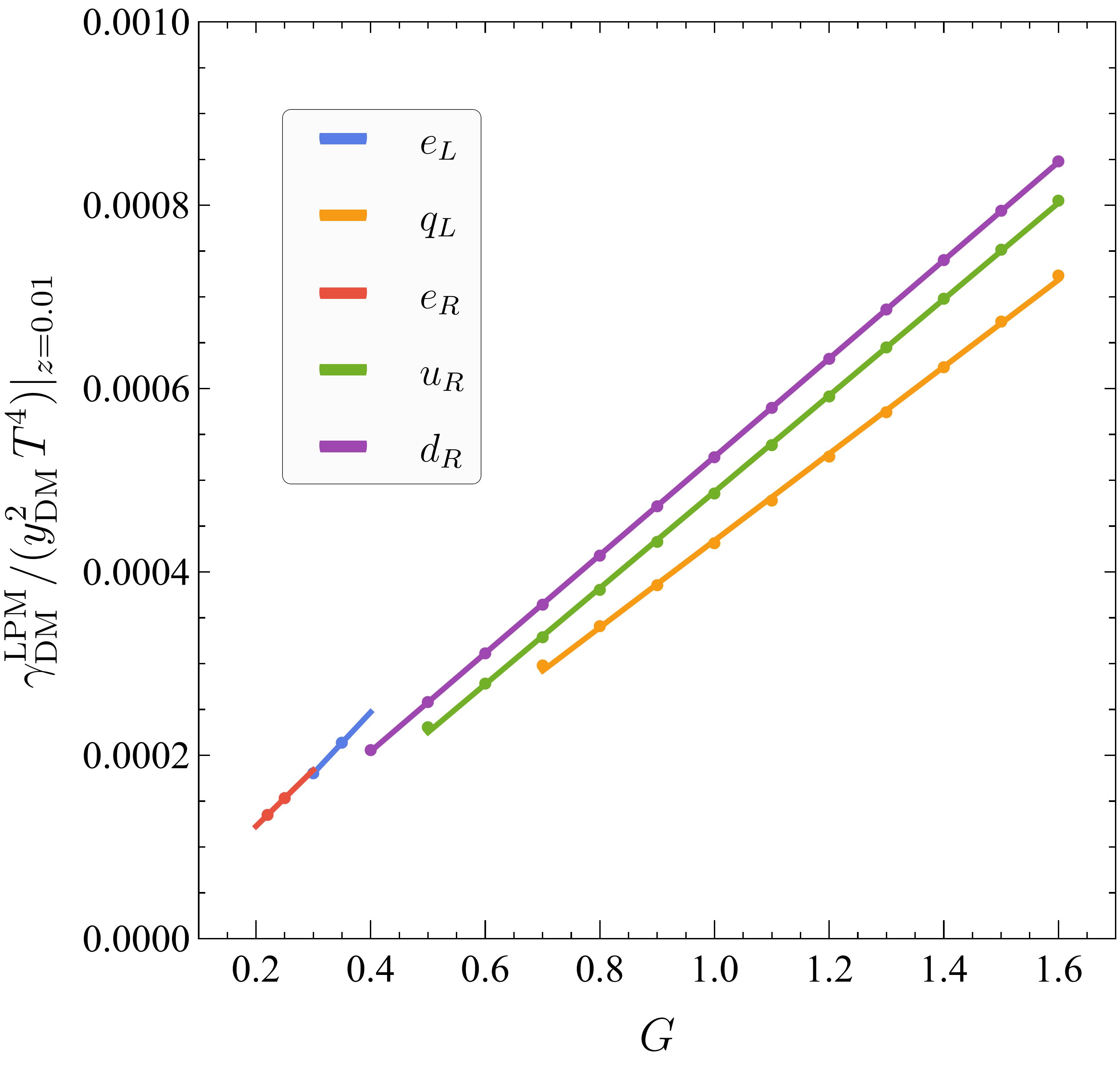}
    \caption{Values of the LPM rate evaluated at $z=0.01$ together with the linear fit.}
    \label{fig:Fit}
\end{figure*}

In this section, we present in Fig.~\ref{fig:Fit} the agreement between the linear fit from Eq.~\eqref{eq:linearfit} and the calculated data, showing that a linear fit resembles the behavior satisfactory. The validity of each fit is constrained by the range of $G$ in which the mediator mass $m_F$ is defined. As previously mentioned, these constraints are approximately set such that the mediator mass $m_F$ lies between the mass of the $Z$ boson and the reheating temperature.

\vfill

\clearpage
\FloatBarrier
\bibliographystyle{JHEP}
\bibliography{biblio.bib}

\end{document}